\documentstyle[epsfig]{cupconf}

\ifoldfss
\else
  \ifnfssone
    \newmathalphabet{\mathit}
      \addtoversion{normal}{\mathit}{cmr}{m}{it}
      \addtoversion{bold}{\mathit}{cmr}{bx}{it}
    \newmathalphabet{\mathcal}
      \addtoversion{normal}{\mathcal}{cmsy}{m}{n}
    \else
    \ifnfsstwo
    \fi
  \fi
\fi

\def\hexnumber#1{\ifcase#1 0\or1\or2\or3\or4\or5\or6\or7\or8\or9\or
 A\or B\or C\or D\or E\or F\fi }

\makeatletter
\ifx\CUP@mtlplain@loaded\undefined
\else
\fi
\makeatother
 \makeatletter
 \ifx\CUP@mtlplain@loaded\undefined
   \font\tenbmi=cmmib10 at 10pt
   \font\sevenbmi=cmmib10 at 7pt
   \font\fivebmi=cmmib10 at 5pt

   \newfam\bmifam
   \textfont\bmifam=\tenbmi
   \scriptfont\bmifam=\sevenbmi
   \scriptscriptfont\bmifam=\fivebmi
   
 \fi
 \makeatother
\ifnfsstwo

\fi
\ifnfssone

\fi
\ifoldfss

\fi

\mathchardef\varLambda="0103

\makeatletter
\ifx\CUP@mtlplain@loaded\undefined
\else
\fi
\makeatother
\makeatletter
\ifx\CUP@mtlplain@loaded\undefined
  \font\tenbms=cmbsy10
  \font\sevenbms=cmbsy10 at 7pt
  \font\fivebms=cmbsy10 at 5pt
  \newfam\bmsfam
  \textfont\bmsfam=\tenbms
  \scriptfont\bmsfam=\sevenbms
  \scriptscriptfont\bmsfam=\fivebms

  \edef\bsy@{\hexnumber\bmsfam}
  \mathchardef\bnabla="0\bsy@72
\fi
\makeatother
\title[Internet Resources for Radio Astronomy]{Internet Resources for Radio Astronomy}

\author[H. Andernach]%
{H\ls E\ls I\ls N\ls Z\ns A\ls N\ls D\ls E\ls R\ls N\ls A\ls C\ls H$^1$}

\affiliation{$^1$Depto.~de Astronom\'\i a, IFUG, Universidad de Guanajuato,
Guanajuato, C.P. 36000, Mexico \\
Email:~~{\tt heinz@astro.ugto.mx} }
\def\deg{$^{\circ}$}
\def\ltsim{\raise 2pt \hbox {$<$} \kern-0.8em \lower 2pt \hbox {$\sim$}}
\def\gtsim{\raise 2pt \hbox {$>$} \kern-0.8em \lower 2pt \hbox {$\sim$}}
\def\tts{\small\tt }

\begin{document}
\ifnfssone
\else
  \ifnfsstwo
  \else
    \ifoldfss
      \let\mathcal\cal
      \let\mathrm\rm
      \let\mathsf\sf
    \fi
  \fi
\fi

%\tableofcontents
%\newpage
%\setcounter{page}{1}

\maketitle

\vspace*{-5.3cm}
\begin{center}
\begin{footnotesize}\baselineskip 10 pt\noindent
To appear in ``Astrophysics with Large Databases in the Internet Age'' \\
Proc.~~{\it IX$^{th}$ Canary Islands Winter School on Astrophysics} \\
Tenerife, Spain, ~Nov\,17--28, 1997 \\
eds.~M.~Kidger,~I.~P\'erez-Fournon,~\& F.~S\'anchez,
Cambridge University Press, 1998
\end{footnotesize}
\end{center}
\vspace*{4.1cm}

\begin{abstract}
A subjective overview of Internet resources for radio-astronomical
information is presented. Basic observing techniques and their implications
for the interpretation of publicly available radio data are described,
followed by a discussion of existing radio surveys, their level of optical
identification, and nomenclature of radio sources. Various collections of
source catalogues and databases for integrated radio source parameters are
reviewed and compared, as well as the WWW interfaces to interrogate the
current and ongoing large-area surveys.  Links to radio observatories with
archives of raw (uv-) data are presented, as well as services providing
images, both of individual objects or extracts (``cutouts'') from
large-scale surveys.  While the emphasis is on radio continuum data, a
brief list of sites providing spectral line data, and atomic or molecular
information is included. The major radio telescopes and surveys under
construction or planning are outlined. A summary is given of a search for
previously unknown optically bright radio sources, as performed by the
students as an exercise, using Internet resources only.  Over 200 different
links are mentioned and were verified, but despite the attempt to make this
report up-to-date, it can only provide a snapshot of the current situation.
\end{abstract}

\firstsection % if your document starts with a section,
\section{Introduction}    \label{introduction }  % lesson 1  section 1

Radio astronomy is now about 65 years old, but is far from retiring.
Karl Jansky made the first detection of cosmic static in 1932,
which he correctly identified with emission from our own Milky Way.
A few years later Grote Reber made the first rough map of the northern sky
at metre wavelengths, demonstrating the concentration of emission
towards the Galactic Plane. During World War II the Sun was discovered
as the second cosmic radio source. It was not until the late 1940s that
the angular resolution was improved sufficiently to allow
the first extragalactic sources be identified: Centaurus A (NGC~5128)
and Virgo A (M~87). Interestingly, the term {\it radio astronomy} was first
used only in 1948 (\cite{hhmm96}, p.\,453, item 2).
During the 1950s it became obvious that
not only were relativistic electrons responsible for the emission,
but also that radio galaxies were reservoirs of unprecedented amounts of
energy. Even more impressive radio luminosities were derived once the
quasars at ever-higher redshifts were found to be the counterparts
of many radio sources. In the 1950s radio astronomers also began to map
the distribution of neutral hydrogen in our Galaxy and find further
evidence for its spiral structure.

Radio astronomy provided crucial observational data for cosmology
from early on, initially based on counts of sources and on their
(extremely isotropic) distribution on the sky, and since
1965 with the discovery and precise measurement of the cosmic
microwave background (CMB).
Only now are the deepest large-area surveys of discrete radio sources
beginning to provide evidence for anisotropies
in the source distribution, and such surveys continue to be
vital for finding the most distant objects in the Universe and
studying their physical environment as it was billions of years ago.
If this were not enough, today's radio astronomy not only provides
the highest angular resolution
achieved in astronomy (fractions of a milliarcecond, or {\tts mas}),
but it also rivals the astrometric precision of optical astronomy
($\sim$2 mas; \cite{sovers98}). The {\it relative} positions
of neighbouring sources can even be measured to a precision of
a few micro-arcsec ($\mu$as), which allows detection of relative motions
of $\sim$20~$\mu$as per year. This is comparable to
the angular ``velocity'' of the growth of human fingernails as seen
from the distance of the Moon.

The ``radio window'' of the electromagnetic spectrum for observations
from the ground is limited at lower frequencies mainly by the
ionosphere, making observations below $\sim$30\,MHz difficult near
maxima of solar activity. While Reber was able to measure
the emission from the Galactic Centre at 0.9\,MHz from
southern Tasmania during solar minimum in 1995, observations below about 1\,MHz
are generally only possible from space.
The most complete knowledge of the radio sky has been achieved
in the frequency range between 300 ($\lambda$=1\,m) and 5000\,MHz
($\lambda$=6\,cm). At higher frequencies both meteorological conditions
as well as receiver sensitivity become problems, and we have good
data in this range only for the strongest sources in the sky.
Beyond about 1000\,GHz ($\lambda<$=0.3\,mm) we reach the far infrared.
Like the optical astronomers, who named their wavebands with certain letters
(e.g.\ U, B, V, R, I, ...), radio astronomers took over the system
introduced by radio engineers. Jargon like P-, L-, S-, C-, X-, U-, K- or
Q-band can still be found in modern literature and stands for radio bands
near 0.33, 1.4, 2.3, 4.9, 8.4, 15, 23 and 40\,GHz (see \cite{refradio75}).
The \cite{craf97} gives a detailed description of the allocation and use
of the various frequency bands allocated to astronomers (excluding the
letter codes).

Unlike optical astronomers with their photographic plates, radio
astronomers have used electronic equipment from the outset.
Given that they had nothing like the ``finding charts'' used in
optical astronomy to orient themselves in the radio sky, they were used
to working with maps showing coordinates, which were rarely
seen in optical research papers. Nevertheless, the display and description
of radio maps in older literature shows some rare features.
Probably due to the recording devices like analogue charts used up to
the early 1980s, the terms ``following'' and ``preceding'', were
frequently used rather than ``east'' and ``west''. Thus, e.g.\ ``Nf''
stands for ``NE'', or ``Sp'' for ``SW''. Sometimes the aspect ratio
of radio maps was deliberately changed from being equi-angular,
just to make the telescope beam appear round (\cite{1970MNRAS.149..319G}).
Neither were radio astronomers at the forefront of archiving their
results and offering publicly available databases. Happily all this has
changed dramatically during the past decade, and the present report
hopes to give a convincing taste of this.

As these lectures are aimed at professional astronomers,
I do not discuss services explicitly dedicated to amateurs.
I leave it here with a mention of the well-organised WWW site of
the ``Society of Amateur Radio Astronomers''
(SARA; {\tts irsociety.com/sara.html}). Note that in all addresses on the
World-Wide-Web (WWW) mentioned here (the so-called ``URL''s)
I shall omit the leading characters ``{\tt http://}''
unless other strings like ``{\tt ftp://}'' need to be specified.
The URLs listed have only been verified to be correct as of May 1998.
% For the sake of brevity, and following modern bibliographic services,
% I quote articles in major astronomical journals by their
% 19-digit ``refcode'' or ``bibcode''
% (see sect.\,6 of my tutorial in this book).

\section{Observing Techniques and Map Interpretation}  \label{obstechniques} % lesson 1  section 2

Some theoretical background of radio radiation, interferometry and receiver
technology has been given in G.~Miley's contribution to these proceedings.
In this section I shall briefly compare the advantages and limitations
of both single dishes and radio interferometers, and mention some tools
to overcome or alleviate some of their limitations. For a discussion of various
types of radio telescopes see \cite{christhoeg85}. Here I limit myself
to those items which appear most important to take into account
when trying to make use of, and to interpret, radio maps drawn from
public archives.

\subsection{Single Dishes versus Interferometers}       \label{dishversinterf}

The basic relation between the angular resolution $\theta$ and the
aperture (or diameter)  $D$ of a telescope is  $\theta\approx\lambda/D$
radians, where $\lambda$ is the wavelength of observation.
For the radio domain $\lambda$ is $\sim$10$^6$ times larger than
in the optical, which would imply that one has to build a radio telescope
a million times larger than an optical
one to obtain the same angular resolution. In the early days of radio
astronomy, when the observing equipment was based on radar dishes
no longer required by the military after World War II,
typical angular resolutions achieved were of the order of degrees.
Consequently interferometry developed into an important and successful
technique by the early 1950s (although arrays of dipoles, or Yagi
antennas were used, rather than parabolic dishes, because the
former were more suited to the metre-wave band used in the early
experiments). Improved economic conditions and
technological advance also permitted a significant increase in the size
of single dishes.  However, the sheer weight of the reflector and its support
structure has set a practical limit of about 100 metres for fully steerable
parabolic single dishes. Examples are the Effelsberg 100-m dish
({\tts www.mpifr-bonn.mpg.de/effberg.html}) near
Bad M\"unstereifel in Germany, completed in 1972, and the
Green Bank Telescope (GBT; \ref{future}) in West Virginia, USA, to be
completed in early 2000.  The spherical 305-m antenna near Arecibo
(Puerto Rico; {\tts www.naic.edu/}) is the largest single dish
available at present. However, it is not steerable; it is built in
a natural and close-to-spherical depression in the ground, and has a limiting
angular resolution of $\sim$1$'$ at the highest operating frequency (8\,GHz).
Apart from increasing the dish size, one may also increase the observing
frequency to improve the angular resolution. However, the $D$ in the above
formula is the aperture within which the antenna surface is accurate to
better than $\sim$0.1$\lambda$, and the technical
limitations imply that the bigger the antenna, the less accurate the surface.
In practice this means that a single dish never achieves a resolution
of better than $\sim$10$''$--20$''$, even at sub-mm wavelengths
(cf.\ Fig.~6.8 in \cite{rohwil96}).

Single dishes do not offer the possibility of instantaneous imaging
as with interferometers by Fourier transform of the visibilities.
Instead, several other methods of observation can be used with single
dishes. If one is interested merely in integrated parameters (flux,
polarisation, variability) of a (known) point source, one can use
``cross-scans'' centred on the source. If one is very sure about
the size and location of the source (and its neighbourhood) one can
even use ``on--off'' scans, i.e.\ point on the source for a while,
then point to a neighbouring patch of ``empty sky'' for comparison.
This is usually done using a pair of feeds and measuring their
difference signal.
However, to take a real image with a single dish it is necessary to raster
the field of interest, by moving the telescope e.g.\ along right
ascension (RA), back and forth, each scan shifted in declination (DEC)
with respect to the other by an amount of no more than $\sim$40\% of the
half-power beam width (HPBW) if the map is to be fully sampled.
At decimetre wavelengths this has the advantage
of being able to cover a much larger area than with a single ``pointing''
of an interferometer (unless the interferometer elements are very
small, thus requiring large amounts of integration time). The biggest
advantage of this raster method is that it allows the map size to be adjusted
to the size of the source of interest, which can be several
degrees in the case of large radio galaxies or supernova remnants (SNRs).
Using this technique a single dish is capable of tracing (in principle)
all large-scale features of very extended radio sources. One may
say that it ``samples'' spatial frequencies in a range from the
the map size down to the beam width. This depends critically on the way
in which a baseline is fitted to the individual scans.
The simplest way is to assume the absence of sources
at the map edges, set the intensity level to zero there, and interpolate
linearly between the two opposite edges of the map.
A higher-order baseline is able to remove the variable atmospheric effects
more efficiently, but it may also remove real underlying source structure.
For example, the radio extent of a galaxy may be significantly underestimated
if the map was made too small. Rastering the galaxy in two opposite
directions may help finding emission close to the map edges using
the so-called ``basket-weaving'' technique (\cite{1979A&A....74..361S}).
Different methods in baseline subtraction and cut-offs in source size
have led to two different versions of source catalogues
(\cite{1991ApJS...75....1B} and \cite{1991ApJS...75.1011G}),
both drawn from the 4.85-GHz Green Bank survey.  The fact that the
surface density of these sources does not change towards
the Galactic plane, while in the very similar southern PMN survey
(\cite{1993PASAu..10..320T}) it {\it does}, is entirely due to
differences in the data reduction method (\S\ref{majorradiosurveys}).

\begin{figure*}[!h]
%\vspace*{2mm}

\hspace*{-3mm}
\mbox{
\epsfig{file=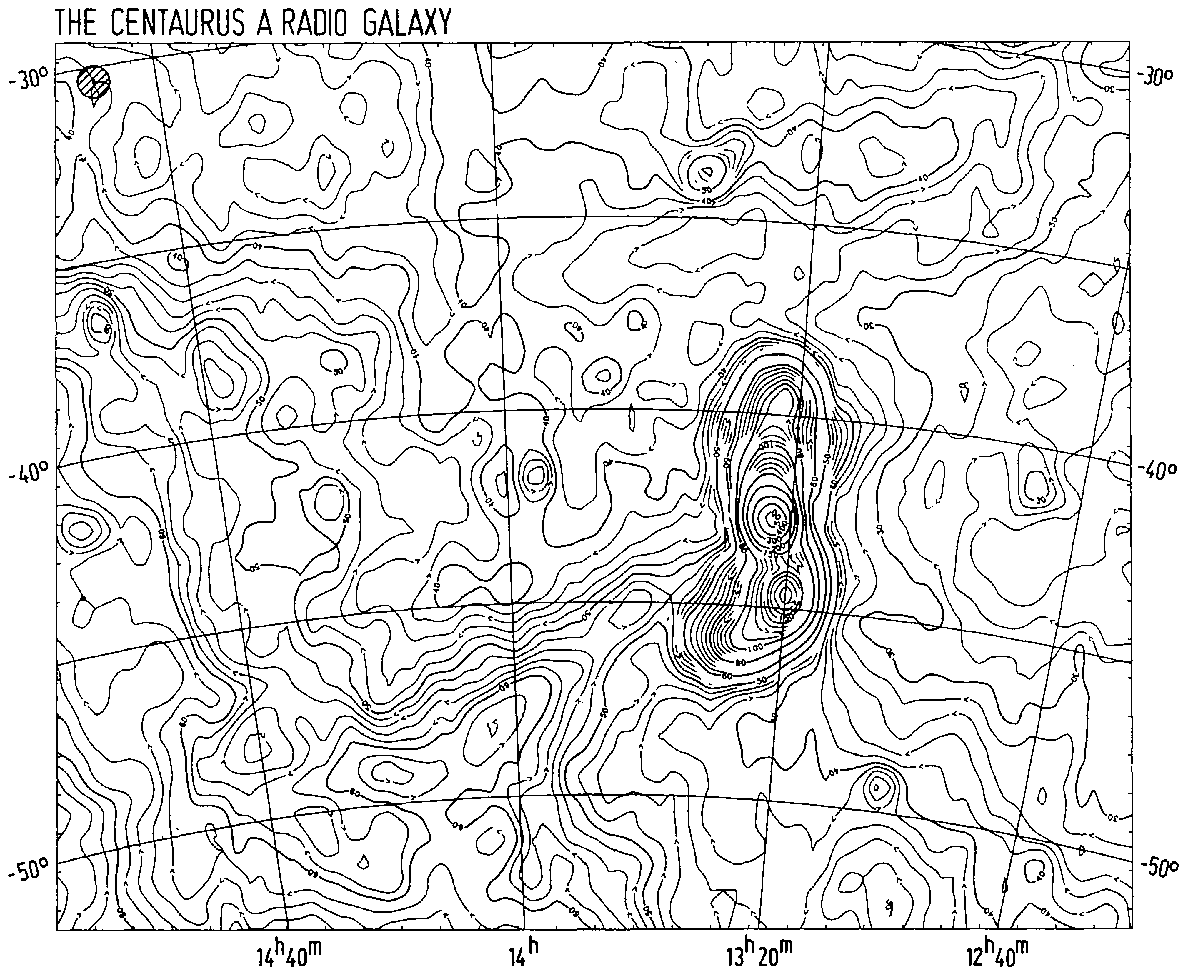,width=7.3cm}
\hspace*{-3mm}

\epsfig{file=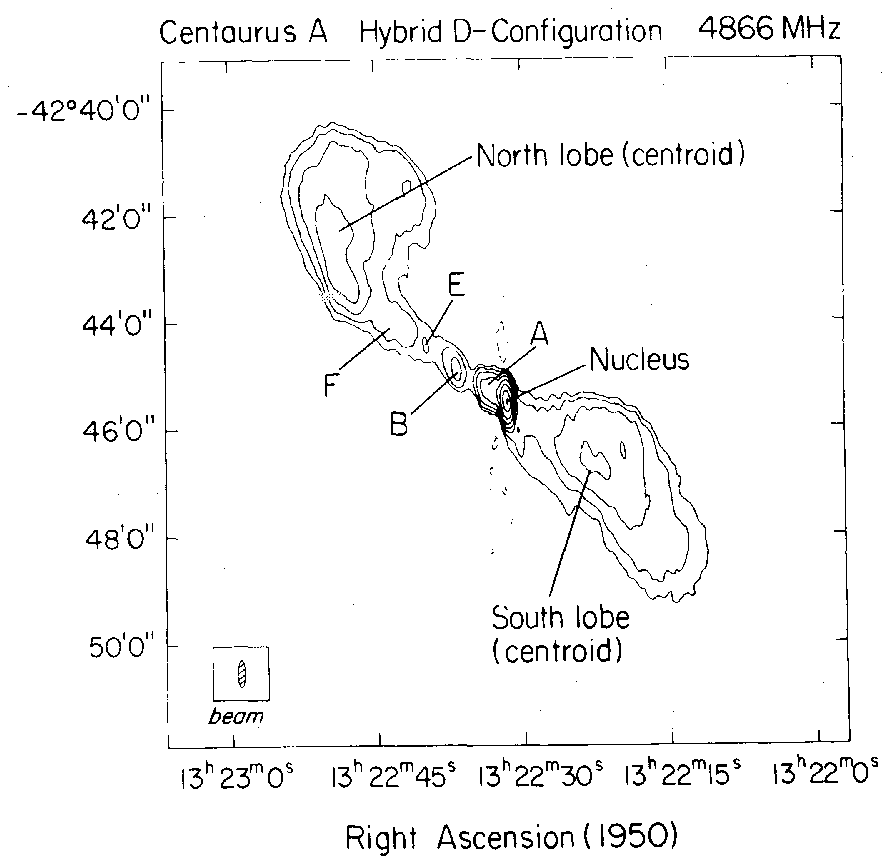,width=6.3cm}
}
\caption{Map of the Centaurus~A region from the 408\,MHz
all-sky survey (Haslam et al.\ (1981), % \cite{1981A&A...100..209H}
showing the full north-south extent of $\sim$10\deg\ of the
radio structure and an emission feature due south east, apparently ``connecting''
Cen~A with the plane of our Galaxy (see Combi et al.\ 1998).  % \cite{1998A&A...333..298C}
Right: A 1.4\,GHz map obtained with the VLA (from Burns et al.\ 1983) % \cite{1983ApJ...273..128B}
showing the inner 10$'$ of Cen~A. Without a single-dish map the full size
of Cen~A would not have been recognised.} \label{CenAfig}
\end{figure*}

In contrast to single dishes, interferometers often have excellent
angular resolution (again $\theta\approx\lambda/D$,
but now $D$ is the maximum distance between any pair of antennas
in the array). However, the field of view is
FOV$\approx\lambda/d$, where $d$ is the size of an individual antenna.
Thus, the smaller the individual antennas, the larger the field of view,
but also the worse the sensitivity. Very large numbers of antennas increase
the design cost for the array and the on-line correlator to process
the signals from a large number of interferometer pairs.
An additional aspect of interferometers is their reduced
sensitivity to extended source components, which depends essentially
on the smallest distance, say $D_{\rm min}$, between two antennas in
the interferometer array. This is often called the {\it minimum spacing}
or {\it shortest baseline}.  Roughly speaking, source components
larger than $\sim\lambda/D_{\rm min}$ radians will be attenuated
by more than 50\% of their flux, and thus practically be lost.
Figure~\ref{CenAfig} gives an extreme example of this,
showing two images of the radio galaxy with the largest apparent size
in the sky (10\deg). It is instructive to compare this with
a high-frequency single-dish map in \cite{1993A&A...269...29J}.

The limitation in sensitivity for extended structure is even more
severe for {\it Very Long Baseline Interferometry} (VLBI) which uses
intercontinental baselines providing $\sim$10$^{-3}$ arcsec (1\,mas)
resolution. The minimum baseline is often several hundred km,
making the largest detectable component much smaller than an arcsec.

\indent
\cite{vri98} created a WWW tool that simulates how radio interferometers
work. This {\it Virtual Radio Interferometer}
(VRI; {\tts \verb*Cwww.jb.man.ac.uk/~dm/vri/C})
comes with the ``VRI Guide'' describing the basic concepts
of radio interferometry.  The applet simulates how the placement of
the antennas affects the uv-coverage of a given array and
illustrates the Fourier transform relationship between the
accumulated radio visibilities and the resultant image.

The comparatively low angular resolution of single dish radio
telescopes naturally suggests their use at relatively high frequencies.
However, at centimetre wavelengths atmospheric effects
(e.g.\ passing clouds) will introduce
additional emission or absorption while scanning, leaving
a stripy pattern along the scanning direction (so-called
``scanning effects''). Rastering the same field along DEC
rather than RA, would lead to a pattern perpendicular to the first one.
A comparison and subsequent combination of the two maps,
either in the real or the Fourier plane, can efficiently
suppress these patterns and lead to a sensitive map of the
region (\cite{1988A&A...190..353E}).

A further efficient method to reduce atmospheric effects in
single-dish radio maps is the so-called ``multi-feed technique''.
The trick is to use pairs of feeds in the focal plane of a
single dish. At any instant each feed receives the emission from
a different part of the sky (their angular separation,
or ``beam throw'', is usually 5--10 beam sizes). Since
they largely overlap within the atmosphere, they are affected
by virtually the same atmospheric effects, which then cancel out
in the difference signal between the two feeds. The resulting
map shows a positive and negative image of the same source,
but displaced by the beam throw. This can then be converted
to a single positive image as described in detail
by \cite{1979A&A....76...92E}. One limitation of the method
is that source components larger than a few times the largest
beam throw involved will be lost.
The method has become so widely used that an entire symposium
has been dedicated to it (\cite{1995ASPC...75}).

  From the above it should be clear that single dishes and interferometers
actually complement each other well, and in order to map both the
small- and large-scale structures of a source it may be necessary
to use both. Various methods for combining single-dish and
interferometer data have been devised, and examples of results can be
found in \cite{1984A&AS...55..179B}, \cite{1990A&A...232..207L},
\cite{1992ApJ...387..591J}, Landecker et al.\ (1992), % \cite{1992A&A...258..495L}, \linebreak[4]
\cite{1992A&AS...92...63N} or \cite{1995ApJ...453..293L}.
The {\it Astronomical Image Processing System} (AIPS; {\tts www.cv.nrao.edu/aips}),
a widely used reduction package in radio astronomy,
provides the task {\tts IMERG} (cf.\
{\tts www.cv.nrao.edu/aips/cook.html}) for this purpose.
The software package {\it Miriad}
({\tts www.atnf.csiro.au/computing/software/miriad}) for
reduction of radio interferometry data offers two programs
({\it immerge} and {\it mosmem}) to realise this combination of
single dish and interferometer data (\S\ref{mosaicing}).
The first one works in the Fourier plane and uses the single dish and
mosaic data for the short and long spacings, respectively.
The second one compares the single dish and mosaic images and
finds the ``Maximum Entropy'' image consistent with both.

\subsection{Special Techniques in Radio Interferometry}   \label{specialtec}

A multitude of ``cosmetic treatments'' of interferometer data
have been developed, both for the ``uv-'' or visibility data and
for the maps (i.e.\ before and after the Fourier transform),
mostly resulting from 20 years of experience with the most versatile
and sensitive radio interferometers currently available,
the {\it Very Large Array} (VLA) and its more recent VLBI counterparts
the {\it European VLBI Network} (EVN), and the
{\it Very Large Baseline Array} (VLBA); see their WWW pages at
{\tts www.nrao.edu/vla/html/VLAhome.shtml},
{\tts www.nfra.nl/jive/evn/evn.html}, and
{\tts www.nrao.edu/vlba/html/VLBA.html}.
The volumes edited by \cite{1989ASPC....6,1991ASPC...19}, and \cite{1995ASPC...82}
give an excellent introduction to these effects, the procedures for treating
them, as well as their limitations. The more prominent topics are
bandwidth and time-average smearing, aliasing, tapering,
uv-filtering, CLEANing, self-calibration, spectral-line imaging,
wide-field imaging, multi-frequency synthesis, etc.

\subsection{Mosaicing}    \label{mosaicing}

One way to extend the field of view of interferometers is to take
``snapshots'' of several individual fields with adjacent pointing
centres (or {\it phase centres}) spaced by no further than
about one (and preferably half a) ``primary beam'', i.e.\ the HPBW
of the individual array element.
For sources larger than the primary beam of the single interferometer
elements the method recovers interferometer spacings down to
about half a dish diameter shorter than those directly measured,
while for sources
that fit into the primary beam mosaicing (also spelled ``mosaicking'')
will recover spacings down to half the dish diameter
(\cite{1988A&A...202..316C}, or \cite{cornwell89}).
The data corresponding to shorter spacings can be taken either from
other single-dish observations, or from the array itself, using it
in a single-dish mode.
The ``Berkeley Illinois Maryland Association''
(BIMA; {\tts bima.astro.umd.edu/bima/})
has developed a {\it homogeneous array} capability, which is
the central design issue for the planned NRAO Millimeter Array
(MMA; {\tts www.mma.nrao.edu/}).
The strategy involves mosaic observations with the BIMA compact array
during a normal 6--8 hour track, coupled with single-antenna observations
with all array antennas mapping the same extended field
(see \cite{pound97} or {\tts bima.astro.umd.edu:80/bima/memo/memo57.ps}).

% The technique has been used to map C[tex2html_wrap_inline305]O in NGC 2024.
% A mosaic of 7 fields, the map contains all spatial frequencies from 8 arcsec
%to 4 arcmin, and includes both cross-correlation and auto-correlation data.

Approximately 15\% of the observing time on the {\it Australia Telescope
Compact Array} (ATCA; {\tts www.narrabri.atnf.csiro.au/}) is spent on
observing mosaics. A new pointing centre may be observed every 25 seconds,
with only a few seconds of this time consumed by slewing and other
overheads. The largest mosaic produced on the ATCA by 1997 is a 1344
pointing-centre spectral-line observation of the Large Magellanic Cloud.
Joint imaging and deconvolution of this data produced a
1997$\times$2230$\times$120 pixel cube
(see {\tts www.atnf.csiro.au/research/lmc\_h1/}).
Mosaicing is heavily used in the current large-scale radio surveys
like NVSS, FIRST, and WENSS (\S\ref{modsurv}).

\subsection{Map Interpretation}          \label{mapinterpretation}

The {\it dynamic range} of a map is usually defined as the ratio of the
peak brightness to that of the ``lowest reliable brightness level'', or
alternatively to that of the rms noise of a source-free region of the image.
For both interferometers and single dishes the dynamic range is often limited by
sidelobes occurring near strong sources, either due to limited uv-coverage,
and/or as part of the diffraction pattern of the antenna.
Sometimes the dynamic range, but more often the ratio between the
peak brightness of the sidelobe and the peak brightness of the source,
is given in {\tt dB}, this being ten times the decimal logarithm of the
ratio.
In interferometer maps these sidelobes can usually be reduced
using the CLEAN method, although more sophisticated
methods are required for the strongest sources
(cf.\ \cite{1982Natur.299..597N}, \cite{perley89}), for
which dynamic ranges of up to 5$\times$10$^5$ can be achieved
(\cite{debrusij93}). For an Alt-Az single dish
the sidelobe pattern rotates with time on the sky, so a simple
average of maps rastered at different times can reduce the sidelobe
level. But again, to achieve dynamic ranges of better than a few thousand
the individual scans have to be corrected independently before
they can be averaged (\cite{1995ASPC...75..318K}).

{\it Confusion} occurs when there is more than one source in the
telescope beam. For a beam area $\Omega_{\rm b}$, the {\it confusion
limit} S$_{\rm c}$ is the flux density at which this happens as one
considers fainter and fainter sources. For an integral source count
N(S), i.e.\ the number of sources per sterad brighter than flux density
S, the number of sources in a telescope beam $\Omega_{\rm b}$ is $\Omega_{\rm b}$\,N(S).
S$_{\rm c}$ is then given by $\Omega_{\rm b}$\,N(S$_{\rm c})\approx$1. A radio survey is
said to be {\it confusion-limited} if the expected minimum detectable flux density
S$_{\rm min}$ is lower than S$_{\rm c}$. Clearly, the confusion limit
decreases with increasing observing frequency and with smaller telescope
beamwidth.  Apart from estimating the confusion limit theoretically from
source counts obtained with a telescope of much lower confusion level
(see \cite{1974ApJ...188..279C}),
one can also derive the confusion limit {\it empirically} by subsequent
weighted averaging of N maps with (comparable) noise level $\sigma_i$,
and with each of them {\it not} confusion-limited. The weight of each map
should be proportional to $\sigma_{i}^{-2}$. In the absence
of confusion, the {\it expected} noise, $\sigma_{\rm N,exp}$, of the
average map should then be

% \[ \sigma_{\rm N,exp} = 1/\sqrt{\sum_{i=1}^{N} \sigma_{\rm i}^{-2}} \]
\[ \sigma_{\rm N,exp} = \left( \sum_{i=1}^{N} \sigma_{\rm i}^{-2} \right)^{-1/2} \]

\noindent
If this is confirmed by experiment, we can say that the ``confusion noise''
is negligible, or at least that $\sigma_{\rm c} \ll \sigma_{\rm N}$.  However,
if $\sigma_{\rm N}$ approaches a saturation limit with increasing N,
then the confusion noise, $\sigma_{\rm c}$, can be estimated according to
$\sigma_{\rm c}^2 = \sigma_{\rm obs}^2 - \sigma_{\rm exp}^2$~.
% \[ \sigma_{\rm c}^2 = \sigma_{\rm obs}^2 - \sigma_{\rm exp}^2 \]
% \noindent
As an example, the confusion limit of a 30-m dish at 1.5\,GHz ($\lambda$=20\,cm)
and a beam width of HPBW=34$'$ is $\sim$400\,mJy. For a 100-m telescope
at 2.7, 5 and 10.7\,GHz ($\lambda$=11\,cm, 6\,cm and 2.8\,cm;
HPBW=4.4$'$, 2.5$'$ and 1.2$'$), the
confusion limits are $\sim$2, 0.5, and\, $\ltsim$\,0.1\,mJy. For
the VLA D-array at 1.4\,GHz (HPBW=50$''$) it is $\sim$\,0.1\,mJy.
For radio interferometers the confusion noise is generally
negligible owing to their high angular resolution, except for deep
maps at low frequencies where confusion due to sidelobes becomes
significant (e.g.\ for WENSS and SUMSS, see \S\ref{modsurv}).
Note the semantic difference
between ``confusion noise'' and ``confusion limit''. They can be related
by saying that in a confusion-limited survey, point sources can be
reliably detected only above the confusion limit, or 2--3 times the
confusion noise, while coherent extended structures can be reliably
detected down to lower limits, e.g.\ by convolution of the map to lower
angular resolution. There is virtually no confusion limit for
polarised intensity, as the polarisation position angles of
randomly distributed, faint background sources tend to cancel out
any net polarisation (see \cite{rohwil96}, p.\,216 for more details).
Examples of confusion-limited surveys are the large-scale
low frequency surveys e.g.\ at 408\,MHz (\cite{1982A&AS...47....1H}),
at 34.5\,MHz (\cite{1990JApA...11..323D}), and at 1.4\,GHz
(\cite{1986AJ....91.1051C}). Of course, confusion
becomes even more severe in crowded areas like the Galactic plane
(\cite{1988ApJS...68..715K}).

When estimating the error in flux density of sources (or their
significance) several factors have to be taken into account.
The error in absolute calibration, $\Delta_{\rm cal}$, depends on the
accuracy of the adopted flux density scale and is usually of the order
of a few per cent. Suitable absolute calibration sources for single-dish
observations are listed in \cite{1977A&A....61...99B} and
\cite{1994A&A...284..331O} for intermediate frequencies,
and in \cite{1990MNRAS.243..637R} for low frequencies. Note that
for the southern hemisphere older flux scales are still in use, e.g.\
\cite{1975AuJPA..38....1W}.
Lists of calibrator sources for intermediate-resolution \linebreak[4] interferometric
observations (such as the VLA) can be found at the URL~ \linebreak[4]
{\tts \verb*Cwww.nrao.edu/~gtaylor/calib.htmlC}, and those for very-high
resolution observations (such as the VLBA) at~
{\tts magnolia.nrao.edu/vlba\_calib/vlbaCalib.txt}.
When comparing different source lists it is important to note that,
especially at frequencies below $\sim$400\,MHz, there are still different
``flux scales'' being used which may differ by $\gtsim$10\%,
and even more below $\sim$100\,MHz.
The ``zero-level error'' is important mainly for single-dish
maps and is given by $\Delta_{\circ}$ =$m\,\sigma$/$\sqrt{n}$,
where $m$ is the number of beam areas contained in the source integration area,
$n$ is the number of beam areas in the area of noise determination, and
$\sigma$ is the noise level determined in regions ``free of emission''
(and includes contributions from the receiver, the atmosphere,
and confusion).  The error due to noise in the integration area
is  $\Delta_{\sigma} = \sigma \sqrt{m}$.
The three errors combine to give a total flux density error of
$\Delta S = \Delta_{\rm cal} + \sqrt{\Delta_{\circ}^2 + \Delta_{\sigma}^2}$
~(\cite{1981A&A....94...29K}).
Clearly, the relative error grows with the extent of a source. This also
implies that the upper limit to the flux density of a non-detected source
depends on the size assumed for it\,: while a point source of ten times the
noise level will clearly be detected, a source of the same flux, but
extending over many antenna beams may well remain undetected.
In interferometer observations the non-zero size of the shortest baseline
limits the sensitivity to extended sources.
At frequencies $\gtsim$10\,GHz the atmospheric absorption starts to
become important, and the measured flux S will depend on elevation
$\epsilon$ approximately according
to~~ S$=$S$_{\circ}$~exp($-\tau$~csc\,$\epsilon$),
where S$_{\circ}$ is the extra-atmospheric flux density, and $\tau$ the
optical depth of the atmosphere.
E.g., at 10.7\,GHz and at sea level, typical values of $\tau$ are
0.05--0.10, i.e.\ 5--10\% of the flux is absorbed even
when pointing at the zenith. These increase with frequency, but decrease
with altitude of the observatory.
Uncertainties in the zenith-distance dependence
may well dominate other sources of error above $\sim$50\,GHz.

When estimating flux densities from interferometer maps, the maps
should have been corrected for the polar diagram (or ``primary beam'')
of the individual antennas, which implies a decreasing sensitivity
with increasing distance from the pointing direction.
This so-called ``primary-beam correction'' divides the map by the
attenuation factor at each map point and thus raises both the intensity
of sources, and the map noise, with increasing distance from the phase centre.
Some older source catalogues, mainly obtained with the {\it Westerbork
Synthesis Radio Telescope} (WSRT; e.g.\ \cite{1987A&AS...71...25O}, or
\cite{1988A&AS...74..315R}) give both the  (uncorrected) ``map flux''
and the (primary-beam corrected) ``sky flux''.
The increasing uncertainty of the exact primary beam shape with
distance from the phase centre may dominate the flux density error
on the periphery of the field of view.

Care should be taken in the
interpretation of structural source parameters in catalogues.
Some catalogues list the ``map-fitted'' source size, $\theta_{\rm m}$,
as drawn directly from a Gaussian fit of the map.  Others quote
the ``deconvolved'' or ``intrinsic'' source size, $\theta_{\rm s}$.
All of these are model-dependent and usually assume both the source and
the telescope beam to be Gaussian (with full-width at half maximum,
FWHM=$\theta_{\rm b}$), in which case we have~
$\theta_{\rm b}^2 + \theta_{\rm s}^2 = \theta_{\rm m}^2$.
Values of ``0.0'' in the size column of catalogues are often found for
``unresolved'' sources. Rather than zero, the intrinsic size is smaller
than a certain fraction of the telescope beam width. The fraction
decreases with increasing signal-to-noise (S/N) ratio of the source.
Estimation of errors in the structure parameters derived from
2-dimensional radio maps is discussed in \cite{1997PASP..109..166C}.
Sometimes flux densities are quoted which are smaller than the error,
or even negative (e.g.\ \cite{1978ApJS...36...53D}, and
\cite{1996A&A...313..417K}).  These should actually be converted to,
and interpreted as {\it upper limits} to the flux density.

\subsection{Intercomparison of Different Observations and Pitfalls}  \label{comparison}

Two main emission mechanisms are at work in radio sources (e.g.\ \cite{pach70}).
The {\it non-thermal} synchrotron emission of relativistic electrons
gyrating in a magnetic field is responsible for supernova remnants,
the jets and lobes of radio galaxies and much of the diffuse emission
in spiral galaxies (including ours) and their haloes. The {\it thermal}
free-free or {\it bremsstrahlung} of an ionised gas cloud dominates
e.g.\ in H\,II regions, planetary nebulae, and in spiral galaxies at
high radio frequencies. In addition, individual stars may show
``magneto-bremsstrahlung'', which is synchrotron emission from either
mildly relativistic electrons (``gyrosynchrotron'' emission) or
from less relativistic electrons (``cyclotron'' or ``gyroresonance'' emission).
The historical confirmation of synchrotron radiation
came from the detection of its polarisation.
In contrast, thermal radiation is unpolarised, and characterised by a very
different spectral shape than that of synchrotron radiation.
Thus, in order to distinguish between these mechanisms, multi-frequency
comparisons are needed. This is trivial for unresolved
sources, but for extended sources care has to be taken to include
the entire emission, i.e.\ {\it integrated} over the source area.
Peak fluxes or fluxes from high-resolution interferometric
observations will usually underestimate their total flux.
Very-low frequency observations may overestimate the flux by picking up
radiation from neighbouring (or ``blending'') sources within
their wide telescope beams. Compilations of integrated spectra
of large numbers of extragalactic sources have been prepared e.g.\ by
\cite{kuhr79}, \cite{1992ApJS...81...83H}, and \cite{1997BSAO...42....5B}
(see {\tts cats.sao.ru/cats\_spectra.html}).

An important diagnostic of the energy transfer within radio sources
is a two-dimension\-al comparison of maps observed at different
frequencies. Ideally, with many such frequencies, a spectral fit
can be made at each resolution element across the source and
parameters like the relativistic electron density and radiation
lifetime, magnetic field strength, separation of thermal and
non-thermal contribution, etc.\ can be estimated
(cf.\ \cite{1989A&A...211..280K} or \cite{1994ApJ...426..116K}). However,
care must be taken that the observing instruments at the different
frequencies were sensitive to the same range of ``spatial frequencies''
present in the source. Thus interferometer data which are to be compared
with single-dish data should be sensitive to components comparable to
the entire size of the source.
The VLA has a set of antenna configurations
with different baseline lengths that can be matched to a subset of observing
frequencies in order to record a similar set of spatial frequencies at
widely different wavelengths -- these are called ``scaled arrays''. For
example, the B-configuration at 1.4\,GHz and the C-configuration at 4.8\,GHz
form one such pair of arrays.
Recent examples of such comparisons for very extended radio galaxies
can be found in \cite{1997A&AS..123..423M} or \cite{1998A&A...331..901S}.
Maps of the spectral indices of Galactic radio emission
between 408 and 1420\,MHz have even been prepared for the entire
northern sky (\cite{1988A&AS...74....7R}). Here the major limitation
is the uncertainty in the absolute flux calibration.

\subsection{Linear Polarisation of Radio Emission} \label{linpol}

As explained in G.~Miley's lectures for this winter school,
the linear polarisation characteristics of radio emission give us
information about the magneto-ionic medium, both within the emitting
source {\it and} along the line of sight between the source and
the telescope. The plane of polarisation (the ``polarisation
position angle'') will rotate while passing through such media,
and the fraction of polarisation (or ``polarisation percentage'') will
be reduced. This ``depolarisation'' may occur due to cancellation
of different polarisation vectors within the antenna beam, or due
to destructive addition of waves having passed through different
amounts of this ``Faraday'' rotation of the plane of polarisation,
or also due to significant rotation of polarisation vectors across
the bandwidth for sources of high rotation measure (RM).
More detailed discussions of the various effects affecting polarised
radio radiation can be found in Pacholczyk (1970, 1977), % \cite{pach70,pach77},
\cite{1966ARA&A...4..245G}, \cite{1966MNRAS.133...67B}, and
\cite{1980AJ.....85..368C}.

During the reduction of polarisation maps, it is
important to estimate the ionospheric contribution to the
Faraday rotation, which increases in importance at lower
frequencies, and may show large variations at sunrise or
sunset. Methods to correct for the ionospheric rotation
depend on model assumptions and are not straightforward.
E.g., within the AIPS package the ``Sunspot'' model may be used in the
task {\tts FARAD}. It relies on the mean monthly sunspot number as
input, available from the US National Geophysical Data Centre at~
{\tts www.ngdc.noaa.gov/stp/stp.html}.
The actual numbers are in files available from~
{\tts ftp://ftp.ngdc.noaa.gov/STP/SOLAR\_DATA/SUNSPOT\_NUMBERS/}
(one per year: filenames are year numbers).
Ionospheric data have been collected at Boulder, Colorado,
up to 1990 and are distributed with the AIPS software, mainly to
be used with VLA observations.  Starting from 1990, a dual-frequency GPS
receiver at the  \linebreak[4] VLA site has been used to estimate ionospheric
conditions, but the data are not yet \linebreak[4] available 
(contact~ {\tts cflatter@nrao.edu}). 
Raw GPS data are available from \linebreak[4]
{\tts ftp://bodhi.jpl.nasa.gov/pub/pro/y1998/} and from 
{\tts ftp://cors.ngs.noaa.gov/rinex/}. \linebreak[4] The AIPS task {\tts GPSDL}
for conversion to total electron content (TEC) and rotation measure
(RM) is being adapted to work with these data.

A comparison of polarisation maps at different frequencies allows one
to derive two-dimensional maps of RM and depolarisation (DP, the
ratio of polarisation percentages between two frequencies).
This requires the maps to be sensitive to the same range of
spatial frequencies.
Generally such comparisons will be meaningful only if
the polarisation angle varies linearly with $\lambda^2$,
as it indeed does when using sufficiently high resolution
(e.g.\ \cite{1987ApJ...316..611D}).
The $\lambda^2$ law may be used to extrapolate the
electric field vector of the radiation to $\lambda=$0.
This direction is called the ``intrinsic'' or ``zero-wavelength''
polarisation angle ($\chi_{\circ}$), and the direction of the
homogeneous component of the magnetic field at this position is
then perpendicular to $\chi_{\circ}$ (for optically thin relativistic
plasmas).
Even then a careful analysis has to be made as to which part of
RM and DP is intrinsic to the source, which is due to a ``cocoon''
or intracluster medium surrounding the source, and which is due to our
own Galaxy. The usual method to estimate the latter contribution
is to average the integrated RM of the five or ten extragalactic
radio sources nearest in position to the source being studied.
Surprisingly, the most complete compilations of RM values of
extragalactic radio sources date back many years
(\cite{1980A&AS...39..379T}, \cite{1981ApJS...45...97S}, or
\cite{1988Ap&SS.141..303B}).

An example of an overinterpretation of these older low-resolution
polarisation data is the recent claim (\cite{1997PhRvL..78.3043N})
that the Universe shows a birefringence for polarised radiation,
i.e.\ a rotation of the polarisation angle not due to any known
physical law, and proportional to the cosmological distance of
the objects emitting linearly polarised radiation (i.e.\ radio
galaxies and quasars).
The analysis was based on 20-year old low-resolution data for
integrated linear polarisation (\cite{1980MNRAS.190..205C}),
and the finding was that the difference angle between the intrinsic
($\lambda$=0) polarisation angle and the major axis of
the radio structure of the chosen radio galaxies was increasing
with redshift.
However, it is now known that the distribution of polarisation angles
at the smallest angular scales is very complex, so that the
integrated polarisation angle may have little or no relation
with the exact orientation of the radio source axis.
Although the claim of birefringence has been contested by radio astronomers
(\cite{1997PhRvL..79.1801W}), and more than a handful of contributions
about the issue have appeared on the LANL/SISSA preprint server
({\tts astro-ph/9704197, 9704263, 9704285, 9705142,
9705243, 9706126, 9707326, 9708114}) the original authors
continue to defend and refine their statistical methods
({\tts astro-ph/9803164}).  Surprisingly, these articles
neither explicitly list the data actually used, nor do they discuss
their quality or their appropriateness for the problem (cf.\ the
comments in sect.\,7.2 of \cite{1998PASP..110..223T}).

\subsection{Cross-Identification Strategies} \label{optidstrat}

While the nature of the radio emission can be inferred from
the spectral and polarisation characteristics, physical parameters
can be derived only if the distance to the source is known.
This requires identification of the source with an optical object
(or an IR source for very high redshift objects) so that an
optical spectrum may be taken and the redshift determined.
By adopting a cosmological model, the distance of extragalactic objects
can then be inferred.  For sources in our own Galaxy kinematical
models of spiral structure can be used to estimate the
distance from the radial velocity, even without optical information,
e.g.\ using the H\,I line (\S\ref{spectrallines}). More indirect
estimates can also be used, e.g.\ emission measures for
pulsars, apparent sizes for H\,I clouds, etc.

The strategies for optical identification of extragalactic
radio sources are very varied. The easiest case is when the
radio position falls within the optical extent of a galaxy.
Also, a detailed radio map of an extended radio galaxy usually
suggests the position of the most likely optical counterpart from
the symmetry of the radio source. Most often two extended radio
lobes straddle a point-like radio core which coincides with
the optical object. However, various types of asymmetries
may complicate the relation between radio morphology and
location of the parent galaxy (see e.g.\ Figs.~6 and 7 of
\cite{1980ARA&A..18..165M}). These may be wiggles due to
precession of the radio jet axis, or bends due to the
movement of the radio galaxy through an intracluster medium
(see {\tts www.jb.man.ac.uk/atlas/icon.html} for a fine
collection of real maps).  For fainter and less extended
sources the literature contains many different methods to
determine the likelihood of a radio-optical association
(\cite{1975AN....296..197N}, \cite{1975AN....296...65R},
\cite{1977A&AS...27..171P}, \cite{1977A&AS...28..211d}).
The last of these papers proposes the dimensionless variable~~
$r = \sqrt{ \left( \Delta\alpha / \sigma_{\alpha} \right) ^2 +
              \left( \Delta\delta / \sigma_{\delta} \right) ^2}$
% \[ r = \sqrt{ \left( \Delta\alpha / \sigma_{\alpha} \right) ^2 +
%               \left( \Delta\delta / \sigma_{\delta} \right) ^2}  \]
%\noindent
~~where $\Delta\alpha$ and $\Delta\delta$ are the
positional differences between radio and optical position,
and $\sigma_{\alpha}$ and $\sigma_{\delta}$ are the combined
radio and optical positional errors in RA ($\alpha$) and
DEC ($\delta$), respectively. The likelihood ratio, $LR$, between
the probability for a real association and that of a
chance coincidence is then~
$ LR(r) = \left(1 /\ 2\,\lambda  \right)\
     exp \left( r^2 \left( 2\lambda - 1 \right) / 2 \right) $,~
% \[ LR(r) = \left(1 /\ 2\,\lambda  \right)\
%    exp \left( r^2 \left( 2\lambda - 1 \right) / 2 \right) \]
% \noindent
where $\lambda = \pi\ \sigma_{\alpha}\ \sigma_{\delta}\ \rho_{\rm opt}$,
with $\rho_{\rm opt}$ being the density of optical objects. The
value of $\rho_{\rm opt}$ will depend on the Galactic latitude and
the magnitude limit of the optical image.
Usually, for small sources, $LR\gtsim$2 is regarded as sufficient
to accept the identification, although the exact threshold is
a matter of ``taste''.
A method that also takes into account also the extent of the radio
sources, and those of the sources to be compared with (be it at
optical or other wavelengths), has been described in
\cite{1989ApJ...339...12H}). A further generalisation to
elliptical error boxes, inclined at any position angle (like
those of the IRAS satellite), is discussed in
\cite{1995AJ....109.2318C}.

A very crude assessment of the number of chance coincidences
from two random sets of $ N_1$ and $ N_2$ sources
distributed all over the sky is~~
$N_{cc} = N_1\ N_2\ \theta^2 / 4 $
%  \[ N_{cc} = N_1\ N_2\ \pi \theta^2 / 4\ \pi  \]
~~chance pairs within an angular separation of less than
$\theta$ (in radians). In practice the decision on the
maximum $\theta$ acceptable for a true association can be drawn
from a histogram of the number of pairs within $\theta$, as a
function of $\theta$. If there is any correlation
between the two sets of objects, the histogram should have a
more or less pronounced and narrow peak of true coincidences
at small $\theta$,
then fall off with increasing $\theta$ up to a minimum at
$\theta_{\rm crit}$, before rising again proportional to
$\theta^2$ due to pure chance coincidences.
The maximum acceptable $\theta$ is then usually chosen near
$\theta_{\rm crit}$ (cf.\ \cite{1997AJ....113.2000B} or
\cite{1998A&AS..129...87B}).
% \cite{boller98}).
At very faint (sub-mJy) flux levels, radio sources tend to be
small ($\ll$10$''$), so that there is
virtually no doubt about the optical counterpart, although very deep
optical images, preferably from the Hubble Space Telescope (HST),
are needed to detect them (\cite{1997ApJ...475L..5F}).

However, the radio morphology of extended radio galaxies
may be such that only the two outer ``hot spots'' are detected
without any trace of a connection between them. In such a case only
a more sensitive radio map will reveal the position of the true
optical counterpart,
by detecting either the radio core between these hot spots, or some
``radio trails'' stretching from the lobes towards the parent galaxy.
The paradigm is that radio

%\vspace*{2mm}

\begin{figure*}[!ht] %
\hspace*{10mm}
\epsfig{file=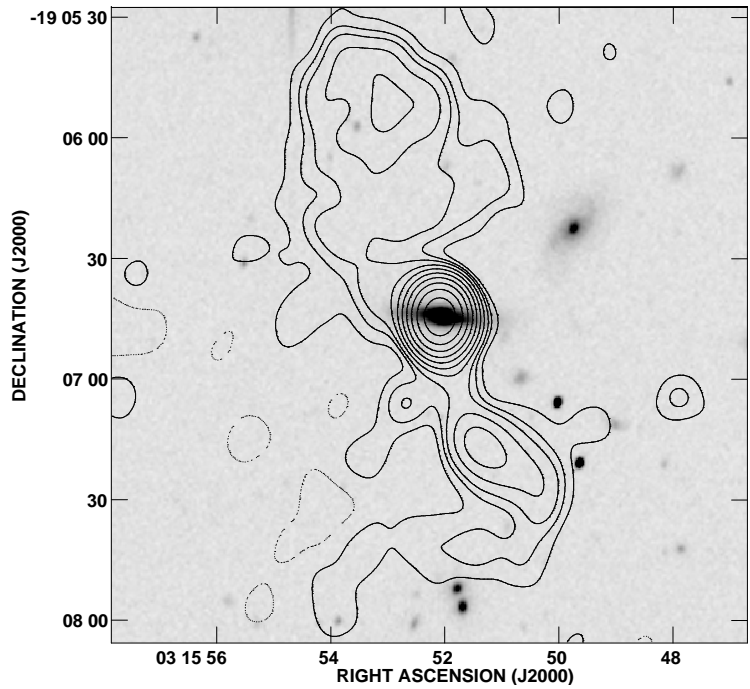,width=10.cm,bbllx=40pt,bblly=148pt,bburx=576pt,bbury=632pt,clip=}
% \centerline {\epsfxsize=9cm \epsfbox[40 148 576 675] {ledlow98_f1.ps}}
% \vspace{-2mm}
\caption{VLA contours at 1.5\,GHz of B1313$-$192 in the galaxy cluster
A\,428, overlaid on an R-band image. The radio source extends
$\approx$100\,h$^{-1}_{\rm 75}$kpc north and south of the host galaxy, which is
disk-like rather than elliptical (from Ledlow et al.\ 1998, % \cite{1998ApJ...495..227L}
courtesy M.~Ledlow).} \label{ledlow98_f1}
\end{figure*}

\vspace*{-6mm}
\begin{figure*}[!hb] %
\hspace*{11mm}
\epsfig{file=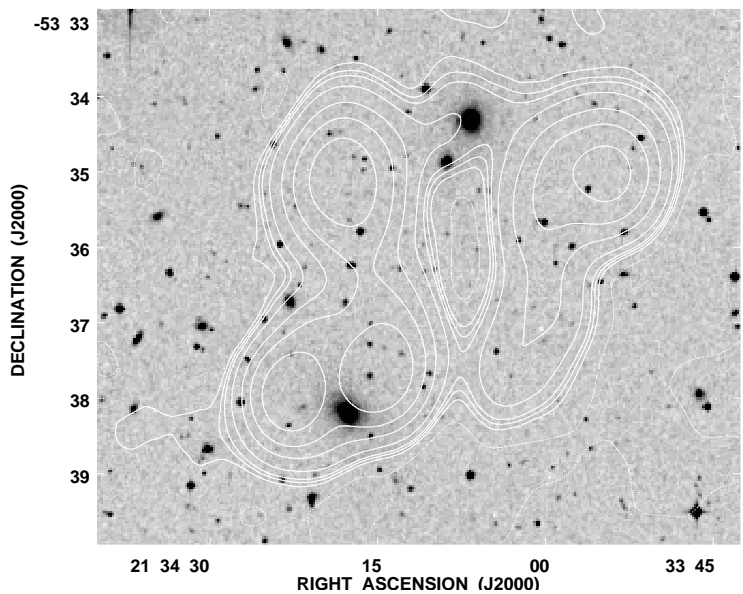,width=10.cm,bbllx=36pt,bblly=165pt,bburx=581pt,bbury=587pt,clip=}
\caption{408\,MHz contours from the Molonglo Observatory
Synthesis Telescope (MOST) of a complex radio source in the
galaxy cluster A\,3785, overlaid on the Digitized Sky Survey.
The source is a superposition of two wide-angle
tailed (WAT) sources associated with the two brightest galaxies in the
image, as confirmed by higher-resolution ATCA maps (from Haigh et al.\ 1997,
% \cite{1997PASA...14..221H}
ctsy.\ A.~Haigh)} \label{haigh_fig3}
\end{figure*}

\newpage
\noindent
%The paradigm is that radio galaxies are generally ellipticals,
galaxies are generally ellipticals,
while spirals only show weak radio emission dominated by the disk,
but with occasional contributions from low-power active nuclei (AGN).
Recently an unusual exception has been discovered: a disk galaxy
hosting a large double-lobed radio source (Figure~\ref{ledlow98_f1}),
almost perpendicular to its disk, and several times the
optical galaxy size (\cite{1998ApJ...495..227L}).

An approach to semi-automated optical identification of
radio sources using the Digitized Sky Survey is described in
\cite{1997PASA...14..221H}. However, Figure~\ref{haigh_fig3} shows one
of the more complicated examples from this paper. Note also that
the concentric contours near the centre of the radio source
encircle a {\it local minimum}, and not a maximum. To avoid such
ambiguities some software packages
(e.g.\ ``NOD2'', \cite{1974A&AS...15..333H}) produce arrowed
contours indicating the direction of the local gradient in the
map.

Morphological considerations can sometimes lead to interesting
misinterpretations. A linear feature detected in a Galactic plane
survey with the Effelsberg 100-m dish had been interpreted as
probably being an optically obscured radio galaxy behind our Galaxy
(\cite{1985A&A...143..478S}).
It was not until five years later (\cite{1990A&A...232..207L})
that interferometer maps taken with the Dominion Radio Astrophysical
Observatory (DRAO; {\tts www.drao.nrc.ca}) revealed
that the linear feature was
merely the straighter part of the shell of a weak and extended
supernova remnant (G\,65.1$+$0.6).

One of the most difficult classes of source to identify optically
are the so-called ``relic'' radio sources, typically occurring in
clusters of galaxies, with a very steep radio continuum spectrum,
and without clear traces of association with any optical
galaxy in their host cluster. Examples can be found in
\cite{1991A&A...252..528G}, \cite{1997NewA....2..501F},
or \cite{1997MNRAS.290..577R}. See {\tts astro-ph/9805367} for the
latest speculation on their origin.

Generally source catalogues are produced only for detections
above the 3--5\,$\sigma$ level. However, \cite{lewis95} and
\cite{1996ApJ...461..127M} have shown that a cross-identification
between catalogues at different wavelengths allows the
``detection'' of real sources even down to the 2\,$\sigma$ level.

\section{Radio Continuum Surveys}  \label{radiosurveys}  % lesson 2  section 3

\subsection{Historical Evolution}  \label{historysurveys}

Our own Galaxy and the Sun were the first cosmic radio sources to
be detected due the work of K.~Jansky, G.~Reber, G.~Southworth, and J.~Hey
in the 1930s and 1940s. Several other regions in the sky had been
found to emit strong discrete radio emission, but in these early days
the angular resolution of radio telescopes was far too poor to uniquely
identify the sources with something ``known'', i.e.\ with an optical object,
as there were simply too many of the latter within the error box of the
radio position. It was not until 1949 that \cite{1949Natur.164..101B}
% (\cite{1949Natur.164..101B})
identified three further sources with optical objects.
They associated Tau~A with the ``Crab Nebula'', a supernova remnant
in our Galaxy, Vir~A with M~87, the central galaxy in the
Virgo cluster, and Cen~A with NGC~5128, a bright nearby elliptical
galaxy with a prominent dust lane.
By 1955, with the publication of the ``2C'' survey
(\cite{1955MmRAS..67..106S}) the majority of radio sources were
still thought to be Galactic stars, albeit faint ones, since
no correlation with bright stars was observed. However,
in the previous year, the bright radio source Cyg~A had been identified
with a very faint ($\sim$16$^{m}$) and distant (z=0.057) optical galaxy
(\cite{1954ApJ...119..206B}).

Excellent accounts of early radio astronomy can be found
in the volumes by Hey (1971, 1973), \cite{grahamsmith74, edgemulkay76},
 \cite{sulliv82}, Sullivan III (1984), \cite{kellerm84,robertson92}, and
in \cite{hhmm96}, the latter two describing the Australian point of view.
The growth in the number of discrete source lists from 1946 to the late
1960s is given in Appendix 4 of \cite{pach70}.
Many of the major source surveys carried out during
the late 1970s and early 1980s
(6C, UTR, TXS, B2, MRC, WSRT, GB, PKS, S1--S5)
are described in \cite{IAUS74}. The proceedings volume by \cite{condonlockman90}
includes descriptions of several large-scale surveys in the continuum,
H\,I, recombination lines, and searches for pulsars and variable sources.

\subsection{Radio Source Nomenclature\,: The Good, the Bad and the Ugly} \label{nomenclature}

As an aside, Appendix 4 of \cite{pach70} explains the difficulty
(and liberty!) with which radio sources were designated originally.
In the early 1950s, with only a few dozen radio sources known,
one could still afford to name them after the constellation in which
they were located followed by an upper case letter in alphabetic sequence,
to distinguish between sources in the same constellation.
This method was abandoned before even a couple of sources received the
letter B. Curiously,
even in 1991, the source PKS\,B1343$-$601 was suggested
{\it a posteriori} to be named ``Cen\,B'' as it is the second
strongest source in Centaurus (\cite{1991PASAu...9..255M}).
Apparently the name has been adopted (see \cite{1998ApJ...499..713T}).
Sequential numbers like 3C~NNN were used in the late 1950s and
early 1960s, sorting the sources in RA (of a given equinox,
like B1950 at that time and until rather recently).  But when
the numbers exceeded a few thousand, with the 4C survey
(\cite{1965MmRAS..69..183P} and \cite{1967MmRAS..71...49G}) a naming like
4C~DD.NN was introduced, where DD indicates the declination strip
in which the source was detected and NN is a sequence number
increasing with RA of the source, thus giving a rough idea
of the source location (although the total number of sources
in one strip obviously depends on the declination). A real
breakthrough in naming was made with the Parkes (PKS) catalogue
(\cite{1964AuJPh..17..340B}) where the ``IAU convention''
of coordinate-based names was introduced. Thus e.g.\ a name
PKS~1234$-$239 would imply that the source lies in the range
RA= 12$^h$34$^m$...12$^h$35$^m$ and
DEC=$-$23\deg\,54$'$...$-$24\deg\,0$'$. Note that
to construct the source name the exact position of the source
is truncated, not rounded. An even number of digits for RA or
DEC would indicate integer hours, minutes or seconds
(respectively of time and arc), while odd numbers of digits would indicate
the truncated (i.e.\ downward-rounded) tenth of the unit of
the preceding pair of digits. Since the coordinates are
equinox-dependent and virtually all previous coordinate-based
names were based on B1950, it has become obligatory to
precede the coordinate-based name with the letter J if they
are based on the J2000 equinox. Thus e.g.\ PKS~B0000$-$506
is the same as PKS~J0002$-$5024, and the additional digit in
DEC merely reflects the need for more precision nowadays.
Vice versa, the {\it lack} of a fourth digit in the B1950
name reflects the recommendation to never change a {\it name}
of a source even if its position becomes better known later.
The current sensitivity of surveys and the resulting surface
density of sources implies much longer names to be unique.
Examples are NVSS~B102023+252903  or FIRST~J102310.0+251352
(which are actually the same source!). Authors should follow
IAU recommendations for object names (\S\ref{futurenomencl}).
The origin of existing names, their acronyms and recommended formats
can be traced with the on-line ``Dictionary of Nomenclature
of Celestial Objects'' ({\tts vizier.u-strasbg.fr/cgi-bin/Dic}; \cite{lortet94}).
A query for the word ``radio'' (option ``Related to words'')
will display the whole variety of naming systems used
in radio astronomy, and will yield what is perhaps the most complete
list of radio source literature available from a single WWW site.
Authors of future radio source lists, and project leaders of
large-scale surveys, are encouraged to consult the latter URL
and register a suitable acronym for their survey well in
advance of publication, so as to guarantee its uniqueness,
which is important for its future recognition in public
databases.

\subsection{Major Radio Surveys}   \label{majorradiosurveys}

Radio surveys may be categorised into {\it imaging} and
{\it discrete source} surveys. {\it Imaging surveys} were mostly done
with single dishes and were dedicated to mapping the extended emission
of our Galaxy (e.g.\ \cite{1982A&AS...47....1H}, \cite{1990JApA...11..323D})
or just the Galactic plane (\cite{1984A&AS...58..197R},
\cite{1985A&AS...62..105J}).
Only some of them are useful for extracting lists of discrete
sources (e.g.\ \cite{1997A&AS..126..413R}).
The semi-automatic procedure of source extraction implies that
the derived catalogues are usually limited to sources with a size
of at most a few beamwidths of the survey. The highest-resolution
radio imaging survey covering the full sky, and containing Galactic
foreground emission on all scales, is still the 408\,MHz survey
(\cite{1982A&AS...47....1H}) with HPBW$\sim$50$'$.
Four telescopes were used and it has taken 15 years from
the first observations to its publication.
Its 1.4\,GHz counterpart in the northern hemisphere
(\cite{1982A&AS...48..219R}, \cite{1986A&AS...63..205R})
is being completed in the south with data from the
30-m dish at Instituto Argentino de Radioastronom\'{\i}a, Argentina.

The {\it discrete source surveys} may be done either with
interferometers or with single dishes. Except for the most
recent surveys (FIRST, NVSS and WENSS, see \S\ref{modsurv})
the interferometer surveys tend to cover only small parts
of the sky, typically a single field of view of the array,
but often with very high sensitivities reaching a few $\mu$Jy in the
deepest surveys.
The source catalogues extracted from discrete source surveys
with single dishes depend somewhat on the detection algorithm
used to find sources from two-dimensional maps.  There are examples
where two different source catalogues were published, based on the
same original maps.
Both the ``87GB'' (\cite{1991ApJS...75.1011G}) and ``BWE''
(\cite{1991ApJS...75....1B}) catalogues were drawn from the
same 4.85\,GHz maps (\cite{1989AJ.....97.1064C}) obtained with
the Green Bank 300-ft telescope. The authors of the two
catalogues (published on 510 pages of the same volume of ApJS),
arrived at 54,579 and 53,522 sources, respectively.
While the 87GB gives the peak flux, size and orientation of the source,
the BWE gives the integrated flux only, plus a spectral index
between 1.4 and 4.85\,GHz from a comparison with another catalogue.
Thus, while being slightly different, both catalogues complement
each other. The same happened in the southern hemisphere,
using the same 4.85\,GHz receiver on the Parkes 64-m antenna:
the ``PMN'' (\cite{1994ApJS...90..179G}) and ``PMNM''
catalogues (\cite{1994ApJS...90..173G}) were constructed from the
same underlying raw scan data, but using different source extraction
algorithms, as well as imposing different limits in both
signal-to-noise for catalogue source detection,
and in the maximum source size. The larger size limit for
sources listed in the PMN catalogue, as compared to the
northern 87GB, becomes obvious in an all-sky plot of sources
from both catalogues\,: the Galactic plane is visible only in the
southern hemisphere (\cite{1993PASAu..10..320T}),
simply due to the large number of extended sources near the plane
which have been discarded in the northern catalogues
(\cite{1991ApJS...75....1B}).  \cite{1998MNRAS.297..545B} have also
found a 2\%--8\% mismatch between 87GB and PMN.
Eventually, a further coverage of the
northern sky made in 1986 (not available as a separate paper) has been
averaged with the 1987 maps (which were the basis for 87GB) to
yield the more sensitive GB6 catalogue (\cite{1996ApJS..103..427G}).
Thus, a significant difference in source peak flux density between 87GB and GB6
may indicate variability, and \cite{1998ASPC..144..283G} have indeed
confirmed over 1400 variables.

If single-dish survey maps (or raster scans) are sufficiently large,
they may be used to reveal the structure of Galactic foreground emission
and discrete features like e.g.\ the ``loops'' or ``spurs'' embedded
in this emission. These are thought to be nearby supernova remnants,
an idea supported by additional evidence from X-rays
(Egger \& Aschenbach 1995) % \cite{1995A&A...294L..25E}
and older polarisation surveys (\cite{1983BASI...11....1S}).
Surveys of the linear polarisation of Galactic emission will not be
dealt with here.  As pointed out by \cite{salter88}, an all-sky survey
of linear polarisation, at a consistent resolution and frequency,
is still badly needed.
No major polarisation surveys have been published since the
compendium of \cite{1976A&AS...26..129B}, except for small parts of the
Galactic plane (\cite{1987A&AS...69..451J}).
This is analogous to a lack of recent surveys for discrete source
polarisation (\S\ref{linpol}). Apart from helping to discern thermal
from non-thermal features, polarisation maps have led to
the discovery of surprising features which are not present in the
total intensity maps (\cite{1993A&A...268..215W}, \cite{1998Natur.393..660G}).
Although the NVSS (\S\ref{modsurv}) is not suitable to map the Galactic
foreground emission and its polarisation, it offers linear polarisation
data for $\sim$2 million radio sources. Many thousands of them will
have sufficient polarisation fractions to be followed up at other
frequencies, and to study their Faraday rotation and depolarisation behaviour.

\subsection{Surveys from Low to High Frequencies: Coverage and Content}  \label{surveycontent}

There is no concise list of all radio surveys ever made. Purton \& Durrell
(1991) used 233 different articles on radio source surveys, published 1954--1991,
to prepare a list of 386 distinct regions of sky covered by these surveys
({\tts cats.sao.ru/doc/SURSEARCH.html}).
While the source lists themselves were not available to these
authors, the list was the basis for a software allowing queries
to determine which surveys cover a given region of sky.
A method to retrieve references to radio surveys by acronym
has been mentioned in \S\ref{nomenclature}. In \S\ref{radcats} a
quantitative summary is given of what is available electronically.

In this section I shall present the ``peak of the iceberg'':
in Table~1, I have listed the largest surveys of discrete radio sources
which have led to source catalogues available in electronic form.
The list is sorted by frequency band (col.~1), and the emphasis is
on finder surveys with more than $\sim$800 sources {\it and} more
than $\sim$0.3 sources/deg$^{2}$. However, some other
surveys were included if they constitute a significant contribution
to our knowledge of the source population at a given frequency,
like e.g.\ re-observations of sources originally found at other frequencies.
It is supposedly complete for source catalogues with $\gtsim$\,2000 entries,
whereas below that limit a few source lists may be missing for not
fulfilling the above criteria.
Further columns give the acronym of the survey or observing instrument,
the year(s) of publication, the approximate range of RA and DEC covered
(or Galactic longitude {\tt l} and latitude {\tt b} for Galactic plane
surveys), the angular resolution in arcmin,
the approximate limiting flux density in mJy, the total number of sources
listed in the catalogue, the average surface density of sources
per square degree, and a reference number
which is resolved into its ``bibcode'' in the Notes to the table.
Three famous series of surveys are excluded from Table~1, as
they are not contiguous large-area surveys, but are dedicated to many
individual fields, either for Galactic or for cosmological studies
(e.g.\ source counts at faint flux levels).
These are the source lists from various individual pointings of the
interferometers at DRAO Penticton (P), Westerbork (W) and the
Cambridge One-Mile telescope (5C).

Both single-dish and interferometer surveys are included in Table~1.
While interferometers usually provide much higher absolute positional
accuracy, there is one major interferometer survey (TXS at 365\,MHz;
\cite{1996AJ....111.1945D}, {\tts utrao.as.utexas.edu/txs.html}),
for which one fifth of its $\sim$67,000 catalogued source positions
suffer from possible ``lobe-shifts''.
These sources have a certain likelihood to be located at an alternative,
but precisely determined position, about 1$'$ from the listed position.
It is not clear {\it a priori} which of the two positions is
the true one, but the ambiguity can usually be solved by comparison
with other sufficiently high resolution maps (see
Fig.~B1 of \cite{1998MNRAS.294..607V} for an example). For a reliable
cross-identification with other catalogues these alternative positions
obviously have to be taken into account.

\newpage
\centerline{Table 1.~~Major Surveys of Discrete Radio Sources~$\dagger$}

% \vspace*{-1mm}
% \scriptsize
\noindent
\rule[1.5mm]{134mm}{0.3mm}

\vspace*{-0.5mm}
%\begin{verbatim}
%--------------------------------------------------------------------------------
{\renewcommand{\baselinestretch}{0.8}
\footnotesize % \baselineskip 11 pt
\noindent
{\verb+Freq   Name  Year    RA(h)   Decl(deg)    HPBW  S_min    N of    n/   Ref Electr+} \\
{\verb+(MHz)       of publ  or l(d)  or b(d)      (')  (mJy)  objects sq.deg     Status+} \\
%\vspace*{-0.1mm}
\rule[1mm]{134mm}{0.3mm}
%--------------------------------------------------------------------------------
\begin{verbatim}
10-25 UTR-2  78-95    0-24     > -13      25-60 10000    1754   0.2   54  A C
  31  NEK     88   350<l<250  |b|<~2.5   13x 11  4000     703   0.7   51  A C
  38  8C     90/95    0-24     > +60       4.5   1000    5859   1.7    1  A C n
  80# CUL1   73/95    0-24    -48,+35      3.7   2000     999   0.04  41  A C
  80# CUL2   73/95    0-24    -48,+35      3.7   2000    1748   0.06  42  A C
  82  IPS     87      0-24    -10,+83    27x350   500    1789   0.08  52    C
 151  6CI     85      0-24     > +80       4.5    200    1761   5.7    2  A C
 151  6CII    88    8.5-17.5  +30,+51      4.5    200    8278   4.1    3  A C
 151  6CIII   90    5.5-18.3  +48,+68      4.5    200    8749   4.5    4  A C
 151  6CIV    91      0-24    +67,+82      4.5    200    5421   3.8   28  A C
 151  6CVa    93    1.6- 6.2  +48,+68      4.5   ~300    2229   3.0   39  A C
 151  6CVb    93   17.3-20.4  +48,+68      4.5   ~300    1229   2.6   39  A C
 151  6CVI    93   22.6- 9.1  +30,+51      4.5   ~300    6752   2.7   40  A C
 151* 7CI     90   (10.5+41)  (6.5+45)     1.2     80    4723   9.7   21    C
 151  7CII    95     15-19    +54,+76      1.2   ~100    2702   6.5   49  A C n
 151  7CIII   96      9-16    +20,+35      1.2   ~150    5526   4.0   56  A C N
 151  7C(G)   98   80<l<180   |b|<5.5      1.2   ~100    6262   4.8   55    C n
 160# CUL3   77/95    0-24    -48,+35      1.9   1200    2045   0.08  43  A C
 178  4C     65/67    0-24     -7,+80    ~23.    2000    4844   0.2   57  A C N M
 232  MIYUN   96      0-24    +30,+90      3.8   ~100   34426   3.3   24  A C
 325  WENSS  97/98    0-24    +30,+90      0.9     18  229420  ~22.   58    C
 327* WSRT   91/93  5 fields (+40,+72)    ~1.0      3    4157  ~50.   32  A C
 327  WSRTGP  96    43<l<91   |b|<1.6     ~1.0    ~10    3984  ~25.   30  A C n
 365  UTRAO36 92      0-24    +31,+41     ~0.1    250    3196   ~2.   38    C
 365  TXS     96      0-24   -35.5,+71.5  ~0.1    250   66841   ~2.   22  A C n
 408  MRC    81/91    0-24    -85,+18.5   ~3.     700   12141   0.5    6  A C N
 408  B2     70-74    0-24    +24,+40     3 x10   250    9929   3.1    7  A C   M
 408  B3      85      0-24    +37,+47     3 x 5   100   13354   5.2    8  A C N
 408  MC1     73      1-17    -22,-19      2.7    100    1545   2.3    9  A C   M
 408  MC4     76      0-18    -74,-62      2.7    130    1257   1.0   10  A C   M
 408  MDS2    84      5-23    -21,-20      2.8     60     799   2.7   11    C
 611  NAIC    75     22-13     -3,+19     12.6    350    3122   0.6   12    C
 608* WSRT   91/93 sev.fields (~40,~72)    0.5      3    1693  ~50.   32  A C
1400  GB      72      7-16    +46,+52    10 x11    90    1086   2.0   13    C
1400  GB2     78      7-17    +32,+40    10 x11    90    2022   2.2   14    C
1400  WB92    92      0-24     -5,+82    10 x11  ~150   31524   0.7   27  A C N
1400  GPSR    90    20<l<120  |b|<0.8      0.08    25    1992   8.9   33  A C
1400  NVSS34  98     0-24    -40,+90       0.9    2.0 1807317  ~55.   60    C
1400  FIRST5  98   7.3,17.4  +22.2,57.6    0.1    1.0  382892  ~90.   59  A C
1400  FIRST5  98   21.3,3.3  -11.5,+1.6    0.1    1.0   54537  ~90.   59  A C
1408  RRF     90   357<l<95.5  |b|<4.0     9.4     98     884   1.1   29  A C
1420  RRF     98   95.5<l<240 -4<|b|<+5    9.4     80    1830   1.5   44  A C
1420* PDF     98   B0112-46   r=1deg       0.1     0.1   1079  ~340.  62    C
1400  ELAISR  98   3 fields   +32,+55      0.25   0.14    867   205.  61    C
1400  GPSR    92   350<l<40   |b|<1.8      0.08    25    1457   8.1   37    C
1500  VLANEP  94   17.4,18.5  63.6,70.4    0.25   0.5    2436  83.    47  A C n
2700  PKS    (90)     0-24    -90,+27     ~8.     ~50    8264   0.3   15  A C N M
2700  F3R     90   357<l<240   |b|< 5      4.3     40    6483   2.7   34  A C
3900  Z       89      0-24     0,+14     1.2x52    50    8503   1.7   16  A C
3900  Z2      91      0-24     0,+14     1.2x52    40    2944   0.6    5  A C
3900  RC     91-93    0-24     4.5,5.5   1.2x52     4    1189   3.2   26    C n
4775# NAIC-GB 83    22.3-13   -3,+19       2.8    ~20    2661   0.6   17    C
4760  GBdeep  86      0-24      ~33        2.8     15     882   6.6   18    C
4850  MG1-4  86-91    var.   -0.5,+51      2.8     40   24180   1.5   20    C n
4850  87GB    91      0-24     0,+75      ~3.5     25   54579   2.7   19  A C N
4850  BWE     91      0-24     0,+75      ~3.5     25   53522   2.7   23  A C N
4850  GB6     96      0-24     0,+75      ~3.5     18   75162   3.7   53  A C
4850  PMNM    94      0-24    -88,-37      4.9     25   15045   1.8   45  A C N
4850  PMN-S   94      0-24   -87.5,-37     4.2     20   23277   2.8   31a A C N
4850  PMN-T   94      0-24    -29,-9.5     4.2     42   13363   2.0   31b A C N
4850  PMN-E   95      0-24    -9.5,+10     4.2     40   11774   1.9   48  A C N
4850  PMN-Z   96      0-24    -37,-29      4.2     72    2400   1.1   50  A C N
4875  ADP79   79   357<l< 60   |b|<1       2.6   ~120    1186   9.4   25    C
5000  HCS79   79   190<l< 40   |b|<2       4.1    260     915   1.1   46  A C
5000  GT      86    40<l<220   |b|<2       2.8     70    1274   1.8   35    C
5000  GPSR    94   350<l< 40   |b|<0.4    ~0.07     3    1272  26.    36  A C
\end{verbatim}
\vspace*{-2.5mm}
\rule{134mm}{0.3mm}
%----------------------------------------------------------------------------------
}

\small
{\renewcommand{\baselinestretch}{0.8}
\vspace*{-0.5mm}
\noindent
$\dagger$~ A total of 66 surveys are listed with altogether 3,058,035 entries. \\
\hspace*{5mm} See the explanations and references in the Notes to this Table.
}

\newpage
{\renewcommand{\baselinestretch}{0.8}
\centerline{References and Notes to Table 1}
\footnotesize
% \vspace*{-2.5mm}
\noindent
\rule{134mm}{0.3mm}
%------------------------------------------------------------------------------------
\begin{verbatim}
 1a 1995MNRAS.274..447Hales+         | 29  1990A&AS...83..539Reich W.+
 1b 1990MNRAS.244..233Rees           | 30  1996ApJS..107..239Taylor+
 2  1985MNRAS.217..717Baldwin+       | 31a 1994ApJS...91..111Wright+
 3  1988MNRAS.234..919Hales+         | 31b 1994ApJS...90..179Griffith+
 4  1990MNRAS.246..256Hales+         | 32  1993BICDS..43...17Wieringa +PhD Leiden
 5  1991SoSAO..68...14Larionov+      | 33  1990ApJS...74..181Zoonematkermani+
 6a 1991Obs...111...72Large+         | 34  1990A&AS...85..805Fuerst+
 6b 1981MNRAS.194..693Large+         | 35  1986AJ.....92..371Gregory & Taylor
 7a 1970A&AS....1..281Colla+         | 36  1994ApJS...91..347Becker+
 7b 1972A&AS....7....1Colla+         | 37  1992ApJS...80..211Helfand+
 7c 1973A&AS...11..291Colla+         | 38  1992ApJS...82....1Bozyan+
 7d 1974A&AS...18..147Fanti+         | 39  1993MNRAS.262.1057Hales+
 8  1985A&AS...59..255Ficarra+       | 40  1993MNRAS.263...25Hales+
 9  1973AuJPA..28....1Davies+        | 41a 1973AuJPA..27....1Slee & Higgins
10  1976AuJPA..40....1Clarke+        | 41b 1995AuJPh..48..143Slee
11  1984PASAu...5..290White          | 42a 1975AuJPA..36....1Slee & Higgins
12  1975NAICR..45.....Durdin+        | 42b 1995AuJPh..48..143Slee
      NAIC Internal Report           | 43a 1977AuJPA..43....1Slee
13  1972AcA....22..227Maslowski      | 43b 1995AuJPh..48..143Slee
14  1978AcA....28..367Machalski      | 44  1997A&AS..126..413Reich, P.+
15  1991PASAu...9..170Otrupcek+Wright| 45a 1994ApJS...90..173Gregory+
16  1989MIRpubl.......Amirkhanyan+   | 45b 1993AJ....106.1095Condon+
      MIR Publ., Moscow              | 46  1979AuJPA..48....1Haynes+
17  1983ApJS...51...67Lawrence+      | 47  1994ApJS...93..145Kollgaard+
18  1986A&AS...65..267Altschuler     | 48  1995ApJS...97..347Griffith+
19  1991ApJS...75.1011Gregory+Condon | 49  1995A&AS..110..419Visser+
20a 1986ApJS...61....1Bennett+       | 50  1996ApJS..103..145Wright+
20b 1990ApJS...72..621Langston+      | 51  1988ApJS...68..715Kassim
20c 1990ApJS...74..129Griffith+      | 52  1987MNRAS.229..589Purvis+
20d 1991ApJS...75..801Griffith+      | 53  1996ApJS..103..427Gregory+
21  1990MNRAS.246..110McGilchrist+   | 54  1995Ap&SS.226..245Braude+ +older refs
22  1996AJ....111.1945Douglas+       | 55  1998MNRAS.294..607Vessey & Green D.A.
23  1991ApJS...75....1Becker+        | 56  1996MNRAS.282..779Waldram+
24  1997A&AS..121...59Zhang+         | 57a 1965MmRAS..69..183Pilkington & Scott
25  1979A&AS...35...23Altenhoff+     | 57b 1967MmRAS..71...49Gower+
26a 1991A&AS...87....1Parijskij+     | 58  1997A&AS..124..259Rengelink+ and WWW
26b 1992A&AS...96..583Parijskij+     | 59  1997ApJ...475..479White+  and WWW
26c 1993A&AS...98..391Parijskij+     | 60  1998AJ....115.1693Condon+ and WWW
27  1992ApJS...79..331White & Becker | 61  1998MNRAS... Ciliegi+ astro-ph/9805353
28  1991MNRAS.251...46Hales+         | 62  1998MNRAS.296..839Hopkins+ +PhD Sydney
\end{verbatim}
%------------------------------------------------------------------------------------
\vspace*{-1mm}
\rule{134mm}{0.3mm}
}

%\vspace*{-1mm}
\small
{\renewcommand{\baselinestretch}{0.85}
\noindent
Notes to Table~1. \#: not a finder survey, but re-observations of previously
catalogued sources.~ \linebreak[4] *: circular field, central coordinates and radius are given.
The catalogue electronic status is coded as follows\,:
A:~available from ADC/CDS (\S\ref{radcats});
C:~(all of them!) searchable simultaneously via CATS (\S\ref{catcoll});
N:~fluxes are in NED; n:~source positions are in
NED (cf.~\S\ref{objdatabases});~ M:~included in MSL (\S\ref{radcats}).
An update of this table is kept at~ {\tts cats.sao.ru/doc/MAJOR\_CATS.html}.
}
\rule{134mm}{0.3mm}

\normalsize
\vspace*{4mm}

\renewcommand{\baselinestretch}{1.0}
%\noindent

The angular resolution of the surveys tends to increase with observing
frequency, while the lowest flux density detected tends to decrease
(but increase again above $\sim$8\,GHz). In fact, until recently the
relation between observing frequency, $\nu$, and limiting flux density,
$S_{\rm lim}$, of large-scale surveys between 10\,MHz and 5\,GHz followed
rather closely the power-law spectrum of an average extragalactic radio source,~
$S\sim\nu^{-0.7}$. This implied a certain bias against the detection
of sources with rare spectra, like e.g.\ the ``compact steep spectrum'' (CSS)
or the ``GHz-peaked spectrum'' (GPS) sources (\cite{1998PASP..110..493O}).
With the new, deep, large-scale radio surveys like WENSS, NVSS and
FIRST (\S\ref{modsurv}), with a sensitivity of 10--50 times better than
previous ones, one should be able to construct much larger samples
of these cosmologically important type of sources
(cf.\ \cite{snellen96}).
A taste of some cosmological applications possible with these new
radio surveys has been given in the proceedings volume by \cite{obscos98}.

Table~1 also shows that there are no appreciable source surveys at frequencies
higher than 5\,GHz, mainly for technical reasons: it takes large amounts
of telescope time to cover a large area of sky to a reasonably low flux limit
with a comparatively small beam. New receiver technology as well as
new scanning techniques will be needed. For example, by continuously (and slowly)
slewing with all elements of an array like the VLA, an adequately
dense grid of phase centres for mosaicing could be simulated using an
appropriate integration time. More probably, the largest gain in
knowledge about the mm-wave radio sky will come from the imminent
space missions for microwave background studies, MAP and PLANCK
(see \S\ref{extendfreq}).  Currently there is no pressing evidence
for ``new'' source populations dominating at mm waves (cf.\
sect.\,3.3 of \cite{1995AJ....109.2318C}), although some examples
among weaker sources were found recently (\cite{1996ApJ...460..225C},
\cite{1998AJ....115.1388C}). Surveys at frequencies well above
5\,GHz are thus important to quantify how such sources would affect
the interpretation of the fluctuations of the microwave background.
Until now, these estimates rely on mere extrapolations of source spectra
at lower frequencies, and certainly the information content of the surveys
in Table~1 has not at all been fully exploited for this purpose.

Table~1 is an updated version of an earlier one (\cite{andern92})
which listed 38 surveys with $\sim$450,000 entries. In 1992 I speculated
that by 2000 the number of measured flux densities would have quadrupled.
The current number (in 1998!) is already seven times the number for 1992.

\subsection{Optical Identification Content} \label{optiddat}

The current information on sources within our Galaxy is
summarized in \S\ref{gplansurv}.
The vast majority of radio sources more than a few degrees
away from the Galactic plane are extragalactic. The latest compilation
of optical identifications of extragalactic radio sources dates
back to 1983 (\cite{1983A&AS...53..219V}, hereafter VV83)
and lists 14,585 entries for 10,173 different sources, based on
917 publications.
About 25\% of these are listed as ``empty'', ``blank'' or ``obscured''
fields (EF, BF, or OF), i.e.\ no optical counterpart has been found
to the limits of detection.
The VV83 compilation has not been updated since 1983, and is not to be
confused with the ``Catalogue of Quasars and Active Galactic Nuclei''
by the same authors. Both compilations are sometimes referred to as
the (``well-known'') ``V\'eron catalogue'', but usually the latter
is meant, and only the latter is being updated
(\cite{1998ESOSR..18....1V} or ``VV98'').
The only other (partial) effort of a compilation similar to VV83 was
PKSCAT90 (\cite{1991PASAu...9..170O}), which was restricted to the 8\,263
fairly strong PKS radio sources and, contrary to initial plans, has not
been updated since 1990. It also lacks quite a few references published
before 1990.

{\it For how many radio sources do we know an optical counterpart\,?}
From Table~1 we may very crudely estimate that currently well over
2 million radio sources are known ($\sim$3.3 million individual measurements
are available electronically).  A compilation of references
(not included in VV83) on optical identifications of
radio sources maintained by the present author currently holds
$\sim$560 references dealing with a total of $\sim$56,000 objects.
This leads the author to estimate that an optical identification (or absence thereof)
has been reported for $\sim$20--40,000 sources.  Note that probably
quite a few of these will either occur in more than one reference or
be empty fields.  Most of the information contained in VV83 is absent
from pertinent object databases (\S\ref{objdatabases}), given that these
started including extragalactic data only since 1983
(SIMBAD) and 1988 (NED). However, most of the optical identifications
published since 1988 can be found in NED. Moreover, numerous
optical identifications of radio sources have been made quietly
(i.e.\ outside any explicit publication) by the NED team. Currently
(May~98) NED contains $\sim$9,800 extragalactic objects which are
also radio sources. Only 57\% of these have a redshift in NED.
Even if we add to this some 2--3,000 optically identified Galactic
sources (\S\ref{gplansurv}) we can state fairly safely that
of all known radio sources, we currently know the optical counterpart
for {\it at most half a percent}, and the distance for no more
than {\it a quarter percent}.
The number of counterparts is likely to increase by thousands
once the new large radio survey catalogues (WENSS, NVSS, FIRST),
as well as new optical galaxy catalogues, e.g.\ from APM
({\tts \verb*Cwww.ast.cam.ac.uk/~apmcatC}), SuperCOSMOS
({\tts www.roe.ac.uk/scosmos.html}) or SDSS (\S\ref{first}),
become available. Clearly, more automated
identification methods and multifibre spectroscopy (like e.g.\ 2dF,
FLAIR, and 6dF, all available from {\tts www.aao.gov.au/})
will be the only way to reduce the growing gap between the number of
catalogued sources and the knowledge about their counterparts.

\subsection{Galactic Plane Surveys and Galactic Sources} \label{gplansurv}

Some of the major discrete source surveys of the Galactic plane
are included in Table~1 (those for which a range in l and b are
listed in columns 4 and 5, and several others covering the plane).
Lists of ``high''-resolution surveys of the Galactic radio continuum
up to 1987 have been given in \cite{1988ApJS...68..715K} and  % \cite{reich91}
\cite{1991IAUS..144..187R}. Due the high density of sources, many of them
with complex structure, the Galactic plane is the most difficult region
for the preparation of discrete source catalogues from maps.
The often unusual shapes of radio continuum sources have led to
designations like the ``snake'', the ``bedspring'' or ``tornado'', the ``mouse''
(cf.\ \cite{1994MNRAS.270..822G}) or a ``chimney'' (\cite{1996Natur.380..687N}).
For extractions of images from some of these surveys see~ \S\ref{images}.

{\it What kind of discrete radio sources can be found in our Galaxy\,?} \\
Of the 100,000 brightest radio sources in the sky, fewer than 20 are stars.
A compilation of radio observations of $\sim$3000 {\bf Galactic stars} has been
maintained until recently by \cite{1995A&AS..109..177W}.
The electronic version is available from ADC/CDS
(catalogue \#\,2199, \S\ref{radcats}) and includes flux densities for about 800
detected stars and upper limits for the rest.
This compilation is not being updated any more. The most recent
push for the detection of new radio stars has just come from
a cross-identification of the FIRST and NVSS catalogues with star catalogues.
In the FIRST survey region the number of known radio stars has
tripled with a few dozen FIRST detections (S$\gtsim$\,1\,mJy at 1.4\,GHz,
\cite{1997AAS...191.1403H}), and 50 (mostly new) radio stars were found
% (\cite{1997BAAS...29.1231H}),
in the NVSS (\cite{1997AAS...191.1402C}), many of them radio variable.
% (\cite{1997BAAS...29.1231C})

A very complete WWW page on {\bf Supernovae} (Sne), including
SNRs, is offered by Marcos J.~Montes at
{\tts \verb*Ccssa.stanford.edu/~marcos/sne.htmlC}.
It provides links to other supernova-related pages, to catalogues of SNe
and SNR, to individual researchers, as well as preprints, meetings and
proceedings on the subject.
D.A.~Green maintains his \linebreak[4]
``Catalogue of Galactic Supernova Remnants''
at {\tts www.mrao.cam.ac.uk/surveys/snrs/}. \linebreak[4]
The catalogue contains details of confirmed Galactic SNRs (almost all
are radio SNRs), and includes bibliographic references, together with lists
of other possible and probable Galactic SNRs.  From a Galactic plane survey
with the RATAN-600 telescope (\cite{1996A&ATr..11..225T}) S.~Trushkin
derived radio profiles along RA at 3.9, 7.7, and 11.1\,GHz for 70 SNRs
at~ {\tts cats.sao.ru/doc/Atlas\_snr.html}
(cf.\ \cite{1996BSAO...41...64T}). Radio continuum spectra
for 192 of the 215 SNRs in Green's catalogue (\cite{trushkin98}) may be
displayed at {\tts cats.sao.ru/cats\_spectra.html}.

{\bf Planetary nebulae} (PNe), the expanding shells of stars in a late stage
of evolution, all emit free-free radio radiation. The deepest
large-scale radio search of PNe has been performed by \cite{1998ApJS..11x..yyyC},
who cross-identified the ``Strasbourg-ESO Catalogue of Galactic Planetary Nebulae''
(SESO, available as ADC/CDS \#\,5084) with the NVSS catalogue. To do
this, some of the poorer optical positions in SESO for the 885 PNe north
of $\delta$=$-$40\deg\ had to be re-measured on the
Digitized Sky Survey (DSS; {\tts archive.stsci.edu/dss/dss\_form.html}).
The authors detect 680 (77\%) PNe brighter than about
S(1.4\,GHz)\,=\,2.5\,mJy/beam.
A database of Galactic Planetary Nebulae is maintained at
Innsbruck ({\tts ast2.uibk.ac.at/}). However, the classification of PNe
is a tricky subject, as shown by several publications over the past
two decades (e.g.\ \cite{1997AN....318...35K}, \cite{1991A&AS...87..499A},
or \cite{1990A&AS...86..219A}).
Thus the presence in a catalogue should not be taken as ultimate proof of
its classification.

{\bf H\,II regions} are clouds of almost fully ionised hydrogen found
throughout most late-type galaxies. Major compilations of H\,II regions
in our Galaxy were published by \cite{1959ApJS....4..257S} (N=313)
and \cite{1974Ap&SS..27....3M} (N=698).
A graphical tool to create charts with objects from 17 catalogues
covering the Galactic Plane, the {\it Milky Way Concordance}
({\tts \verb*Ccfa-www.harvard.edu/~peterb/concordC}),
has already been mentioned in my tutorial in this volume.
Methods to find candidate H\,II regions based on IR colours
of IRAS Point Sources have been given in \cite{1989AJ.....97..786H} and
% \cite{1989ApJ...340..265W}
Wood \& Churchwell (1989), and were further exploited to confirm
ultracompact H\,II regions (UC\,H\,II) via radio continuum observations
(\cite{1994ApJS...91..659K}) or 6.7\,GHz methanol maser searches \linebreak[4]
(\cite{1997MNRAS.291..261W}).  \cite{1997ApJ...488..224K} merged six
previous compilations to construct an all-sky list of 1048 Galactic H\,II
regions, in order to look for radio counterparts in the 87GB and
PMN maps at 4.85\,GHz. They detect
about 760 H\,II regions above the survey threshold of $\sim$30\,mJy
(87GB) and $\sim$60\,mJy (PMN). These authors also point out the
very different characteristics of these surveys, the 87GB being much
poorer in extended Galactic plane sources than the PMN, for the reasons
mentioned above (\S\ref{majorradiosurveys}).

The ``Princeton Pulsar Group'' ({\tts pulsar.princeton.edu/})
offers basic explanations of the pulsar phenomenon, a calculator
to convert between dispersion measure and distance for user-specified
Galactic coordinates, software for analysis of pulsar timing data,
links to pulsar researchers, and even audio-versions of the pulses of a
few pulsars. The largest catalogue of known {\bf pulsars}, originally
published with 558 records by \cite{1993ApJS...88..529T} is also
maintained and searchable there (with currently 706 entries).
Pulsars have very steep radio spectra
(e.g.\ \cite{malofeev96}, or {\tts astro-ph/9801059, 9805241}),
are point-like and polarised, so that pulsar candidates can be found
from these criteria in large source surveys (\cite{1996ASPC..105...15K}).
Data on pulsars, up to pulse profiles of individual pulsars,
from dozens of different papers can be found at the
``European Pulsar Network'' ({\tts www.mpifr-bonn.mpg.de/pulsar/data/}).
They have developed a flexible data format for exchange of
pulsar data (\cite{1998A&AS..128..541L}), which is now used in
an on-line database of pulse profiles as well as an interface for
their simultaneous observations of single pulses. The database can
be searched by various criteria like equatorial and/or Galactic
coordinates, observing frequency, pulsar period and dispersion
measure (DM).
Further links on radio pulsar resources have been compiled at
{\tts pulsar.princeton.edu/rpr.shtml}, including
many recent papers on pulsar research. \cite{1998ApJS..11x..yyyK} have used
the NVSS to search for phase-averaged radio emission from the pulsars
north of $\delta_{2000}$=$-$40\deg\ in the \cite{1993ApJS...88..529T}
pulsar catalogue. They identify 79 of these pulsars with a flux
of S(1.4\,GHz)\,$\gtsim$\,2.5\,mJy, and 15 of them are also in the
WENSS source catalogue.

An excellent description of the various types of Galactic radio
sources, including masers, is given in several of the chapters of \cite{verkel88}.

Last, but not least, Galactic plane radio sources can point us to
galaxies and clusters in the ``{\bf Zone of Avoidance}'' (ZOA).
In fact, in a large number of surveys for discrete radio sources,
the Galactic plane does not show any excess number density of
(usually compact) sources, e.g.\ in TXS (\cite{1996AJ....111.1945D})
or BWE (\cite{1991ApJS...75....1B}). In a 2.7\,GHz survey of the
region~ $-$3\deg$<\ell<$\,240\deg, $|b|<$5\deg, with the
Effelsberg 100-m telescope, the density of unresolved sources
($\ltsim$\,1$'$ intrinsic size) was {\it not} found to vary
with Galactic latitude (\cite{1990A&AS...85..805F}). At much
higher resolution (5$''$), using the VLA to cover the area
$-$10\deg$<\ell<$\,40\deg, $|b|<$1.8\deg, a concentration
of compact sources (size $\ltsim$20$''$) towards the Galactic
plane becomes noticeable, but only for $|b|<$\,0.4\deg\
(\cite{1992ApJS...80..211H}).
\cite{1994ApJS...91..347B} have shown that at 5\,GHz this
distribution has a width of only 10$'$--15$'$.  Most of the
extragalactic sources will be far too optically faint to be ever
identified. Two notable counter-examples are the prototype
``head-tail'' radio source 3C\,129 (z=0.021; \cite{1972Natur.237..269M}),
now known to be a member of the Perseus supercluster of galaxies
(\cite{1987A&A...184...43H}), and ``Centaurus\,B'' (see \S\ref{nomenclature}).
Other examples are two tailed radio
sources, PKS\,B1610$-$608 and PKS\,B1610$-$605 in Abell cluster A\,3627,
which is thought to be the central clump of the ``Great Attractor''
(\cite{1997PASA...14...15K}). A typical wide-angle tailed (WAT) radio
source (G\,357.30$+$01.24) has been found very close to the Galactic
centre, indicating the presence of a cluster of galaxies in that
direction (\cite{1994MNRAS.270..861G}). Due to the extreme optical
obscuration, there is little hope of optically identifying
this cluster and determining its distance.

\subsection{Modern Large-Scale Discrete Source Surveys: NVSS, FIRST, WENSS and SUMSS} \label{modsurv}

Some of the first large-scale contiguous surveys with interferometers
had become available in the 1980s. These were made with arcmin resolution
at low frequencies where the large fields of view required only a
few pointings (e.g.\ 6C or 8C; {\tts www.mrao.cam.ac.uk/surveys/}).
Only in the 1990s, however, has the increase in computing power
allowed such surveys to be made at even sub-arcmin resolution with the most
powerful interferometers like the VLA and the WSRT, requiring up to
a quarter of a million pointings.  Four ongoing or recently finished surveys
in this category are described below.

\subsubsection{The ``WENSS'' Survey at 325--350\,MHz}  \label{wenss}

The ``Westerbork Northern Sky Survey'' (WENSS; {\tts www.strw.leidenuniv.nl/wenss})
is a radio survey made with the WSRT ({\tts www.nfra.nl/wsrt/wsrtpage.htm})
from late 1990 to 1996, at frequencies 325 and 610\,MHz ($\lambda$ 92 and 49\,cm).
The entire sky north of declination $+$30\deg\ has been covered with
$\sim$6000 pointings, using a central frequency of 325\,MHz below $\delta$=74\deg,
and 350\,MHz for the polar region.  Only 2000 square degrees ($\sim$20\% of the sky
north of $+$30\deg) were mapped at 610\,MHz, and for the time being only
the 325\,MHz data have been made available to the public.
At 325\,MHz the resolution is 54$''\times$54$''$\,csc($\delta$), and the
positional accuracy for strong sources is 1.5$''$.
The limiting flux density is $\sim$18 mJy (5 $\sigma$) at both
frequencies.
The final products of WENSS, a 325-MHz atlas of 6\deg$\times$6\deg\ maps
centred on the new POSS plate positions (5\deg\ grid)
as well as a source catalogue are now available. FITS maps
can be drawn from the anonymous {\tt ftp} server at
{\tts ftp://vliet.strw.leidenuniv.nl/pub/wenss/HIGHRES/}.
A postage stamp server for extraction of smaller images is
planned for the near future.
Also total intensity maps at a lower resolution of 4.2$'$ will soon be
made available.
Ionospheric instabilities make the generation of polarisation maps
(Stokes Q, U and V) formidably difficult and their production is not
currently foreseen.
% see also {\tts www.cv.nrao.edu/html/newsletter/nraonews75.html}

Presently there are two source catalogues at 325\,MHz.
The main catalogue contains 211,235 sources for 28\deg\ $<\delta <$76\deg\ .
The polar catalogue contains 18,341 sources above 74\deg\ .
These can be browsed at the URL~ {\tts www.strw.leidenuniv.nl/wenss/search.html}.
A detailed description of the survey and the contents of the
source lists are given in \cite{1997A&AS..124..259R}.

Due to its low frequency and sensitivity to extended structure
the WENSS survey is well-suited to detect very extended or low surface
brightness objects like giant radio galaxies (cf.\ \cite{1998A&A...VVV..pppS}),
cluster haloes, and nearby galaxies. Comparing WENSS with other surveys
at higher frequencies allows one to isolate candidates for high-redshift
radio galaxies, GHz-peaked spectrum (GPS) sources, flat spectrum sources
(e.g.\ high-redshift quasars), and pulsars.

\subsubsection{The ``NRAO VLA Sky Survey'' at 1.4\,GHz (NVSS)}  \label{nvss}

The VLA has been used from 1993 to 1997 to map the entire sky north
of $\delta$=$-$40\deg\ (82\% of the sky) in its most compact (D)
configuration, giving an angular resolution of 45$''$ at 1.4\,GHz.
About 220,000 individual snapshots (phase centres) have been observed.
They were of a mere 23.5\,sec duration each, except at low elevation
when they were increased to up to 60~sec to make up for the
loss of sensitivity due to ground radiation and air mass.
A detailed description is given in  \cite{1998AJ....115.1693C}
({\tts ftp://www.cv.nrao.edu/pub/nvss/paper.ps}).
% \cite{nvss98}.
The principal data products are:
\begin{itemize}
\item {A set of 2326 continuum map ``cubes,'' 4\deg$\times$4\deg\ with images
of Stokes parameters I, Q, and U. The noise level is $\sim$0.45 mJy/beam in I,
and 0.29 mJy/beam in Q and U. Positional accuracy varies from $<$\,1$''$
for strong (S$>$15 mJy) point sources to 7$''$ for the faintest ($\sim$2.3 mJy)
detectable sources.}
\item {A catalogue of $\sim$2,000,000 discrete sources detected in the entire
survey}
\item {Processed uv-data (visibilities) for each map cube constructed from over
100 individual pointings, for users wishing to investigate the data underlying
the images.}
\end{itemize}

The NVSS is accessible from {\tts \verb*Cwww.cv.nrao.edu/~jcondon/nvss.htmlC},
and is virtually complete at the time of writing.
The latest version of the NVSS catalogue (\#\,34, May~98) is a single
152\,Mb FITS file with 1.8$\times$10$^{6}$ sources. It can be downloaded via anonymous
{\tt ftp}, but users interested in exploiting the entire catalogue may
consider requesting a tape copy from NRAO. The publicly available program
{\tts NVSSlist} can extract selected portions of the catalogue very rapidly
and is easily installed on the user's local disk
for extensive cross-identification projects.

\begin{figure*}[!ht]
\vspace*{2mm}

\hspace*{-3mm}
\mbox{
%\vspace*{5mm}
\epsfig{file=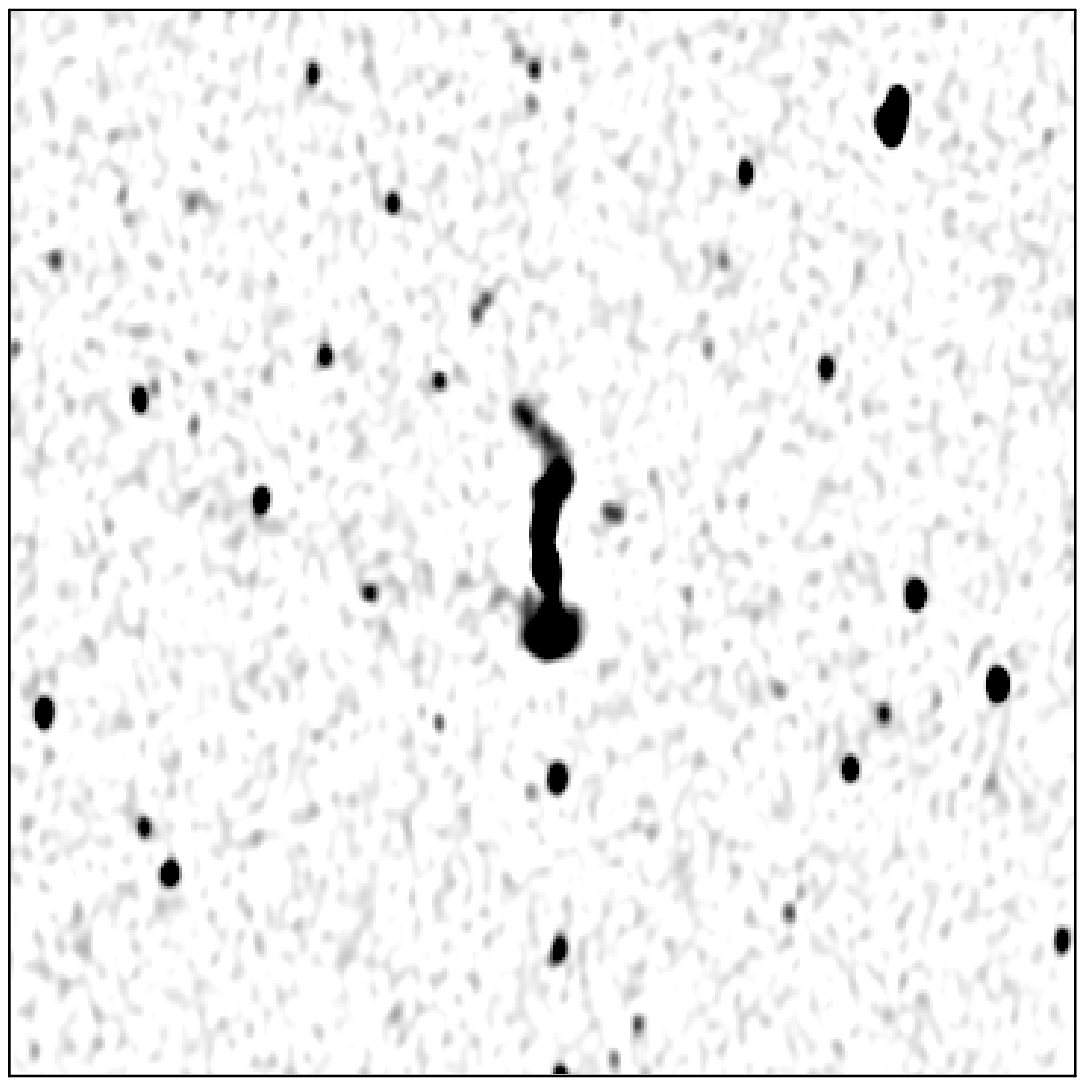,width=6.6cm,angle=-90}
\hspace*{-1mm}

\epsfig{file=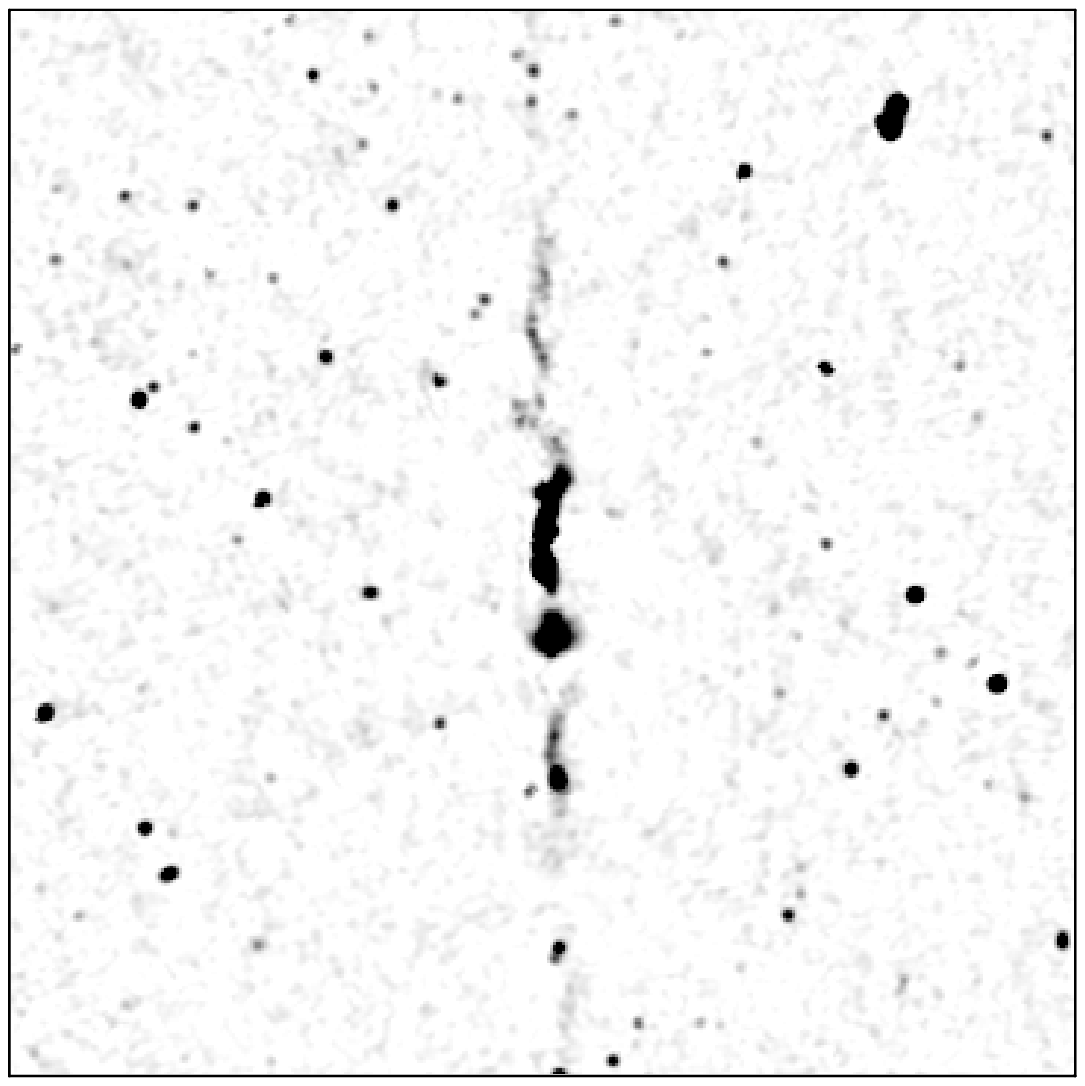,width=6.6cm,angle=-90}}
\caption{Reality and ``ghosts'' in radio maps of 1.5\deg$\times$1.5\deg\
centred on the radio galaxy 3C\,449.
Left: a 325\,MHz WENSS map shows the true extent ($\sim$23$'$) of 3C\,449.
Right: The 1.4\,GHz NVSS map shows additional weak ghost images extending
up to $\sim$40$'$, both north and south of 3C\,449. This occurs for very
short exposures (here 23~sec) when there is extended emission
along the projection of one of the VLA ``arms'' (here the north arm).
Note that neither map shows any indication of the extended
foreground emission found at 1.4\,GHz coincident with an optical
emission nebula stretching from NE to SW over the entire area shown
(Fig.~4c of Andernach et al.\ 1992). % \cite{1992A&AS...93..331A}
However, with longer integrations,
the uv-coverage of interferometers is sufficient to show the feature
(Leahy et al.\ 1998).} \label{3c449fig}
%({\tts www.jb.man.ac.uk/atlas/object/3C449.html})}
\end{figure*}

The catalogue can also be browsed at {\tts www.cv.nrao.edu/NVSS/NVSS.html}
and a ``postage stamp server'' to extract NVSS images is available at
{\tts www.cv.nrao.edu/NVSS/postage.html}.  Images are also
available from Skyview ({\tts skyview.gsfc.nasa.gov/}), but they neither are
as up-to-date as those at NRAO, nor do thay have the same FITS header (\S\ref{software}).
As always, care must be taken in the interpretation of these
images. Short integration times and poor uv-coverage can cause grating
residuals and limited sensitivity to extended structure (see Fig.~\ref{3c449fig}).

\subsubsection{The ``FIRST'' Survey at 1.4\,GHz}   \label{first}

The VLA has been used at 1.4\,GHz ($\lambda$=21.4\,cm) in its
B-configuration for another large-scale survey at 5$''$ resolution.
It is called FIRST (``Faint Images of the Radio Sky at Twenty-centimeters'')
and is designed to produce the radio equivalent of the Palomar Observatory Sky
Survey over 10,000 square degrees of the North Galactic Cap.
An automated mapping pipeline produces images with 1.8$''$ pixels and
a typical rms noise of 0.15 mJy. At the 1~mJy source detection
threshold, there are $\sim$90 sources per square degree, about a third of
which have resolved structure on scales from 2$''$--30$''$.

Individual sources have 90\% confidence error circles of radius
$<$0.5$''$ at the 3~mJy level and 1$''$ at the survey threshold.
Approximately 15\% of the sources have optical counterparts at the
limit of the POSS-I plates (E$\sim$20.0), and  unambiguous
optical identifications are achievable to m$_{\rm v}\sim$24.
The survey area has been chosen to coincide with that of the
Sloan Digital Sky Survey (SDSS; {\tts www-sdss.fnal.gov:8000/}).
This area consists mainly of the north Galactic cap
($|b|>+30$\deg) and a smaller region in the south Galactic
hemisphere.  At the m$_{\rm v}\sim$24 limit of SDSS, about half
of the optical counterparts to FIRST sources will be detected.

The homepage of FIRST is~ {\tts sundog.stsci.edu/}.
By late 1997 the survey had covered about 5000 square degrees.
The catalogue of the entire region (with presently $\sim$437,000 sources)
can be searched interactively at~ {\tts sundog.stsci.edu/cgi-bin/searchfirst}.
A postage stamp server for FIRST images (presently for 3000 square degrees)
is available at~ {\tts third.llnl.gov/cgi-bin/firstcutout}.
For 1998 and 1999, the FIRST survey was granted enough time to cover an
additional 3000 square degrees.

\begin{figure*}[!ht]  % \label{3c40fig}
%\vspace*{3mm}

\hspace*{-8mm}
\vspace*{4mm}
\mbox{

\epsfig{file=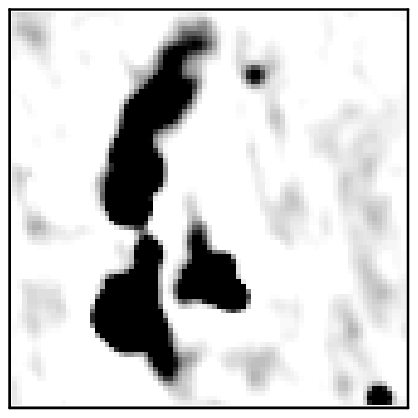,width=7.5cm,angle=-90}
\hspace*{-7mm}

\vspace*{-10mm}
\epsfig{file=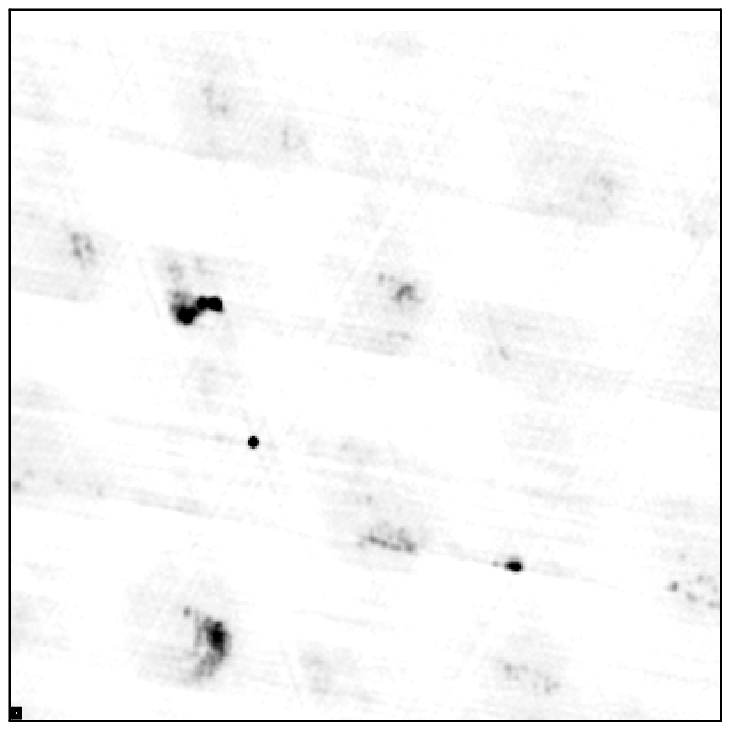,width=7.0cm,angle=-90}
%\vspace*{0.8cm}
}
\caption{1.4\,GHz maps of the radio galaxy 3C~40 (PKS~B0123$-$016).
Left: NVSS map of 20$'\times$20$'$.
The point-like component at the gravity centre of the radio complex
coincides with NGC~547, a dominant dumb-bell galaxy in the core of the
A\,194 galaxy cluster (cf.\ Fig.~2a in my tutorial)\,;
the head-tail like source $\sim$5$'$ due SW is NGC~541.
Right: FIRST map of 12$'\times$12$'$: only the strongest parts
of each component are detected and show fine structure, but appear
unrelated. The radio core of NGC~547 is unresolved. Some
fainter components appear to be artefacts.} \label{3c40fig}
\end{figure*}

The availability of the full NVSS data products has reduced the
enthusiasm of parts of the community to support the finishing of
FIRST's goals. However, only the FIRST survey (and less so the NVSS)
provides positions accurate enough for reliable optical identifications,
particularly for the cosmologically interesting faint and compact sources.
On the other hand, in Figure~\ref{3c40fig} I have shown an extreme example
of the advantage of NVSS for studies of extended sources. In fact,
the Figure shows the complementary properties of NVSS and FIRST.
A very extended source, perhaps just recognisable with NVSS,
will be broken up by FIRST into apparently unrelated components.
Thus, it would be worthwhile to look into the feasibility of merging the uv data
of NVSS and FIRST to create maps at 10$''$--15$''$ resolution in
the region covered by both surveys.

\subsubsection{The ``SUMSS'' 843\,MHz Survey with ``MOST''}   \label{sumss}

Since 1994 the ``Molonglo Observatory Synthesis Telescope'' (MOST)
has been upgraded from the previous 70$'$ field of view to a
2.7\deg\ diameter field of view. As MOST's aperture is almost filled,
the image contains Fourier components with
a wide range of angular scales, and has low sidelobes.
In mid-1997, the MOST started the
``Sydney University Molonglo Sky Survey'' (SUMSS; \cite{hunst98}).
The entire sky south of DEC=$-$30\deg\ and $|b|>$10\deg\
will be mapped at 843\,MHz, a total of 8000 square degrees
covered by 2713 different 12-h synthesis field centres. It
complements northern surveys like WENSS and NVSS, and it overlaps with
NVSS in a 10\deg\ strip in declination ($-$30\deg\ to $-$40\deg),
so as to allow spectral comparisons.
SUMSS is effectively a continuation of NVSS to the southern
hemisphere (see Table~\ref{majsurvtab}). However, with its much better
uv coverage it surpasses both WENSS and NVSS in sensitivity to
low surface brightness features, and it can fill in some of the ``holes''
in the uv plane where it overlaps with NVSS.  The MOST is also
being used to perform a Galactic plane survey (\S\ref{future}).

SUMSS positions are uncertain by no more than 1$''$ for sources brighter
than 20\,mJy, increasing to $\sim$2$''$ at 10~mJy and 3--5$''$ at 5\,mJy,
so that reliable optical identifications of sources close to the
survey limit may be made, at least at high Galactic latitude.
In fact, the identification rate on the DSS (\S\ref{gplansurv}) is
$\sim$30\% down to b$_{\rm J}\sim$22 (\cite{sadler98}).
Observations are made only at night, so the survey rate is
$\sim$1000 deg$^2$ per year, implying a total period of 8 years for
the data collection.  The south Galactic cap ($b<-$30\deg)
should be completed by mid-2000. The release of the first
mosaic images (4\deg$\times$4\deg) is expected for late 1998.
The SUMSS team at Univ.\ Sydney plans to use the NVSS WWW software,
so that access to SUMSS will look similar to that for NVSS.
For basic information about MOST and SUMSS
see~ {\tts www.physics.usyd.edu.au/astrop/SUMSS/}.

\setcounter{table}{1}
\begin{table*}[!ht]  %  \label{majsurvtab}
\begin{center}
% Table~2. Comparison of the new large-area radio surveys
\caption{Comparison of the new large-area radio surveys} \label{majsurvtab}
\begin{tabular}{lcccc}  \hline
                 & WENSS        & SUMSS   & NVSS   &   FIRST \\ \hline
Frequency        & 325 MHz     & 843 MHz   & 1400 MHz    &  1400 MHz  \\
Area (deg$^2$)  & 10,100       & 8,000    & 33,700      & 10,000  \\
Resolution ($''$) & $54 \times 54\, {\rm csc} (\delta)$ & $43 \times 43\, {\rm csc} (|\delta|)$ & 45 & 5 \\
Detection limit     & 15 mJy    &   $<$5 mJy &   2.5 mJy    &   1.0 mJy  \\
Coverage            & $\delta>+30^\circ$  & $\delta<-30^\circ$, $|b|>10^\circ$ & $\delta>-40^\circ$ & $|b|>30^\circ$, $\delta>-12^\circ$ \\
Sources\,/\,deg$^2$  & 21        & $>$40   &     60     &   90  \\
No.~of sources  &  230,000   &    320,000  &  2,000,000  &  900,000 \\   \hline
\end{tabular}
\end{center}
\end{table*}

\section{Integrated Source Parameters on the Web}  \label{integdata}  % lesson 3 section 4

In this section I shall describe the resources of information on ``integrated''
source parameters like position, flux density at one or more frequencies,
size, polarisation, spectral index, etc.~ This information can be found in
two distinct ways, either from individual source catalogues, each of which
have different formats and types of parameters, or from ``object databases''
like NED, SIMBAD or LEDA (\S\ref{objdatabases}). The latter have the
advantage of providing a
``value-added'' service, as they attempt to cross-identify radio sources
with known objects in the optical or other wavebands. The disadvantage is that this
is a laborious process, implying that radio source catalogues are being
integrated at a slow pace, often several years after their publication.
In fact, many valuable catalogues and compilations never made it into
these databases, and the only way for the user to complement this partial
information is to search the available catalogues separately on other
servers. Due to my own involvement in providing the latter facilities,
I shall briefly review their history.

\subsection{The Evolution of Electronic Source Catalogues} \label{radcats}

Radio astronomers have used electronic equipment from the outset
and already needed powerful computers in the 1960s to make radio maps
of the sky by Fourier transformation of interferometer visibilities.
Surprisingly radio astronomers have {\it not} \,been at the forefront
of archiving their results, not even the initially rather small-sized
catalogues of radio sources. It is hard to believe that the WSRT maintained
one of the earliest electronic and publicly searchable archives of raw
interferometer data (see {\tts www.nfra.nl/scissor/}), but at the same
time the source lists of 65 WSRT single-pointing surveys, published
from 1973 to 1987 with altogether 8200 sources, had not been kept in
electronic form.  Instead, 36 of them with a total of 5250 sources
were recovered in 1995--97 by the present author, using page-scanners
and ``Optical Character Recognition'' (OCR) techniques.

During the 1970s, R.~Dixon at Ohio State Univ.\ maintained what he
called the ``Master Source List'' (MSL). The first version appeared
in print almost 30 years ago (\cite{1970ApJS...20....1D}), and
contained $\sim$25,000 entries for $\sim$12,000 distinct sources.
Each entry contained the RA, DEC and flux density of a source at a
given observing frequency; any further information published
in the original tables was not included. The last version (\#\,43, Nov.~1981)
contained 84,559 entries drawn from 179 references published 1953--1978.
The list gives $\sim$75,000 distinct source names, but the number of
distinct sources is much smaller, though difficult to estimate.
It was typed entirely by hand, for which reason it is affected
by numerous typing errors (\cite{1989BICDS..37..139A}). Also,
it was meant to collect positions and fluxes  only from
new finder surveys, not to update information on already known sources.

Although the 1980s saw a ``renaissance'' of radio surveys
(e.g.\ MRC, B3, 6C, MIT-GB, GT, NEK, IPS in Table~1) that decade was a
truly ``dark age'' for radio source databases (\cite{andern92}).
The MSL, apart from being distributed on tape at cost, was not being
updated any more, and by the end of the 1980s there was not a single radio
source catalogue among the then over 600 catalogues available from the archives
of the two established astronomical data centres, the ``Astronomical Data
Center'' (ADC; {\tts adc.gsfc.nasa.gov/adc.html}) at NASA-GSFC, and
the ``Centre de Donn\'ees astronomiques de Strasbourg'' (CDS;
{\tts cdsweb.u-strasbg.fr/CDS.html}). This may explain why even
in 1990 the MSL was used to search for high-redshift quasars
of low radio luminosity, simply by cross-correlating it with
quasar catalogues (\cite{1991PASP..103...21H}, HDP91 in what follows).
These authors (using a version of MSL including data published up to 1975!)
noted that the MSL had 23 coincidences within 60$''$ from QSOs
in the HB\,89 compilation (\cite{1989ApJS...69....1H}) which were
not listed as ``radio quasars'' in HB\,89. However, HDP91 failed to note
that 13 of these 23 objects were already listed with an optical identification
in VV83, published seven years before! From the absence of {\it weak}~
($\ltsim$\,100\,mJy) radio sources associated with z\,$\gtsim$\,2.5 quasars,
HDP91 concluded that there were no high-z quasars of low radio
luminosity. However, had the authors used the 1989 edition of
VV98 (\cite{1989ESOSR...7....1V}, ADC/CDS \#\,7126) they would have
found about ten quasars weaker than $\sim$\,50\,mJy at 5-GHz,
from references published {\it before} 1989.
This would have proven the existence of the objects searched
for (but not found) by HDP91 from compilations readily available
at that time. The most recent studies by \cite{1997AJ....113.2000B}
and \cite{1996ApJ...473..746H}), however, indicate that these
objects are indeed quite rare.

Alerted by this deficiency of publicly available radio source
catalogues, I initiated, in late 1989, an email campaign among
radio astronomers world-wide. The response from several dozen
individuals (\cite{1990BICDS..38...69A}) was generally favourable,
and I started to actively collect electronic source catalogues
from the authors. By the time of the IAU General Assembly in 1991,
I had collected the tabular data from about 40 publications
totalling several times the number of records in the MSL.
However, it turned out that none of the major radio astronomical
institutes was willing to support the idea of a public radio source
database with manpower, e.g.\ to continue the collection effort and
prepare the software tools. As a result, the {\it EINSTEIN On-line
Service} (EINLINE or EOLS), designed to manage X-ray data from the
{\it EINSTEIN} satellite, offered to serve as a testbed for querying
radio source catalogues. Until mid-1993 some 67 source tables with
$\sim$523,000 entries had been integrated in collaboration with
the present author (\cite{harris95}). These are still searchable
simultaneously via a simple telnet session
({\tts telnet://einline@einline.harvard.edu}). However, in 1994
NASA's funding of EOLS ceased, and no further catalogues have been
integrated since then.  A similar service is available from DIRA2
({\tts www.ira.bo.cnr.it/dira/gb/}), providing 54 radio catalogues
with 2.3 million records, including older versions of the NVSS and
FIRST catalogues, as well as many items from the present author's
collection. However, due to lack of manpower, DIRA's catalogue collection
is now outdated, and many items from Table~1 are missing.
In late 1993, Alan Wright (ATNF) and the present author produced a
stand-alone package (called ``COMRAD'') of 12 major radio source catalogues
with some 303,600 entries. It comes with {\tt dBaseIV} search software
for PCs and can still be downloaded from URL~
{\tts wwwpks.atnf.csiro.au/databases/surveys/comrad/comrad.html}.
Several other sites offer more or less ``random'' and outdated sets
of catalogues (a few radio items included) and are less suitable
when seeking up-to-date and complete information. Among these are
ESO's STARCAT ({\tts arch-http.hq.eso.org/starcat.html}),
ASTROCAT at CADC ({\tts cadcwww.dao.nrc.ca/astrocat}), and
CURSA within the Starlink project
({\tts www.roe.ac.uk/acdwww/cursa/home.html}). CURSA is actually designed
to be copied to the user's machine, and to work with local catalogues
in a CURSA-compatible format.

  From late 1989 until the present, I have continued my activities
of collecting source catalogues, and since 1995, I have also employed
OCR methods to convert printed source lists into electronic form,
among them well-known compilations such as that of \cite{kuhr79}
of 250 pages, for which the electronic version had not survived
the transition through various storage media. Recovery by OCR
requires careful proof-reading, especially for those published in
tiny or poorly printed fonts (e.g.\ Harris \& Miley (1978), % \cite{1978A&AS...34..117H},
\cite{1985A&AS...61..451W}, and \cite{1984A&AS...56..245B}).
In many cases the original publications were impossible to recover
with OCR. For some of these I had kept preprints
(e.g.\ for \cite{1980A&AS...39..379T}) whose larger fonts facilitated
the OCR. Numerous other tables (e.g.\ \cite{1979Ap&SS..64...73B},
\cite{1988A&AS...76...21Q} or \cite{1979A&AS...35...23A})
were patiently retyped by members of the CATS team (see below).
Since about 1996, older source tables are also actively recovered with
OCR methods at CDS. Unfortunately, due to poor proof-reading methods,
errors are found quite frequently in tables prepared via OCR and released
by CDS.
Occasionally, tables were prepared independently by two groups, allowing
the error rate to be further reduced by inter-comparison of the results.
Up to now, radio source tables from 177 articles with a total of 75,000
data records and many thousand lines of text (used as documentation
for the tables) were prepared via OCR, mostly by the present author.
Surprisingly, about half of the tables (including those received
directly from the authors) show some kind of problem (e.g.\ in
nomenclature, internal consistency, or formatting, etc.) that requires
attention, before they are able to
be integrated into a database or searchable catalogue collection.
This shows, unfortunately, that not enough attention is paid to
the data section by the referees of papers. While this section may appear
uninteresting to them, one should keep in mind that in future
re-analyses based on old published data, it is exactly the data section
which remains as a heritage to future researchers, and not the
interpretations given in the original papers.
In the early 1990s, most of the tables were received from the original
authors upon request, but currently about half of the tables
can be collected from the LANL/SISSA electronic preprint server
({\tts xxx.lanl.gov}). However, this has the danger that they may
not be identical to the actual publication. The vast
majority of tabular data sets are received in \TeX\ format,
and their conversion to ASCII requires substantial efforts.

Currently my collection of radio source lists
({\tts \verb*Ccats.sao.ru/~cats/doc/Andernach.htmlC}) contains
source lists from 500 articles, but only $\sim$22\% of the tables
are also available from ADC or CDS (\S\ref{catcoll}).
While the collection started in 1989, half the 500 data sets were
collected or prepared since 1996, and the current growth rate
is $\sim$\,80 data sets per year.  About three dozen further
source lists exist in the CDS archive (\S\ref{catcoll}), most of which
are either from the series of nine AAS CDROMs, issued twice a year
from 1994 to 1997 with tables from the ApJ and AJ journals,
or from recent volumes of the A\&AS journal, thanks to a 1993 agreement
between the Editors of A\&AS and CDS to archive all major
tables of A\&AS at the CDS. Unfortunately
such an agreement does not exist with other journals, for which
reason my collection efforts will probably continue until virtually
all astronomical journals provide electronic editions.
Presently some tabular data (e.g.\ in the electronic A\&A)
are offered only as images, while other journals offer only hypertext
versions of their tables, which frequently need further treatment
to be converted to plain ASCII format, required for their
ingestion into databases.

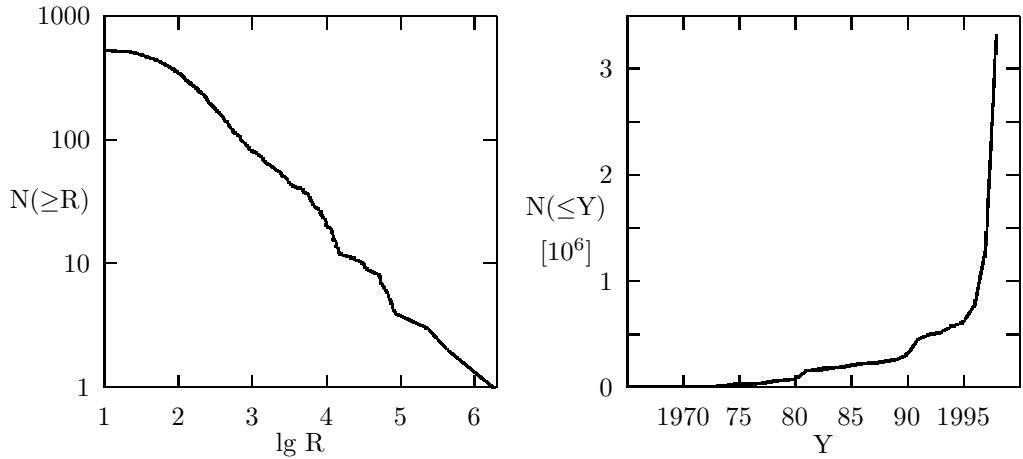
\begin{figure*}[!ht]   % \label{radhistfig}
\vspace*{3mm}
\hspace*{-5mm}
% \mbox{
% %\vspace*{5mm}
% \epsfig{file=tagof1.ps,width=6.7cm,bbllx=140pt,bblly=482pt,bburx=335pt,bbury=655pt}
%
% \hspace*{-2mm}
% \epsfig{file=tagof2.ps,width=6.7cm,bbllx=140pt,bblly=482pt,bburx=335pt,bbury=655pt}
% }

\begin{minipage}[b]{.46\linewidth}

% GNUPLOT: LaTeX picture
\setlength{\unitlength}{0.240900pt}
\ifx\plotpoint\undefined\newsavebox{\plotpoint}\fi
\sbox{\plotpoint}{\rule[-0.200pt]{0.400pt}{0.400pt}}%
\begin{picture}(900,720)(70,0)
\font\gnuplot=cmr10 at 10pt
\gnuplot
\sbox{\plotpoint}{\rule[-0.200pt]{0.400pt}{0.400pt}}%
\put(220.0,113.0){\rule[-0.200pt]{4.818pt}{0.400pt}}
\put(198,113){\makebox(0,0)[r]{1}}
\put(816.0,113.0){\rule[-0.200pt]{4.818pt}{0.400pt}}
\put(220.0,308.0){\rule[-0.200pt]{4.818pt}{0.400pt}}
\put(198,308){\makebox(0,0)[r]{10}}
\put(816.0,308.0){\rule[-0.200pt]{4.818pt}{0.400pt}}
\put(220.0,502.0){\rule[-0.200pt]{4.818pt}{0.400pt}}
\put(198,502){\makebox(0,0)[r]{100}}
\put(816.0,502.0){\rule[-0.200pt]{4.818pt}{0.400pt}}
\put(220.0,697.0){\rule[-0.200pt]{4.818pt}{0.400pt}}
\put(198,697){\makebox(0,0)[r]{1000}}
\put(816.0,697.0){\rule[-0.200pt]{4.818pt}{0.400pt}}
\put(220.0,113.0){\rule[-0.200pt]{0.400pt}{4.818pt}}
\put(220,68){\makebox(0,0){1}}
\put(220.0,677.0){\rule[-0.200pt]{0.400pt}{4.818pt}}
\put(336.0,113.0){\rule[-0.200pt]{0.400pt}{4.818pt}}
\put(336,68){\makebox(0,0){2}}
\put(336.0,677.0){\rule[-0.200pt]{0.400pt}{4.818pt}}
\put(452.0,113.0){\rule[-0.200pt]{0.400pt}{4.818pt}}
\put(452,68){\makebox(0,0){3}}
\put(452.0,677.0){\rule[-0.200pt]{0.400pt}{4.818pt}}
\put(569.0,113.0){\rule[-0.200pt]{0.400pt}{4.818pt}}
\put(569,68){\makebox(0,0){4}}
\put(569.0,677.0){\rule[-0.200pt]{0.400pt}{4.818pt}}
\put(685.0,113.0){\rule[-0.200pt]{0.400pt}{4.818pt}}
\put(685,68){\makebox(0,0){5}}
\put(685.0,677.0){\rule[-0.200pt]{0.400pt}{4.818pt}}
\put(801.0,113.0){\rule[-0.200pt]{0.400pt}{4.818pt}}
\put(801,68){\makebox(0,0){6}}
\put(801.0,677.0){\rule[-0.200pt]{0.400pt}{4.818pt}}
\put(220.0,113.0){\rule[-0.200pt]{148.394pt}{0.400pt}}
\put(836.0,113.0){\rule[-0.200pt]{0.400pt}{140.686pt}}
\put(220.0,697.0){\rule[-0.200pt]{148.394pt}{0.400pt}}
\put(135,405){\makebox(0,0){N$(\ge$R$)$}}
\put(528,23){\makebox(0,0){lg R}}
\put(220.0,113.0){\rule[-0.200pt]{0.400pt}{140.686pt}}
\sbox{\plotpoint}{\rule[-0.400pt]{0.800pt}{0.800pt}}%
\put(831,113){\usebox{\plotpoint}}
\multiput(826.12,114.41)(-0.610,0.502){111}{\rule{1.176pt}{0.121pt}}
\multiput(828.56,111.34)(-69.559,59.000){2}{\rule{0.588pt}{0.800pt}}
\multiput(757.09,172.00)(-0.503,0.529){57}{\rule{0.121pt}{1.050pt}}
\multiput(757.34,172.00)(-32.000,31.821){2}{\rule{0.800pt}{0.525pt}}
\multiput(719.11,207.41)(-1.078,0.504){41}{\rule{1.900pt}{0.122pt}}
\multiput(723.06,204.34)(-47.056,24.000){2}{\rule{0.950pt}{0.800pt}}
\multiput(674.07,230.00)(-0.536,1.913){5}{\rule{0.129pt}{2.733pt}}
\multiput(674.34,230.00)(-6.000,13.327){2}{\rule{0.800pt}{1.367pt}}
\multiput(668.08,249.00)(-0.526,1.176){7}{\rule{0.127pt}{1.914pt}}
\multiput(668.34,249.00)(-7.000,11.027){2}{\rule{0.800pt}{0.957pt}}
\multiput(661.08,264.00)(-0.516,0.800){11}{\rule{0.124pt}{1.444pt}}
\multiput(661.34,264.00)(-9.000,11.002){2}{\rule{0.800pt}{0.722pt}}
\put(651.84,278){\rule{0.800pt}{2.650pt}}
\multiput(652.34,278.00)(-1.000,5.500){2}{\rule{0.800pt}{1.325pt}}
\multiput(644.86,290.40)(-1.155,0.514){13}{\rule{1.960pt}{0.124pt}}
\multiput(648.93,287.34)(-17.932,10.000){2}{\rule{0.980pt}{0.800pt}}
\put(627.34,299){\rule{0.800pt}{2.000pt}}
\multiput(629.34,299.00)(-4.000,4.849){2}{\rule{0.800pt}{1.000pt}}
\multiput(619.53,309.40)(-1.066,0.520){9}{\rule{1.800pt}{0.125pt}}
\multiput(623.26,306.34)(-12.264,8.000){2}{\rule{0.900pt}{0.800pt}}
\multiput(599.73,317.40)(-1.789,0.526){7}{\rule{2.714pt}{0.127pt}}
\multiput(605.37,314.34)(-16.366,7.000){2}{\rule{1.357pt}{0.800pt}}
\put(586.84,323){\rule{0.800pt}{1.686pt}}
\multiput(587.34,323.00)(-1.000,3.500){2}{\rule{0.800pt}{0.843pt}}
\multiput(586.06,330.00)(-0.560,0.592){3}{\rule{0.135pt}{1.160pt}}
\multiput(586.34,330.00)(-5.000,3.592){2}{\rule{0.800pt}{0.580pt}}
\put(579.34,342){\rule{0.800pt}{1.200pt}}
\multiput(581.34,342.00)(-4.000,2.509){2}{\rule{0.800pt}{0.600pt}}
\put(583.0,336.0){\rule[-0.400pt]{0.800pt}{1.445pt}}
\put(576.84,353){\rule{0.800pt}{0.964pt}}
\multiput(577.34,353.00)(-1.000,2.000){2}{\rule{0.800pt}{0.482pt}}
\put(575.84,357){\rule{0.800pt}{1.204pt}}
\multiput(576.34,357.00)(-1.000,2.500){2}{\rule{0.800pt}{0.602pt}}
\put(569,362.34){\rule{1.800pt}{0.800pt}}
\multiput(573.26,360.34)(-4.264,4.000){2}{\rule{0.900pt}{0.800pt}}
\put(566.84,366){\rule{0.800pt}{0.964pt}}
\multiput(567.34,366.00)(-1.000,2.000){2}{\rule{0.800pt}{0.482pt}}
\put(579.0,347.0){\rule[-0.400pt]{0.800pt}{1.445pt}}
\put(565.84,374){\rule{0.800pt}{0.964pt}}
\multiput(566.34,374.00)(-1.000,2.000){2}{\rule{0.800pt}{0.482pt}}
\put(562,378.34){\rule{1.200pt}{0.800pt}}
\multiput(564.51,376.34)(-2.509,4.000){2}{\rule{0.600pt}{0.800pt}}
\put(559.34,382){\rule{0.800pt}{0.723pt}}
\multiput(560.34,382.00)(-2.000,1.500){2}{\rule{0.800pt}{0.361pt}}
\put(557.84,385){\rule{0.800pt}{0.723pt}}
\multiput(558.34,385.00)(-1.000,1.500){2}{\rule{0.800pt}{0.361pt}}
\put(568.0,370.0){\rule[-0.400pt]{0.800pt}{0.964pt}}
\put(552,391.84){\rule{1.686pt}{0.800pt}}
\multiput(555.50,390.34)(-3.500,3.000){2}{\rule{0.843pt}{0.800pt}}
\put(549,394.84){\rule{0.723pt}{0.800pt}}
\multiput(550.50,393.34)(-1.500,3.000){2}{\rule{0.361pt}{0.800pt}}
\put(546.84,398){\rule{0.800pt}{0.723pt}}
\multiput(547.34,398.00)(-1.000,1.500){2}{\rule{0.800pt}{0.361pt}}
\put(545.84,401){\rule{0.800pt}{0.482pt}}
\multiput(546.34,401.00)(-1.000,1.000){2}{\rule{0.800pt}{0.241pt}}
\put(544.34,403){\rule{0.800pt}{0.723pt}}
\multiput(545.34,403.00)(-2.000,1.500){2}{\rule{0.800pt}{0.361pt}}
\put(542.84,406){\rule{0.800pt}{0.723pt}}
\multiput(543.34,406.00)(-1.000,1.500){2}{\rule{0.800pt}{0.361pt}}
\put(542,408.34){\rule{0.482pt}{0.800pt}}
\multiput(543.00,407.34)(-1.000,2.000){2}{\rule{0.241pt}{0.800pt}}
\put(559.0,388.0){\rule[-0.400pt]{0.800pt}{0.964pt}}
\put(539,413.34){\rule{0.723pt}{0.800pt}}
\multiput(540.50,412.34)(-1.500,2.000){2}{\rule{0.361pt}{0.800pt}}
\put(536.84,416){\rule{0.800pt}{0.482pt}}
\multiput(537.34,416.00)(-1.000,1.000){2}{\rule{0.800pt}{0.241pt}}
\put(532,417.84){\rule{1.445pt}{0.800pt}}
\multiput(535.00,416.34)(-3.000,3.000){2}{\rule{0.723pt}{0.800pt}}
\put(529.84,421){\rule{0.800pt}{0.482pt}}
\multiput(530.34,421.00)(-1.000,1.000){2}{\rule{0.800pt}{0.241pt}}
\put(542.0,411.0){\usebox{\plotpoint}}
\put(522,424.34){\rule{2.168pt}{0.800pt}}
\multiput(526.50,423.34)(-4.500,2.000){2}{\rule{1.084pt}{0.800pt}}
\put(515,426.34){\rule{1.686pt}{0.800pt}}
\multiput(518.50,425.34)(-3.500,2.000){2}{\rule{0.843pt}{0.800pt}}
\put(531.0,423.0){\usebox{\plotpoint}}
\put(511,430.34){\rule{0.964pt}{0.800pt}}
\multiput(513.00,429.34)(-2.000,2.000){2}{\rule{0.482pt}{0.800pt}}
\put(508.84,433){\rule{0.800pt}{0.482pt}}
\multiput(509.34,433.00)(-1.000,1.000){2}{\rule{0.800pt}{0.241pt}}
\put(515.0,429.0){\usebox{\plotpoint}}
\put(508,436.34){\rule{0.482pt}{0.800pt}}
\multiput(509.00,435.34)(-1.000,2.000){2}{\rule{0.241pt}{0.800pt}}
\put(507,437.84){\rule{0.241pt}{0.800pt}}
\multiput(507.50,437.34)(-0.500,1.000){2}{\rule{0.120pt}{0.800pt}}
\put(504.84,440){\rule{0.800pt}{0.482pt}}
\multiput(505.34,440.00)(-1.000,1.000){2}{\rule{0.800pt}{0.241pt}}
\put(503,441.34){\rule{0.723pt}{0.800pt}}
\multiput(504.50,440.34)(-1.500,2.000){2}{\rule{0.361pt}{0.800pt}}
\put(500,442.84){\rule{0.723pt}{0.800pt}}
\multiput(501.50,442.34)(-1.500,1.000){2}{\rule{0.361pt}{0.800pt}}
\put(498,444.34){\rule{0.482pt}{0.800pt}}
\multiput(499.00,443.34)(-1.000,2.000){2}{\rule{0.241pt}{0.800pt}}
\put(495.84,447){\rule{0.800pt}{0.482pt}}
\multiput(496.34,447.00)(-1.000,1.000){2}{\rule{0.800pt}{0.241pt}}
\put(510.0,435.0){\usebox{\plotpoint}}
\put(494.84,450){\rule{0.800pt}{0.482pt}}
\multiput(495.34,450.00)(-1.000,1.000){2}{\rule{0.800pt}{0.241pt}}
\put(492,450.84){\rule{0.964pt}{0.800pt}}
\multiput(494.00,450.34)(-2.000,1.000){2}{\rule{0.482pt}{0.800pt}}
\put(490,452.34){\rule{0.482pt}{0.800pt}}
\multiput(491.00,451.34)(-1.000,2.000){2}{\rule{0.241pt}{0.800pt}}
\put(488,453.84){\rule{0.482pt}{0.800pt}}
\multiput(489.00,453.34)(-1.000,1.000){2}{\rule{0.241pt}{0.800pt}}
\put(485.84,456){\rule{0.800pt}{0.482pt}}
\multiput(486.34,456.00)(-1.000,1.000){2}{\rule{0.800pt}{0.241pt}}
\put(483,456.84){\rule{0.964pt}{0.800pt}}
\multiput(485.00,456.34)(-2.000,1.000){2}{\rule{0.482pt}{0.800pt}}
\put(480.84,459){\rule{0.800pt}{0.482pt}}
\multiput(481.34,459.00)(-1.000,1.000){2}{\rule{0.800pt}{0.241pt}}
\put(481,459.84){\rule{0.241pt}{0.800pt}}
\multiput(481.50,459.34)(-0.500,1.000){2}{\rule{0.120pt}{0.800pt}}
\put(479,460.84){\rule{0.482pt}{0.800pt}}
\multiput(480.00,460.34)(-1.000,1.000){2}{\rule{0.241pt}{0.800pt}}
\put(473,462.34){\rule{1.445pt}{0.800pt}}
\multiput(476.00,461.34)(-3.000,2.000){2}{\rule{0.723pt}{0.800pt}}
\put(497.0,449.0){\usebox{\plotpoint}}
\put(472,464.84){\rule{0.241pt}{0.800pt}}
\multiput(472.50,464.34)(-0.500,1.000){2}{\rule{0.120pt}{0.800pt}}
\put(473.0,465.0){\usebox{\plotpoint}}
\put(469,467.34){\rule{0.723pt}{0.800pt}}
\multiput(470.50,466.34)(-1.500,2.000){2}{\rule{0.361pt}{0.800pt}}
\put(472.0,467.0){\usebox{\plotpoint}}
\put(468,469.84){\rule{0.241pt}{0.800pt}}
\multiput(468.50,469.34)(-0.500,1.000){2}{\rule{0.120pt}{0.800pt}}
\put(469.0,470.0){\usebox{\plotpoint}}
\put(465,473.84){\rule{0.723pt}{0.800pt}}
\multiput(466.50,473.34)(-1.500,1.000){2}{\rule{0.361pt}{0.800pt}}
\put(462,474.84){\rule{0.723pt}{0.800pt}}
\multiput(463.50,474.34)(-1.500,1.000){2}{\rule{0.361pt}{0.800pt}}
\put(461,475.84){\rule{0.241pt}{0.800pt}}
\multiput(461.50,475.34)(-0.500,1.000){2}{\rule{0.120pt}{0.800pt}}
\put(468.0,472.0){\usebox{\plotpoint}}
\put(460,478.84){\rule{0.241pt}{0.800pt}}
\multiput(460.50,478.34)(-0.500,1.000){2}{\rule{0.120pt}{0.800pt}}
\put(457,479.84){\rule{0.723pt}{0.800pt}}
\multiput(458.50,479.34)(-1.500,1.000){2}{\rule{0.361pt}{0.800pt}}
\put(454,480.84){\rule{0.723pt}{0.800pt}}
\multiput(455.50,480.34)(-1.500,1.000){2}{\rule{0.361pt}{0.800pt}}
\put(450,482.34){\rule{0.964pt}{0.800pt}}
\multiput(452.00,481.34)(-2.000,2.000){2}{\rule{0.482pt}{0.800pt}}
\put(461.0,478.0){\usebox{\plotpoint}}
\put(448,484.84){\rule{0.482pt}{0.800pt}}
\multiput(449.00,484.34)(-1.000,1.000){2}{\rule{0.241pt}{0.800pt}}
\put(447,485.84){\rule{0.241pt}{0.800pt}}
\multiput(447.50,485.34)(-0.500,1.000){2}{\rule{0.120pt}{0.800pt}}
\put(446,486.84){\rule{0.241pt}{0.800pt}}
\multiput(446.50,486.34)(-0.500,1.000){2}{\rule{0.120pt}{0.800pt}}
\put(450.0,485.0){\usebox{\plotpoint}}
\put(445,489.84){\rule{0.241pt}{0.800pt}}
\multiput(445.50,489.34)(-0.500,1.000){2}{\rule{0.120pt}{0.800pt}}
\put(446.0,489.0){\usebox{\plotpoint}}
\put(444.0,492.0){\usebox{\plotpoint}}
\put(442,491.84){\rule{0.482pt}{0.800pt}}
\multiput(443.00,491.34)(-1.000,1.000){2}{\rule{0.241pt}{0.800pt}}
\put(441,492.84){\rule{0.241pt}{0.800pt}}
\multiput(441.50,492.34)(-0.500,1.000){2}{\rule{0.120pt}{0.800pt}}
\put(444.0,492.0){\usebox{\plotpoint}}
\put(440,494.84){\rule{0.241pt}{0.800pt}}
\multiput(440.50,494.34)(-0.500,1.000){2}{\rule{0.120pt}{0.800pt}}
\put(439,495.84){\rule{0.241pt}{0.800pt}}
\multiput(439.50,495.34)(-0.500,1.000){2}{\rule{0.120pt}{0.800pt}}
\put(438,496.84){\rule{0.241pt}{0.800pt}}
\multiput(438.50,496.34)(-0.500,1.000){2}{\rule{0.120pt}{0.800pt}}
\put(437,497.84){\rule{0.241pt}{0.800pt}}
\multiput(437.50,497.34)(-0.500,1.000){2}{\rule{0.120pt}{0.800pt}}
\put(435,498.84){\rule{0.482pt}{0.800pt}}
\multiput(436.00,498.34)(-1.000,1.000){2}{\rule{0.241pt}{0.800pt}}
\put(441.0,495.0){\usebox{\plotpoint}}
\put(434.0,501.0){\usebox{\plotpoint}}
\put(433,501.84){\rule{0.241pt}{0.800pt}}
\multiput(433.50,501.34)(-0.500,1.000){2}{\rule{0.120pt}{0.800pt}}
\put(434.0,501.0){\usebox{\plotpoint}}
\put(432,504.84){\rule{0.241pt}{0.800pt}}
\multiput(432.50,504.34)(-0.500,1.000){2}{\rule{0.120pt}{0.800pt}}
\put(433.0,504.0){\usebox{\plotpoint}}
\put(429,506.84){\rule{0.723pt}{0.800pt}}
\multiput(430.50,506.34)(-1.500,1.000){2}{\rule{0.361pt}{0.800pt}}
\put(428,507.84){\rule{0.241pt}{0.800pt}}
\multiput(428.50,507.34)(-0.500,1.000){2}{\rule{0.120pt}{0.800pt}}
\put(432.0,507.0){\usebox{\plotpoint}}
\put(428,510){\usebox{\plotpoint}}
\put(427,509.84){\rule{0.241pt}{0.800pt}}
\multiput(427.50,509.34)(-0.500,1.000){2}{\rule{0.120pt}{0.800pt}}
\put(423,510.84){\rule{0.964pt}{0.800pt}}
\multiput(425.00,510.34)(-2.000,1.000){2}{\rule{0.482pt}{0.800pt}}
\put(428.0,510.0){\usebox{\plotpoint}}
\put(423,513){\usebox{\plotpoint}}
\put(422,512.84){\rule{0.241pt}{0.800pt}}
\multiput(422.50,512.34)(-0.500,1.000){2}{\rule{0.120pt}{0.800pt}}
\put(421,513.84){\rule{0.241pt}{0.800pt}}
\multiput(421.50,513.34)(-0.500,1.000){2}{\rule{0.120pt}{0.800pt}}
\put(423.0,513.0){\usebox{\plotpoint}}
\put(421,516){\usebox{\plotpoint}}
\put(420,515.84){\rule{0.241pt}{0.800pt}}
\multiput(420.50,515.34)(-0.500,1.000){2}{\rule{0.120pt}{0.800pt}}
\put(421.0,516.0){\usebox{\plotpoint}}
\put(420,518){\usebox{\plotpoint}}
\put(418,518.84){\rule{0.482pt}{0.800pt}}
\multiput(419.00,518.34)(-1.000,1.000){2}{\rule{0.241pt}{0.800pt}}
\put(420.0,518.0){\usebox{\plotpoint}}
\put(417.0,521.0){\usebox{\plotpoint}}
\put(416,520.84){\rule{0.241pt}{0.800pt}}
\multiput(416.50,520.34)(-0.500,1.000){2}{\rule{0.120pt}{0.800pt}}
\put(417.0,521.0){\usebox{\plotpoint}}
\put(416,523){\usebox{\plotpoint}}
\put(415,521.84){\rule{0.241pt}{0.800pt}}
\multiput(415.50,521.34)(-0.500,1.000){2}{\rule{0.120pt}{0.800pt}}
\put(414,523.84){\rule{0.241pt}{0.800pt}}
\multiput(414.50,523.34)(-0.500,1.000){2}{\rule{0.120pt}{0.800pt}}
\put(415.0,524.0){\usebox{\plotpoint}}
\put(414,526){\usebox{\plotpoint}}
\put(413,526.84){\rule{0.241pt}{0.800pt}}
\multiput(413.50,526.34)(-0.500,1.000){2}{\rule{0.120pt}{0.800pt}}
\put(412,527.84){\rule{0.241pt}{0.800pt}}
\multiput(412.50,527.34)(-0.500,1.000){2}{\rule{0.120pt}{0.800pt}}
\put(414.0,526.0){\usebox{\plotpoint}}
\put(412,530){\usebox{\plotpoint}}
\put(410,528.84){\rule{0.482pt}{0.800pt}}
\multiput(411.00,528.34)(-1.000,1.000){2}{\rule{0.241pt}{0.800pt}}
\put(410,531){\usebox{\plotpoint}}
\put(408,529.84){\rule{0.482pt}{0.800pt}}
\multiput(409.00,529.34)(-1.000,1.000){2}{\rule{0.241pt}{0.800pt}}
\put(408.0,532.0){\usebox{\plotpoint}}
\put(407.0,534.0){\usebox{\plotpoint}}
\put(406,533.84){\rule{0.241pt}{0.800pt}}
\multiput(406.50,533.34)(-0.500,1.000){2}{\rule{0.120pt}{0.800pt}}
\put(407.0,534.0){\usebox{\plotpoint}}
\put(406.0,536.0){\usebox{\plotpoint}}
\put(405.0,538.0){\usebox{\plotpoint}}
\put(404,537.84){\rule{0.241pt}{0.800pt}}
\multiput(404.50,537.34)(-0.500,1.000){2}{\rule{0.120pt}{0.800pt}}
\put(405.0,538.0){\usebox{\plotpoint}}
\put(403.0,540.0){\usebox{\plotpoint}}
\put(403.0,540.0){\usebox{\plotpoint}}
\put(401.0,542.0){\usebox{\plotpoint}}
\put(401.0,542.0){\usebox{\plotpoint}}
\put(400.0,544.0){\usebox{\plotpoint}}
\put(400.0,544.0){\usebox{\plotpoint}}
\put(399.0,545.0){\usebox{\plotpoint}}
\put(399.0,545.0){\usebox{\plotpoint}}
\put(397.0,546.0){\usebox{\plotpoint}}
\put(397.0,546.0){\usebox{\plotpoint}}
\put(395,545.84){\rule{0.241pt}{0.800pt}}
\multiput(395.50,545.34)(-0.500,1.000){2}{\rule{0.120pt}{0.800pt}}
\put(396.0,547.0){\usebox{\plotpoint}}
\put(395,548){\usebox{\plotpoint}}
\put(394,547.84){\rule{0.241pt}{0.800pt}}
\multiput(394.50,547.34)(-0.500,1.000){2}{\rule{0.120pt}{0.800pt}}
\put(395.0,548.0){\usebox{\plotpoint}}
\put(394,550){\usebox{\plotpoint}}
\put(393,548.84){\rule{0.241pt}{0.800pt}}
\multiput(393.50,548.34)(-0.500,1.000){2}{\rule{0.120pt}{0.800pt}}
\put(393,551){\usebox{\plotpoint}}
\put(392,549.84){\rule{0.241pt}{0.800pt}}
\multiput(392.50,549.34)(-0.500,1.000){2}{\rule{0.120pt}{0.800pt}}
\put(391.0,552.0){\usebox{\plotpoint}}
\put(391.0,552.0){\usebox{\plotpoint}}
\put(389,552.84){\rule{0.241pt}{0.800pt}}
\multiput(389.50,552.34)(-0.500,1.000){2}{\rule{0.120pt}{0.800pt}}
\put(390.0,554.0){\usebox{\plotpoint}}
\put(389,555){\usebox{\plotpoint}}
\put(388,553.84){\rule{0.241pt}{0.800pt}}
\multiput(388.50,553.34)(-0.500,1.000){2}{\rule{0.120pt}{0.800pt}}
\put(388,556){\usebox{\plotpoint}}
\put(388.0,556.0){\usebox{\plotpoint}}
\put(386,555.84){\rule{0.241pt}{0.800pt}}
\multiput(386.50,555.34)(-0.500,1.000){2}{\rule{0.120pt}{0.800pt}}
\put(387.0,557.0){\usebox{\plotpoint}}
\put(385.0,558.0){\usebox{\plotpoint}}
\put(385.0,558.0){\usebox{\plotpoint}}
\put(384.0,559.0){\usebox{\plotpoint}}
\put(384.0,559.0){\usebox{\plotpoint}}
\put(382,559.84){\rule{0.241pt}{0.800pt}}
\multiput(382.50,559.34)(-0.500,1.000){2}{\rule{0.120pt}{0.800pt}}
\put(383.0,561.0){\usebox{\plotpoint}}
\put(382,562){\usebox{\plotpoint}}
\put(382.0,562.0){\usebox{\plotpoint}}
\put(381.0,564.0){\usebox{\plotpoint}}
\put(381.0,564.0){\usebox{\plotpoint}}
\put(380.0,565.0){\usebox{\plotpoint}}
\put(379,565.84){\rule{0.241pt}{0.800pt}}
\multiput(379.50,565.34)(-0.500,1.000){2}{\rule{0.120pt}{0.800pt}}
\put(380.0,565.0){\usebox{\plotpoint}}
\put(379,568){\usebox{\plotpoint}}
\put(378,567.84){\rule{0.241pt}{0.800pt}}
\multiput(378.50,567.34)(-0.500,1.000){2}{\rule{0.120pt}{0.800pt}}
\put(379.0,568.0){\usebox{\plotpoint}}
\put(376,568.84){\rule{0.241pt}{0.800pt}}
\multiput(376.50,568.34)(-0.500,1.000){2}{\rule{0.120pt}{0.800pt}}
\put(377.0,570.0){\usebox{\plotpoint}}
\put(376,571){\usebox{\plotpoint}}
\put(376,571){\usebox{\plotpoint}}
\put(376.0,571.0){\usebox{\plotpoint}}
\put(374.0,573.0){\usebox{\plotpoint}}
\put(373,572.84){\rule{0.241pt}{0.800pt}}
\multiput(373.50,572.34)(-0.500,1.000){2}{\rule{0.120pt}{0.800pt}}
\put(374.0,573.0){\usebox{\plotpoint}}
\put(371,573.84){\rule{0.241pt}{0.800pt}}
\multiput(371.50,573.34)(-0.500,1.000){2}{\rule{0.120pt}{0.800pt}}
\put(372.0,575.0){\usebox{\plotpoint}}
\put(371,576){\usebox{\plotpoint}}
\put(371,576){\usebox{\plotpoint}}
\put(371.0,576.0){\usebox{\plotpoint}}
\put(369.0,577.0){\usebox{\plotpoint}}
\put(369.0,577.0){\usebox{\plotpoint}}
\put(367.0,578.0){\usebox{\plotpoint}}
\put(366,577.84){\rule{0.241pt}{0.800pt}}
\multiput(366.50,577.34)(-0.500,1.000){2}{\rule{0.120pt}{0.800pt}}
\put(367.0,578.0){\usebox{\plotpoint}}
\put(366,580){\usebox{\plotpoint}}
\put(366,580){\usebox{\plotpoint}}
\put(365,579.84){\rule{0.241pt}{0.800pt}}
\multiput(365.50,579.34)(-0.500,1.000){2}{\rule{0.120pt}{0.800pt}}
\put(366.0,580.0){\usebox{\plotpoint}}
\put(365,582){\usebox{\plotpoint}}
\put(364.0,582.0){\usebox{\plotpoint}}
\put(364.0,582.0){\usebox{\plotpoint}}
\put(363.0,583.0){\usebox{\plotpoint}}
\put(362,582.84){\rule{0.241pt}{0.800pt}}
\multiput(362.50,582.34)(-0.500,1.000){2}{\rule{0.120pt}{0.800pt}}
\put(363.0,583.0){\usebox{\plotpoint}}
\put(360.0,585.0){\usebox{\plotpoint}}
\put(360.0,585.0){\usebox{\plotpoint}}
\put(359.0,586.0){\usebox{\plotpoint}}
\put(359.0,586.0){\usebox{\plotpoint}}
\put(358.0,587.0){\usebox{\plotpoint}}
\put(357,586.84){\rule{0.241pt}{0.800pt}}
\multiput(357.50,586.34)(-0.500,1.000){2}{\rule{0.120pt}{0.800pt}}
\put(358.0,587.0){\usebox{\plotpoint}}
\put(355,587.84){\rule{0.241pt}{0.800pt}}
\multiput(355.50,587.34)(-0.500,1.000){2}{\rule{0.120pt}{0.800pt}}
\put(356.0,589.0){\usebox{\plotpoint}}
\put(355,590){\usebox{\plotpoint}}
\put(355,590){\usebox{\plotpoint}}
\put(354,588.84){\rule{0.241pt}{0.800pt}}
\multiput(354.50,588.34)(-0.500,1.000){2}{\rule{0.120pt}{0.800pt}}
\put(354,591){\usebox{\plotpoint}}
\put(354,591){\usebox{\plotpoint}}
\put(354,591){\usebox{\plotpoint}}
\put(353,589.84){\rule{0.241pt}{0.800pt}}
\multiput(353.50,589.34)(-0.500,1.000){2}{\rule{0.120pt}{0.800pt}}
\put(351,590.84){\rule{0.241pt}{0.800pt}}
\multiput(351.50,590.34)(-0.500,1.000){2}{\rule{0.120pt}{0.800pt}}
\put(352.0,592.0){\usebox{\plotpoint}}
\put(351,593){\usebox{\plotpoint}}
\put(351,593){\usebox{\plotpoint}}
\put(351.0,593.0){\usebox{\plotpoint}}
\put(349.0,594.0){\usebox{\plotpoint}}
\put(349.0,594.0){\usebox{\plotpoint}}
\put(347.0,595.0){\usebox{\plotpoint}}
\put(347.0,595.0){\usebox{\plotpoint}}
\put(346.0,597.0){\usebox{\plotpoint}}
\put(346.0,597.0){\usebox{\plotpoint}}
\put(345.0,598.0){\usebox{\plotpoint}}
\put(345.0,598.0){\usebox{\plotpoint}}
\put(343,598.84){\rule{0.241pt}{0.800pt}}
\multiput(343.50,598.34)(-0.500,1.000){2}{\rule{0.120pt}{0.800pt}}
\put(344.0,600.0){\usebox{\plotpoint}}
\put(343,601){\usebox{\plotpoint}}
\put(343,601){\usebox{\plotpoint}}
\put(342.0,601.0){\usebox{\plotpoint}}
\put(341,600.84){\rule{0.241pt}{0.800pt}}
\multiput(341.50,600.34)(-0.500,1.000){2}{\rule{0.120pt}{0.800pt}}
\put(342.0,601.0){\usebox{\plotpoint}}
\put(341,603){\usebox{\plotpoint}}
\put(341,603){\usebox{\plotpoint}}
\put(341.0,603.0){\usebox{\plotpoint}}
\put(340.0,604.0){\usebox{\plotpoint}}
\put(340.0,604.0){\usebox{\plotpoint}}
\put(338.0,605.0){\usebox{\plotpoint}}
\put(338.0,605.0){\usebox{\plotpoint}}
\put(336.0,606.0){\usebox{\plotpoint}}
\put(335,605.84){\rule{0.241pt}{0.800pt}}
\multiput(335.50,605.34)(-0.500,1.000){2}{\rule{0.120pt}{0.800pt}}
\put(336.0,606.0){\usebox{\plotpoint}}
\put(334.0,608.0){\usebox{\plotpoint}}
\put(334.0,608.0){\usebox{\plotpoint}}
\put(331,607.84){\rule{0.482pt}{0.800pt}}
\multiput(332.00,607.34)(-1.000,1.000){2}{\rule{0.241pt}{0.800pt}}
\put(333.0,609.0){\usebox{\plotpoint}}
\put(331,610){\usebox{\plotpoint}}
\put(331,610){\usebox{\plotpoint}}
\put(330.0,610.0){\usebox{\plotpoint}}
\put(329,609.84){\rule{0.241pt}{0.800pt}}
\multiput(329.50,609.34)(-0.500,1.000){2}{\rule{0.120pt}{0.800pt}}
\put(330.0,610.0){\usebox{\plotpoint}}
\put(329,612){\usebox{\plotpoint}}
\put(329,612){\usebox{\plotpoint}}
\put(327,610.84){\rule{0.241pt}{0.800pt}}
\multiput(327.50,610.34)(-0.500,1.000){2}{\rule{0.120pt}{0.800pt}}
\put(328.0,612.0){\usebox{\plotpoint}}
\put(327,613){\usebox{\plotpoint}}
\put(326.0,613.0){\usebox{\plotpoint}}
\put(326.0,613.0){\usebox{\plotpoint}}
\put(325.0,614.0){\usebox{\plotpoint}}
\put(325.0,614.0){\usebox{\plotpoint}}
\put(322.0,615.0){\usebox{\plotpoint}}
\put(322.0,615.0){\usebox{\plotpoint}}
\put(321.0,616.0){\usebox{\plotpoint}}
\put(321.0,616.0){\usebox{\plotpoint}}
\put(320.0,617.0){\usebox{\plotpoint}}
\put(320.0,617.0){\usebox{\plotpoint}}
\put(317,616.84){\rule{0.241pt}{0.800pt}}
\multiput(317.50,616.34)(-0.500,1.000){2}{\rule{0.120pt}{0.800pt}}
\put(318.0,618.0){\usebox{\plotpoint}}
\put(315.0,619.0){\usebox{\plotpoint}}
\put(315.0,619.0){\usebox{\plotpoint}}
\put(314.0,620.0){\usebox{\plotpoint}}
\put(313,619.84){\rule{0.241pt}{0.800pt}}
\multiput(313.50,619.34)(-0.500,1.000){2}{\rule{0.120pt}{0.800pt}}
\put(314.0,620.0){\usebox{\plotpoint}}
\put(313,622){\usebox{\plotpoint}}
\put(310.0,622.0){\usebox{\plotpoint}}
\put(310.0,622.0){\usebox{\plotpoint}}
\put(309.0,623.0){\usebox{\plotpoint}}
\put(309.0,623.0){\usebox{\plotpoint}}
\put(307.0,624.0){\usebox{\plotpoint}}
\put(307.0,624.0){\usebox{\plotpoint}}
\put(306.0,625.0){\usebox{\plotpoint}}
\put(306.0,625.0){\usebox{\plotpoint}}
\put(301.0,626.0){\rule[-0.400pt]{1.204pt}{0.800pt}}
\put(301.0,626.0){\usebox{\plotpoint}}
\put(300.0,627.0){\usebox{\plotpoint}}
\put(300.0,627.0){\usebox{\plotpoint}}
\put(298.0,628.0){\usebox{\plotpoint}}
\put(298.0,628.0){\usebox{\plotpoint}}
\put(295,627.84){\rule{0.241pt}{0.800pt}}
\multiput(295.50,627.34)(-0.500,1.000){2}{\rule{0.120pt}{0.800pt}}
\put(296.0,629.0){\usebox{\plotpoint}}
\put(291.0,630.0){\rule[-0.400pt]{0.964pt}{0.800pt}}
\put(291.0,630.0){\usebox{\plotpoint}}
\put(290.0,631.0){\usebox{\plotpoint}}
\put(290.0,631.0){\usebox{\plotpoint}}
\put(286.0,632.0){\rule[-0.400pt]{0.964pt}{0.800pt}}
\put(286.0,632.0){\usebox{\plotpoint}}
\put(283.0,633.0){\usebox{\plotpoint}}
\put(283.0,633.0){\usebox{\plotpoint}}
\put(280.0,634.0){\usebox{\plotpoint}}
\put(280.0,634.0){\usebox{\plotpoint}}
\put(277,633.84){\rule{0.482pt}{0.800pt}}
\multiput(278.00,633.34)(-1.000,1.000){2}{\rule{0.241pt}{0.800pt}}
\put(279.0,635.0){\usebox{\plotpoint}}
\put(277,636){\usebox{\plotpoint}}
\put(275.0,636.0){\usebox{\plotpoint}}
\put(275.0,636.0){\usebox{\plotpoint}}
\put(268,635.84){\rule{0.482pt}{0.800pt}}
\multiput(269.00,635.34)(-1.000,1.000){2}{\rule{0.241pt}{0.800pt}}
\put(270.0,637.0){\rule[-0.400pt]{1.204pt}{0.800pt}}
\put(268,638){\usebox{\plotpoint}}
\put(268,638){\usebox{\plotpoint}}
\put(268,638){\usebox{\plotpoint}}
\put(266.0,638.0){\usebox{\plotpoint}}
\put(266.0,638.0){\usebox{\plotpoint}}
\put(260,637.84){\rule{0.482pt}{0.800pt}}
\multiput(261.00,637.34)(-1.000,1.000){2}{\rule{0.241pt}{0.800pt}}
\put(262.0,639.0){\rule[-0.400pt]{0.964pt}{0.800pt}}
\put(260,640){\usebox{\plotpoint}}
\put(260,640){\usebox{\plotpoint}}
\put(255.0,640.0){\rule[-0.400pt]{1.204pt}{0.800pt}}
\put(255.0,640.0){\usebox{\plotpoint}}
\put(225,639.84){\rule{3.614pt}{0.800pt}}
\multiput(232.50,639.34)(-7.500,1.000){2}{\rule{1.807pt}{0.800pt}}
\put(240.0,641.0){\rule[-0.400pt]{3.613pt}{0.800pt}}
\put(225,642){\usebox{\plotpoint}}
\put(220.0,642.0){\rule[-0.400pt]{1.204pt}{0.800pt}}
\end{picture}

\end{minipage} \hfill
\begin{minipage}[b]{.46\linewidth}

% GNUPLOT: LaTeX picture
\setlength{\unitlength}{0.240900pt}
\ifx\plotpoint\undefined\newsavebox{\plotpoint}\fi
\sbox{\plotpoint}{\rule[-0.200pt]{0.400pt}{0.400pt}}%
\begin{picture}(900,720)(110,0)
\font\gnuplot=cmr10 at 10pt
\gnuplot
\sbox{\plotpoint}{\rule[-0.200pt]{0.400pt}{0.400pt}}%
\put(220.0,113.0){\rule[-0.200pt]{148.394pt}{0.400pt}}
\put(220.0,113.0){\rule[-0.200pt]{4.818pt}{0.400pt}}
\put(198,113){\makebox(0,0)[r]{0}}
\put(816.0,113.0){\rule[-0.200pt]{4.818pt}{0.400pt}}
\put(220.0,196.0){\rule[-0.200pt]{4.818pt}{0.400pt}}
\put(198,196){\makebox(0,0)[r]{}}
\put(816.0,196.0){\rule[-0.200pt]{4.818pt}{0.400pt}}
\put(220.0,280.0){\rule[-0.200pt]{4.818pt}{0.400pt}}
\put(198,280){\makebox(0,0)[r]{1}}
\put(816.0,280.0){\rule[-0.200pt]{4.818pt}{0.400pt}}
\put(220.0,363.0){\rule[-0.200pt]{4.818pt}{0.400pt}}
\put(198,363){\makebox(0,0)[r]{}}
\put(816.0,363.0){\rule[-0.200pt]{4.818pt}{0.400pt}}
\put(220.0,447.0){\rule[-0.200pt]{4.818pt}{0.400pt}}
\put(198,447){\makebox(0,0)[r]{2}}
\put(816.0,447.0){\rule[-0.200pt]{4.818pt}{0.400pt}}
\put(220.0,530.0){\rule[-0.200pt]{4.818pt}{0.400pt}}
\put(198,530){\makebox(0,0)[r]{}}
\put(816.0,530.0){\rule[-0.200pt]{4.818pt}{0.400pt}}
\put(220.0,614.0){\rule[-0.200pt]{4.818pt}{0.400pt}}
\put(198,614){\makebox(0,0)[r]{3}}
\put(816.0,614.0){\rule[-0.200pt]{4.818pt}{0.400pt}}
\put(220.0,697.0){\rule[-0.200pt]{4.818pt}{0.400pt}}
\put(198,697){\makebox(0,0)[r]{}}
\put(816.0,697.0){\rule[-0.200pt]{4.818pt}{0.400pt}}
\put(220.0,113.0){\rule[-0.200pt]{0.400pt}{4.818pt}}
%\put(220,68){\makebox(0,0){1965}}
\put(220.0,677.0){\rule[-0.200pt]{0.400pt}{4.818pt}}
\put(308.0,113.0){\rule[-0.200pt]{0.400pt}{4.818pt}}
\put(308,68){\makebox(0,0){1970}}
\put(308.0,677.0){\rule[-0.200pt]{0.400pt}{4.818pt}}
\put(396.0,113.0){\rule[-0.200pt]{0.400pt}{4.818pt}}
\put(396,68){\makebox(0,0){75}}
\put(396.0,677.0){\rule[-0.200pt]{0.400pt}{4.818pt}}
\put(484.0,113.0){\rule[-0.200pt]{0.400pt}{4.818pt}}
\put(484,68){\makebox(0,0){80}}
\put(484.0,677.0){\rule[-0.200pt]{0.400pt}{4.818pt}}
\put(572.0,113.0){\rule[-0.200pt]{0.400pt}{4.818pt}}
\put(572,68){\makebox(0,0){85}}
\put(572.0,677.0){\rule[-0.200pt]{0.400pt}{4.818pt}}
\put(660.0,113.0){\rule[-0.200pt]{0.400pt}{4.818pt}}
\put(660,68){\makebox(0,0){90}}
\put(660.0,677.0){\rule[-0.200pt]{0.400pt}{4.818pt}}
\put(748.0,113.0){\rule[-0.200pt]{0.400pt}{4.818pt}}
\put(748,68){\makebox(0,0){1995}}
\put(748.0,677.0){\rule[-0.200pt]{0.400pt}{4.818pt}}
\put(836.0,113.0){\rule[-0.200pt]{0.400pt}{4.818pt}}
%\put(836,68){\makebox(0,0){2000}}
\put(836.0,677.0){\rule[-0.200pt]{0.400pt}{4.818pt}}
\put(220.0,113.0){\rule[-0.200pt]{148.394pt}{0.400pt}}
\put(836.0,113.0){\rule[-0.200pt]{0.400pt}{140.686pt}}
\put(220.0,697.0){\rule[-0.200pt]{148.394pt}{0.400pt}}
\put(125,395){\makebox(0,0){N$(\le$Y$)$}}
\put(125,325){\makebox(0,0){$[$10$^{6}]$}}
\put(528,23){\makebox(0,0){Y}}
\put(220.0,113.0){\rule[-0.200pt]{0.400pt}{140.686pt}}
\sbox{\plotpoint}{\rule[-0.400pt]{0.800pt}{0.800pt}}%
\put(220,114){\usebox{\plotpoint}}
\put(361,113.34){\rule{4.095pt}{0.800pt}}
\multiput(361.00,112.34)(8.500,2.000){2}{\rule{2.048pt}{0.800pt}}
\put(378,115.34){\rule{4.336pt}{0.800pt}}
\multiput(378.00,114.34)(9.000,2.000){2}{\rule{2.168pt}{0.800pt}}
\put(220.0,114.0){\rule[-0.400pt]{33.967pt}{0.800pt}}
\put(414,116.84){\rule{4.095pt}{0.800pt}}
\multiput(414.00,116.34)(8.500,1.000){2}{\rule{2.048pt}{0.800pt}}
\put(431,118.84){\rule{4.336pt}{0.800pt}}
\multiput(431.00,117.34)(9.000,3.000){2}{\rule{2.168pt}{0.800pt}}
\put(449,121.34){\rule{4.095pt}{0.800pt}}
\multiput(449.00,120.34)(8.500,2.000){2}{\rule{2.048pt}{0.800pt}}
\put(466,122.84){\rule{4.336pt}{0.800pt}}
\multiput(466.00,122.34)(9.000,1.000){2}{\rule{2.168pt}{0.800pt}}
\multiput(484.00,126.41)(0.645,0.509){21}{\rule{1.229pt}{0.123pt}}
\multiput(484.00,123.34)(15.450,14.000){2}{\rule{0.614pt}{0.800pt}}
\put(502,137.84){\rule{4.095pt}{0.800pt}}
\multiput(502.00,137.34)(8.500,1.000){2}{\rule{2.048pt}{0.800pt}}
\put(519,139.84){\rule{4.336pt}{0.800pt}}
\multiput(519.00,138.34)(9.000,3.000){2}{\rule{2.168pt}{0.800pt}}
\put(537,141.84){\rule{4.095pt}{0.800pt}}
\multiput(537.00,141.34)(8.500,1.000){2}{\rule{2.048pt}{0.800pt}}
\put(554,143.84){\rule{4.336pt}{0.800pt}}
\multiput(554.00,142.34)(9.000,3.000){2}{\rule{2.168pt}{0.800pt}}
\put(572,146.84){\rule{4.336pt}{0.800pt}}
\multiput(572.00,145.34)(9.000,3.000){2}{\rule{2.168pt}{0.800pt}}
\put(590,148.84){\rule{4.095pt}{0.800pt}}
\multiput(590.00,148.34)(8.500,1.000){2}{\rule{2.048pt}{0.800pt}}
\put(607,150.34){\rule{4.336pt}{0.800pt}}
\multiput(607.00,149.34)(9.000,2.000){2}{\rule{2.168pt}{0.800pt}}
\put(625,152.84){\rule{4.095pt}{0.800pt}}
\multiput(625.00,151.34)(8.500,3.000){2}{\rule{2.048pt}{0.800pt}}
\multiput(642.00,157.40)(1.212,0.520){9}{\rule{2.000pt}{0.125pt}}
\multiput(642.00,154.34)(13.849,8.000){2}{\rule{1.000pt}{0.800pt}}
\multiput(661.41,164.00)(0.506,0.698){29}{\rule{0.122pt}{1.311pt}}
\multiput(658.34,164.00)(18.000,22.279){2}{\rule{0.800pt}{0.656pt}}
\multiput(678.00,190.40)(1.351,0.526){7}{\rule{2.143pt}{0.127pt}}
\multiput(678.00,187.34)(12.552,7.000){2}{\rule{1.071pt}{0.800pt}}
\put(695,195.84){\rule{4.336pt}{0.800pt}}
\multiput(695.00,194.34)(9.000,3.000){2}{\rule{2.168pt}{0.800pt}}
\multiput(713.00,200.40)(0.877,0.514){13}{\rule{1.560pt}{0.124pt}}
\multiput(713.00,197.34)(13.762,10.000){2}{\rule{0.780pt}{0.800pt}}
\multiput(730.00,210.39)(1.802,0.536){5}{\rule{2.600pt}{0.129pt}}
\multiput(730.00,207.34)(12.604,6.000){2}{\rule{1.300pt}{0.800pt}}
\multiput(749.41,215.00)(0.506,0.757){29}{\rule{0.122pt}{1.400pt}}
\multiput(746.34,215.00)(18.000,24.094){2}{\rule{0.800pt}{0.700pt}}
\multiput(767.41,242.00)(0.507,2.568){27}{\rule{0.122pt}{4.153pt}}
\multiput(764.34,242.00)(17.000,75.380){2}{\rule{0.800pt}{2.076pt}}
\multiput(784.41,326.00)(0.506,9.931){29}{\rule{0.122pt}{15.400pt}}
\multiput(781.34,326.00)(18.000,310.037){2}{\rule{0.800pt}{7.700pt}}
\put(396.0,118.0){\rule[-0.400pt]{4.336pt}{0.800pt}}
\end{picture}

\end{minipage}

%\vspace*{0.7cm}
\caption{~Left:~ Size~ distribution~ of~ radio~ source~ catalogues~
available~ in~ electronic~ form. R is the number of records in
a source catalogue, and N($\ge$R) is the number of
radio source catalogues with R records or more.
The bottom right corner corresponds to the NVSS catalogue. % \linebreak[4]
Right: The growth in time of the number of continuum radio source
measurements. Y is the year of publication of a radio source
catalogue, and N($\le$Y) is the
cumulative number of records (in millions) contained in
catalogues published up to and including year Y.} \label{radhistfig}
\end{figure*}

The size distribution of electronic radio source catalogues
(including my collection and that of CDS) is plotted in the
left panel of Figure~\ref{radhistfig}. For catalogues with more
than $\sim$200 entries the curve follows a power law
with index near $-0.6$, a manifestation of {\it Zipf's law} in
bibliometrics (\cite{zipflaw}). The decline for smaller catalogues
is due to the fact that many of them simply do not exist in
electronic form.  In the right panel of Fig.~\ref{radhistfig}
the growth over time of the cumulative number of records
of these catalogues is plotted.  The three major increases are due to the MSL
in 1981, to the 87GB/GB6/PMN surveys in 1991, and, more recently,
to the release of NVSS, FIRST and WENSS in 1996/97.

Already, since the early 1990s, the author's collection of
radio catalogues has been the most comprehensive one stored at a single site.
However, the problem of making this heterogeneous set of tables
searchable with a common user interface was only solved in 1996,
when the author started collaborating with a group of radio astronomers
at the Special Astrophysical Observatory (SAO, Russia), who had built such
an interface for their  ``Astrophysical {\bf CAT}alogs support {\bf S}ystem''
(CATS; \S\ref{catcoll}). Their common interests in radio astronomy
stimulated the ingestion of a large number of items from the collection.
By late 1996, CATS had surpassed EOLS in size and scope, and in mid-1997 an
email service was opened by CATS, allowing one to query about 200 different
source lists simultaneously for any number of user-specified sky
positions, with just a single and simple email request.
% {\tts cats.sao.ru}

\subsection{Searching in Radio Catalogues: VizieR and CATS}  \label{catcoll}

The largest collections of astronomical catalogues, and published
tabular data in general, are maintained at the CDS and ADC.
The ``Astronomer's Bazaar'' at CDS \linebreak[4]
({\tts cdsweb.u-strasbg.fr/Cats.html}) has over 2200 catalogues and
tables for downloading via anonymous {\tt ftp}.
The full list of items ({\tts ftp://cdsarc.u-strasbg.fr/pub/cats/cats.all})
may be queried for specific catalogues by author name, keyword,
wavelength range, or by name of (space) mission.
At NASA's ADC ({\tts adc.gsfc.nasa.gov/adc.html})
a similar service exists.  Despite the claims that ``mirror copies''
exist in Japan, India and Russia,
CDS and ADC are the only ones keeping their archives {\it current}.
Both have their own catalogue browsers:
{\tt VizieR} at CDS ({\tts vizier.u-strasbg.fr/cgi-bin/VizieR}),
and {\tts Catseye} at ADC
({\tts tarantella.gsfc.nasa.gov/catseye/ViewerTopPage.html}),
but currently none of them allows one to query large numbers of
catalogues at the same time, although such a system is in preparation
within {\tt VizieR} at the CDS. Presently $\sim$\,200 catalogues appear
when VizieR is queried for the waveband ``radio''. This includes
many lists of H\,II regions, masers, etc., but excludes many of the
major radio continuum surveys listed in Table~1.
For radio source catalogues, the CATS system currently has the largest
collection, and CATS is definitely preferable when radio continuum data
are needed.

The CATS system ({\tts cats.sao.ru}) currently permits searches through
about 200 radio source catalogues from about 150 different references,
with altogether over 3 million entries, including current versions of the NVSS,
FIRST and WENSS catalogues. Many further radio source lists are
available via anonymous FTP, as they have not yet been integrated into the
search facility (e.g.\ when only source names, and not positions, are given
in the available electronic version of the catalogue).
Documentation is available for most of the source lists, and in many cases
even large parts of the original paper text were prepared from page scans.

Catalogues in CATS may be selected individually from
{\tts cats.sao.ru/cats\_search.html}, or globally by wavelength range.
One may even select {\it all} searchable catalogues in CATS
(including optical, IR, X-ray), making up  over 4 million entries.
They may be searched interactively on the WWW, or by sending a batch
job via email. To receive the instructions about the exact format for
such a batch request, send an empty email to {\tt cats@sao.ru}
(no subject required). The output can be delivered as a homogeneous
table of sources from the different catalogues, or each catalogue
in its native format. The latter assures that all columns as originally
published (but not included in the homogenised table format) may be
retrieved, although currently the user has to check the individual
catalogue documentation to find out what each column means.
With the {\tt select} option one may retrieve sources from a single sky region,
either a rectangle or a circle in different coordinate systems
(equatorial B1950 or J2000, or Galactic), while the {\tt match} option
allows a whole list of regions to be searched in order to find
all the objects in each region. It is then the responsibility of the user
to find out which of these data represent the object (or parts of an object,
depending on the telescope characteristics) and may be used for
inter-comparisons.

\begin{figure*}[!h]
\hspace*{-1mm}
%\vspace*{-4mm}
\mbox{
\epsfig{file=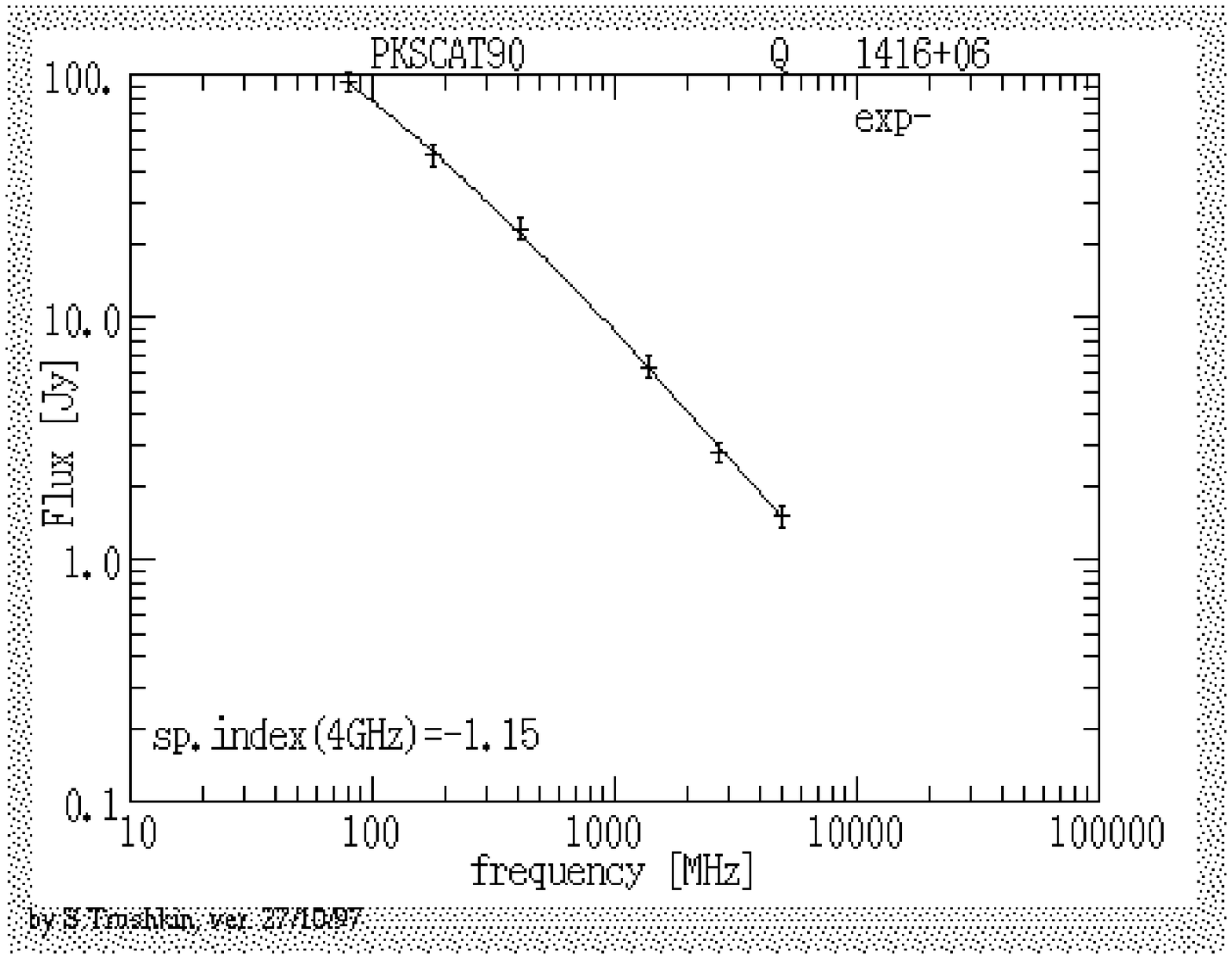,width=6.6cm}
%\hspace*{-6mm}
\epsfig{file=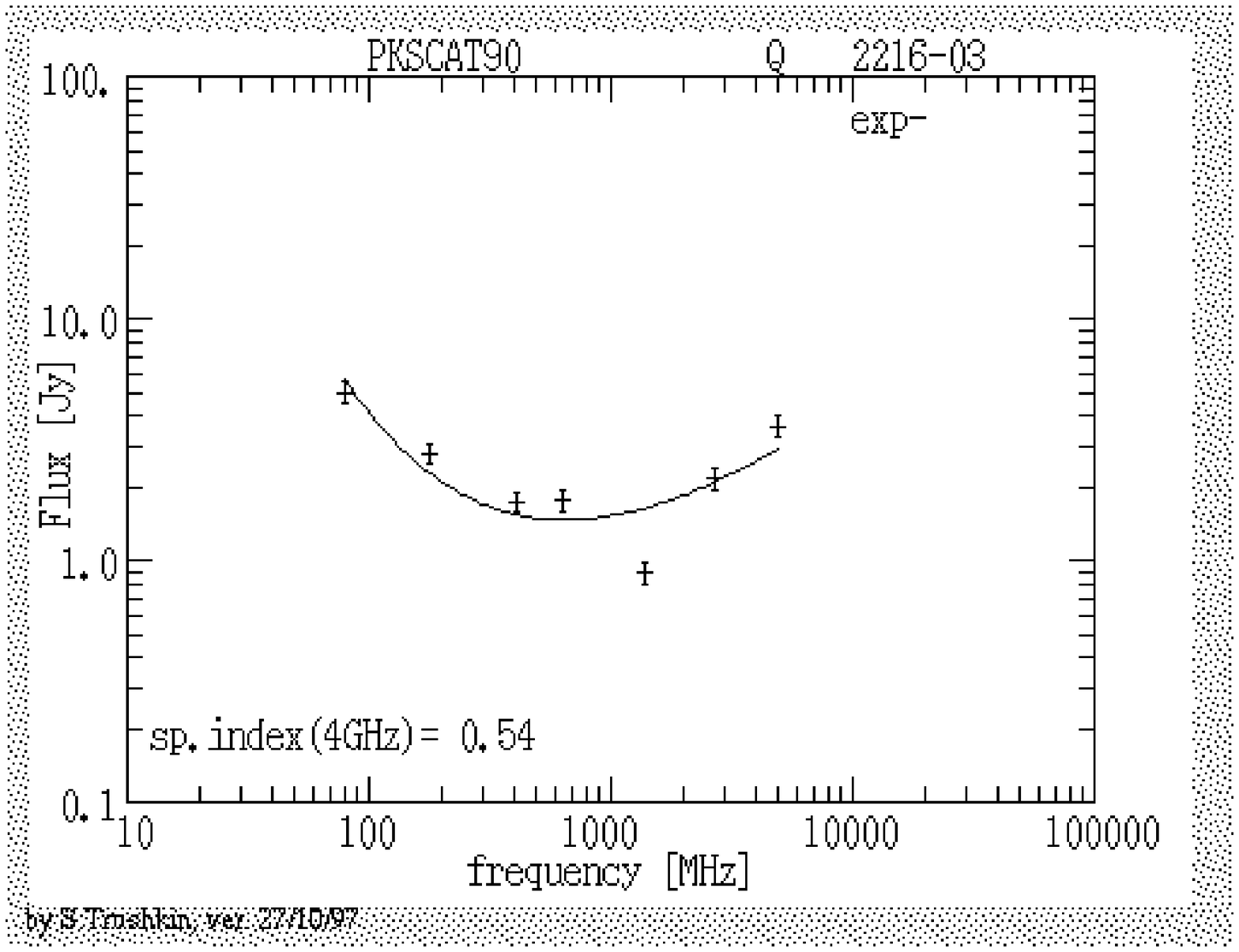,width=6.6cm}
}
%\hspace*{1mm}
%\vspace*{-2mm}
\mbox{
\epsfig{file=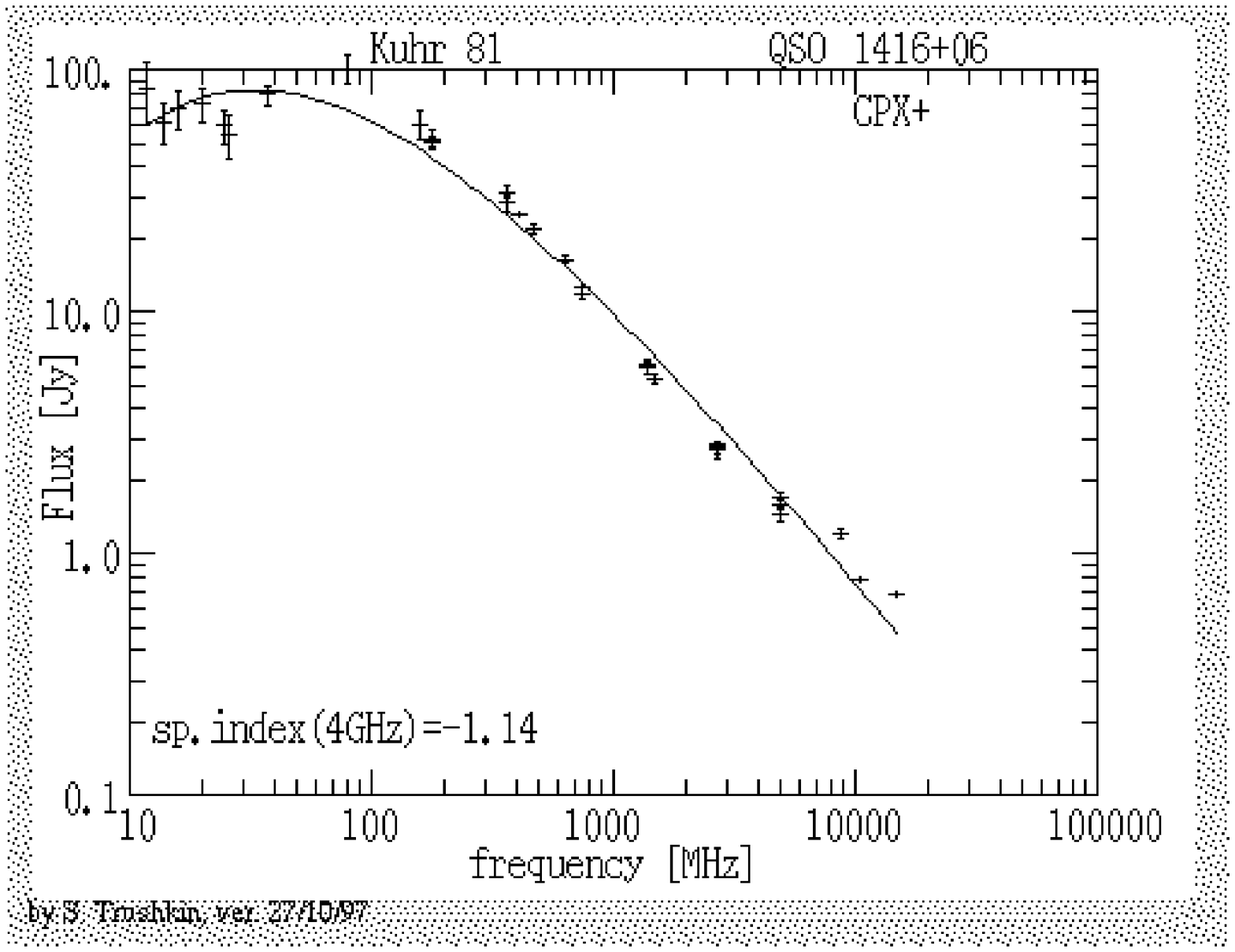,width=6.6cm}
%\hspace*{-6mm}
\epsfig{file=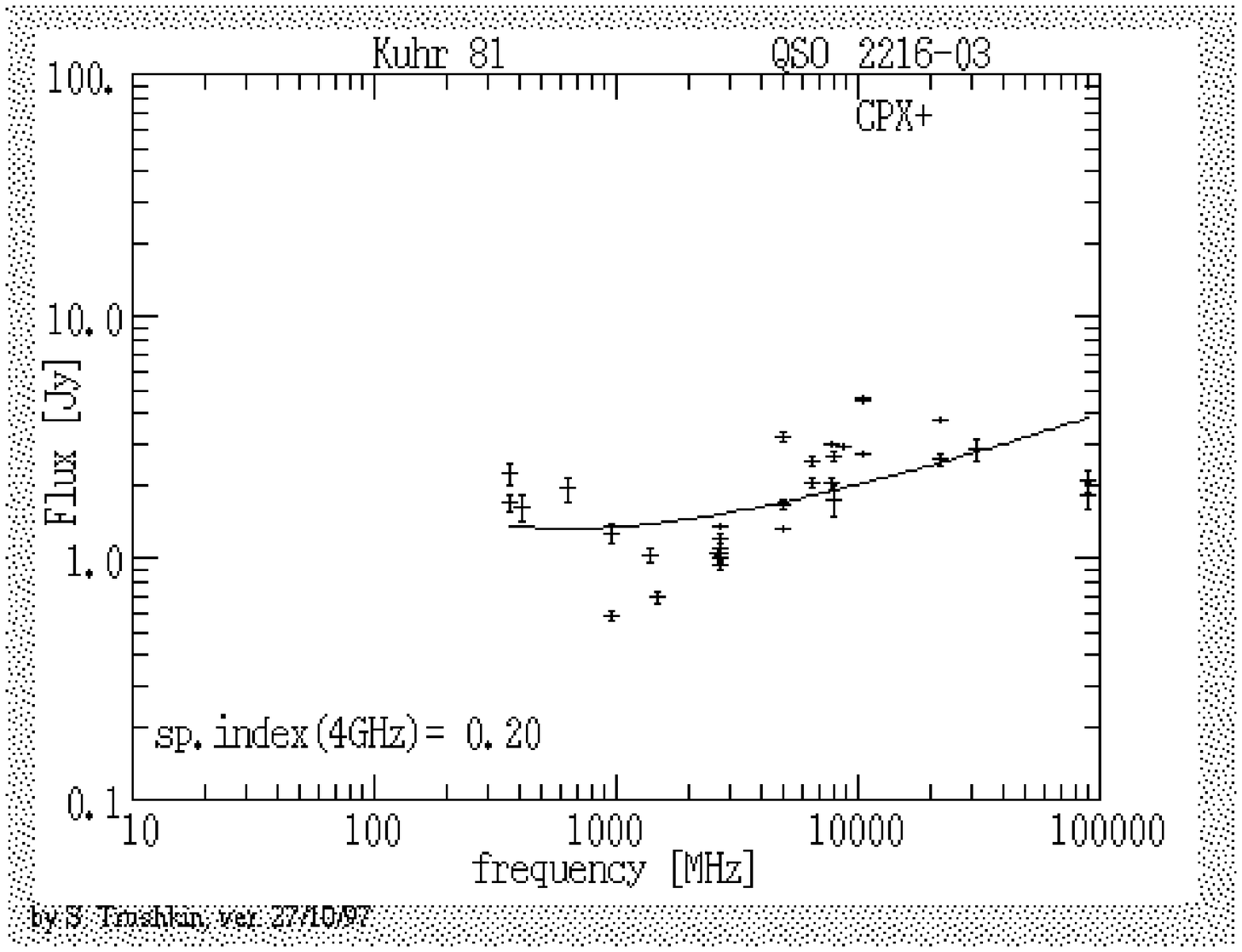,width=6.6cm}
}
%\vspace{-6mm}
\caption{Radio Spectra of the two PKS sources PKS\,1416+06 (left) and
PKS\,2216-03 (right) plotted with CATS.
Upper row: data from PKSCAT90 (one epoch per frequency);
lower row: data from the multi-epoch compilation by K\"uhr et al.\ (1981). % \cite{1981A&AS...45..367K}
} \label{catspecpl}
\end{figure*}

CATS offers a few other useful features. For several multi-frequency
radio catalogues (or rather compilations of radio sources)
CATS allows radio spectra to be plotted on-the-fly, e.g.\ for
\cite{kuhr79}, \cite{1981A&AS...45..367K}, (\cite{1991PASAu...9..170O}, =PKSCAT90),
\cite{1996BSAO...41...64T}, \cite{1980A&AS...42..227K},
\cite{1997BSAO...42....5B}. Various options for fitting these spectra
and weighting the individual flux errors are provided.
Examples for two sources are shown in Fig.~\ref{catspecpl}.
Note that PKSCAT90 includes data obtained at only one epoch
per frequency, while the K\"uhr compilations include several
epochs at a given frequency. Therefore the variability of QSO\,2216$-$03
(=PKS\,B2216$-$038) becomes obvious only in the lower right panel of
Fig.~\ref{catspecpl}.

Note that CATS (at least at present) is a searchable collection of
catalogues, and not a relational database, i.e.\ no cross-identifications
have yet been made between catalogues (except for a few, which resulted
in yet other catalogues). However, given its vastly larger collection
of radio source data, it is an indispensable tool that complements the
information on radio sources found e.g.\ in NED or SIMBAD.

In future it is planned that the user may display both the sky distribution
and a spectral energy distribution (radio spectrum) of all entries found
for a (sufficiently narrow) positional search. The sky plot will indicate
the angular resolution, the positional error box, and (if available)
the shape of each catalogued source, so that the user may interactively
discard possibly unrelated sources, and arrive at the radio spectrum
of the object of interest, as mentioned by \cite{verkho97}.

\subsection{Object Databases: NED, SIMBAD, and LEDA}  \label{objdatabases}

These databases have already been described in my tutorial
for this winter school, so I shall concentrate here on their relevance for radio
astronomy. All three databases were originally built around catalogues of
optical objects (galaxies in the case of NED and LEDA, and stars in the case
of SIMBAD). It is quite natural that information on otherwise unidentified
radio sources is not their priority. Also, being an extragalactic database,
NED tends to provide more information on radio sources than SIMBAD,
which was originally dedicated to stars, which constitute only a negligible
population of radio continuum sources in the sky.
The fact that before being included into NED or SIMBAD, every new (radio
or other) source has to be checked for its possible identification
with another object already in these databases, implies that the integration
of large catalogues may take years from their publication. The rightmost
column of Table~1 gives an idea of this problem. A further obstacle for
database managers is that they have to actively collect the published
data from the authors or other resources. If you wish to see your data
in databases soonest, the best thing is to send them (preferably in plain ASCII
format) to the database managers directly after publication.

SIMBAD is accessible via password from {\tts simbad.u-strasbg.fr/Simbad},
and has its priority in maintaining a good bibliography for astronomical objects
(not necessarily those detected as a radio source only).
NED can be accessed freely through the URL {\tts nedwww.ipac.caltech.edu}
and tends to make an effort to also populate its various ``data frames''
(like optical magnitude, fluxes at various frequencies, etc.) with
recently published measurements.

Searches by object name rely on rather strict rules. In databases these
may not always conform to IAU recommendations
({\tts cdsweb.u-strasbg.fr/iau-spec.html}), mainly due to deviations from
these recommendations by individual authors. In case of doubt about
the exact name of a source, it is wise
to start searching the databases by position.

The ``Lyon-Meudon Extragalactic Database'' (LEDA; {\tts www-obs.univ-lyon1.fr/leda})
is primarily intended for studying the kinematics of the local
Universe, and as such has little interest
in radio continuum data on galaxies. However, LEDA is the ideal place to
look for integrated neutral hydrogen (H\,I) data of nearby
(z\,$\ltsim$\,0.2) galaxies. These H\,I measurements play an important role
for distance estimates of galaxies, independent of their radial velocities.
This allows their ``peculiar motions'' to be calculated, i.e.\ the
deviations of their radial velocities from the Hubble law.

Finally, one should keep in mind that SIMBAD and NED started to include
references on extragalactic objects only since 1983 and 1988
respectively, although a few major references before these dates have
now been included. In the following I give just one example in which
the consequences of this have {\it not} been considered by users
of NED or SIMBAD. The X-ray source RX\,J15237+6339 was identified (from NED)
with the radio source 4C$+$63.22 by \cite{1994A&A...285..812B},
and these authors comment that ``One object (4C$+$63.22)
is classified as `Radio Source' only in the NED data base, so,
strictly speaking, it belongs to the class of unidentified objects.''
However, according to VV83 (their ref.\,603=\cite{1980MNRAS.191..607P})
the source 4C$+$63.22 had actually been identified with an 18.5\,m galaxy.
This object is within 5$''$ of the brightest source in the
NVSS catalogue within a radius of 90$''$. However, the NVSS map
shows a large triple radio galaxy with a North-South extent of $\sim$\,4$'$.
Later re-observation at 5\,GHz with the VLA at 1.3$''$ resolution
(\cite{1997A&AS..122..235L}) detected only a radio core
coincident with a 16.6\,m object which the 1980 authors had already
mentioned in their notes as 1$'$ offset from their prime candidate
identification for the source 1522$+$638. I should add that the
data table of the 1980 paper was published on microfiche.

\section{Miscellaneous Databases and Surveys of Radio Sources}

Several WWW sites offering topical databases of special
types of radio source research will be mentioned briefly
in this section.

\subsection{Clusters of Galaxies}    \label{clusterdatabase}

A collection of well-chosen radio source catalogues has been used,
together with optical sky surveys like the DSS, to develop a database
of radio-optical information on Abell/ACO (\cite{1989ApJS...70....1A})
clusters of galaxies. It is managed by A.~Gubanov at the St.-Petersburg State
University in collaboration with the present author
(Gubanov \& Andernach 1997).    % \cite{1997BaltA...6..263G}
At the URL~ {\tts future.astro.spbu.ru/Clusters.html}, the user
may interactively prepare schematic radio-optical overlays,
charts from the FIRST, NVSS, APM or DSS survey data, or retrieve
references for cluster data. Radio continuum spectra of cluster radio
galaxies may be displayed and fitted with user-specified functions.
Source luminosities may be derived assuming cluster membership.

\subsection{VLBI and Astrometric Surveys} \label{vlbi}

The VLBA Calibrator Survey ({\tts magnolia.nrao.edu/vlba\_calib/})
is an ongoing project to provide
phase-reference calibrators for VLBA experiments. When completed it will
contain astrometric ($\sim$1\,mas) positions and 2.7 and 8.4 GHz images
of the $\sim$3000 sources in the JVAS catalogue (\S\ref{glenses}).

The ``Pearson-Readhead'' (PR) and ``Caltech -- Jodrell\,Bank'' (CJ)
imaging data base is a VLBI source archive at Caltech
({\tts \verb*Castro.caltech.edu/~tjp/cj/C}).
It offers images of over 300 VLBI sources at
$\delta>$35\deg\ observed in the PR (\cite{1988ApJ...328..114P}),
CJ1 (\cite{1995ApJS...99..297X}), and CJ2 (\cite{1995ApJS..100....1H})
surveys. Many of these sources are
excellent calibrators for the VLA and VLBA.
It has mostly 5\,GHz (6\,cm) and some 1.67\,GHz (18\,cm) VLBI images,
as well as 1.4 and 5-GHz VLA images of extragalactic sources.
Contour maps are publicly available as PostScript files.

A sample of 132 compact sources have been observed in ``snapshot''
mode with the VLBA at 15\,GHz (2\,cm; \cite{1998AJ....115.1295K}).
At present it contains images at one epoch ({\tts www.cv.nrao.edu/2cmsurvey/}
~and~ {\tts www.mpifr-bonn.mpg.de/zensus/2cmsurvey/}), but it eventually will have
multi-epoch sub-milliarcsecond data.

The ``Radio Reference Frame Image Database'' (RRFID) is maintained at the
U.S.\ Naval Observatory (USNO). Data obtained with the VLBA and the
``Global Geodetic Network'' at 2.3, 8.4, and 15\,GHz (13, 3.6 and 2\,cm)
are publicly available as over 1400 images for more than 400 sources,
at {\tts maia.usno.navy.mil/rorf/rrfid.html}.
In April 1998 first-epoch imaging of northern hemisphere radio
reference frame sources was completed.  Images at both 2.3 and 8.5\,GHz
now exist for $\sim$97\,\% of the ``Radio Optical Reference Frame'' (RORF)
sources (\cite{1995AJ....110..880J}) north of $\delta$=$-$20\deg,
which is $\sim$90\% of the ``International Celestial Reference Frame'' (ICRF)
sources in this region of sky.
A number of links are available from the RRFID page, in particular to the
RORF data base of sources
({\tts maia.usno.navy.mil/rorf/rorf.html}).

The European VLBI Network (EVN; {\tts www.nfra.nl/jive/evn/evn.html})
was formed in 1980 by the major European radio astronomy institutions,
and is an array of radio telescopes spread over Europe, the Ukraine,
and China. A catalogue of observations carried out so far can be
retrieved from~ {\tts terra.bo.cnr.it/ira/dira/vlbinet.dat}. The
column explanation is available at~
{\tts www.ira.bo.cnr.it/dira/vlbinet.doc}.

\subsection{Gravitational Lens Surveys}   \label{glenses}

About 2500 compact northern sources stronger than 200\,mJy at 5\,GHz have
been mapped with the VLA at 8.4\,GHz in the
``Jodrell-Bank/VLA Astrometric Survey''
(JVAS; \linebreak[4] {\tts \verb*Cwww.jb.man.ac.uk/~njj/glens/jvas.htmlC}).
The goal was (\cite{1992MNRAS.254..655P}) to find phase calibrator sources for
the MERLIN interferometer ({\tts www.jb.man.ac.uk/merlin/}) and
gravitational lens candidates. If the redshift of both the
parent object of the compact source and that of the intervening
galaxy or cluster (causing the lensing effect) can be determined,
and if in addition the compact source shows variability (not uncommon
for such sources), the time delay between radio flares in the different
images of the lensed object may be used to constrain the Hubble
constant.

The ``Cosmic Lens All-Sky Survey'' (CLASS;
{\tts \verb*Castro.caltech.edu/~cdf/class.htmlC}),
is a project to map more than 10,000 radio sources in
order to create the largest and best studied statistical sample of
radio-loud gravitationally lensed systems. Preliminary 8.4-GHz fluxes
and positions are already available from
{\tts \verb*Cwww.jb.man.ac.uk/~njj/glens/class.htmlC} and
{\tts \verb*Cwww.jb.man.ac.uk/~ceres1/list_pub.htmlC}. The whole
database will eventually be made public.

The ``CfA -- Arizona Space Telescope Lens Survey'' (CASTLeS)
provides information and data on gravitational lens systems
at~  {\tts cfa-www.harvard.edu/glensdata/}.
It includes HST and radio images that can be downloaded
via {\tt ftp}. The service distinguishes between
multiply imaged systems and binary quasars.

The ``VLBI Space Observatory Program'' (VSOP or HALCA;
{\tts www.vsop.isas.ac.jp/}) has put an 8-m radio antenna
into a highly elliptical Earth orbit so as to extend terrestrial
interferometer baselines into space.
Several hundred sources in the VSOP Survey Program
({\tts www.vsop.isas.ac.jp/obs/Survey.html}) are listed,
together with their observational status,
at {\tts www.ras.ucalgary.ca/survey.html}.
These were selected to have 5-GHz flux above 1\,Jy
and a radio spectrum flatter than S$\sim\nu^{-0.5}$.
Galactic masers are also being surveyed.
An image gallery, including the first images ever made in
Space-VLBI, may be viewed at {\tts www.vsop.isas.ac.jp/general/Images.html}.
Further images of EVN-HALCA observations are available
at {\tts www.nfra.nl/jive/evn/evn-vsop.html}.
The same page will soon provide access to VLBA images of over 350
extragalactic sources observed with the VLBA at 5\,GHz prior to the VSOP
launch and some results of the pre-launch OH-maser survey.

\subsection{Variable Sources and Monitoring Projects}  \label{variables}

Since 1997, about forty Galactic and extragalactic variable sources have been
monitored with the Green Bank Interferometer (GBI) at 2.25 and 8.3\,GHz
(HPBW 11$''$ and 3$''$, resp.), under NASA's OSIRIS project.
The instrument consists of three 26-m antennas, and the targets
are preferentially X-ray and $\gamma$-ray active.
Radio light curves and tables of flux densities are provided at~
{\tts info.gb.nrao.edu/gbint/GBINT.html}.

The ``University of Michigan Radio Astronomy Observatory'' (UMRAO) database
({\tts www.astro.lsa.umich.edu/obs/radiotel/umrao.html};
\cite{1992ApJ...396..469H}) contains
the ongoing observations of the University of Michigan
26-m telescope at 5, 8.5 and 15\,GHz. A number of strong sources are
frequently (weekly) monitored and some weaker sources a bit less often.
Currently the database offers flux densities, and polarisation percentage
and angle (if available), for over 900 sources. For some objects the
on-line data go back to 1965.

\subsubsection{Solar Radio Data}       \label{solardata}

An explanation of the types of solar bursts and a list of special
events observed can be found at {\tts www.ips.gov.au/culgoora/spectro/}.
Daily flux measurements of the Sun at 2.8\,GHz (10.7\,cm) back
to 1947 are offered at {\tts www.drao.nrc.ca/icarus/www/sol\_home.shtml}.
The Ondrejov Solar Radio Astronomy Group
({\tts \verb*Csunkl.asu.cas.cz/~radioC}) provides an archive
of events detected with a 3.0\,GHz
continuum receiver and two spectrographs covering the range 1.0--4.5\,GHz.
The Metsahovi Radio Station in Finland offers solar radio data
at {\tts kurp.hut.fi/sun}, like e.g.\ radio images of the full Sun
at 22, 37 and 87\,GHz, a catalogue of flares since 1989, or ``track plots''
(light curves) of active regions of the solar surface.
For further data, get in contact with the staff at~
{\tts solar@hut.fi} ~or~ {\tts Seppo.Urpo@hut.fi}.

The ``National Geophysical Data Center'' (NGDC) collects
solar radio data from several dozen stations over the world
at {\tts www.ngdc.noaa.gov/stp/SOLAR/getdata.html}.
This ``Radio Solar Telescope Network'' (RSTN) of 55 stations
has now collected 722 station-years worth of data.
Information about solar bursts, the solar continuum flux, and \linebreak[4]
spectra from RSTN may be retrieved and displayed graphically at the URL \linebreak[4]
{\tts www.ngdc.noaa.gov:8080/production/html/RSTN/rstn\_search\_frames.html}.

Since July 1992, the Nobeyama Radio Observatory (NRO) has been offering
(at {\tts solar.nro.nao.ac.jp/}) a daily 17\,GHz image
of the Sun, taken with its Nobeyama Radio Heliograph.
Also available are daily total flux
measurements at five frequencies between 1~and~17\,GHz observed since
May 1994, as well as some exciting images of solar radio flares.

Cracow Observatory offers daily measurements of solar radio emission
at six decimetric frequencies (410--1450\,MHz) from
July 1994 to the present ({\tts www.oa.uj.edu.pl/sol/}).

Measurements with the Tremsdorf radio telescope of the
Astrophysics Institute Potsdam (AIP), Germany, based on
four solar sweep spectrographs (40--800\,MHz) are available
at {\tts \verb*Caipsoe.aip.de/~detC}.

The Radio Astronomy Group (RAG) of the ETH Z\"urich offers the
data from its various digital radio spectrometers and sweep spectrograph
at {\tts www.astro.phys.ethz.ch/rag}, and also hosts the homepage of the
``Community of European Solar Radio Astronomers'' (CESRA).
The ``Joint Organization for Solar Observations''
(JOSO) offers a comprehensive list of links
to solar telescopes and solar data centres
at {\tts joso.oat.ts.astro.it}.

A wide variety of solar data, including East-West scan images from
the Algonquin Radio Observatory 32-element interferometer,
and reports of ionospheric data, are provided by the University of
Lethbridge, Alberta, Canada ({\tts holly.cc.uleth.ca/solar}, or its
{\tt ftp} server at {\tts ftp://ftp.uleth.ca/pub/solar/}).

Finally, there are many sites about solar-terrestrial processes
and ``Space Weather Reports'', e.g.\ at~
{\tts www.sel.noaa.gov} or {\tts www.ips.gov.au/}.
There are spacecraft
solar radio data at~ {\tts lep694.gsfc.nasa.gov/waves/waves.html}
~and~ {\tts www-istp.gsfc.nasa.gov/}.

\section{Raw Data, Software, Images and Spectral Line Data}   \label{uvimagesspectrallines}  % lesson 4 section 5

\subsection{Radio Observatories and their Archives}  \label{obsarchives}

A list of 70 radio astronomy centres with direct links to their WWW pages
has been compiled at
{\tts www.ls.eso.org/lasilla/Telescopes/SEST/html/radioastronomy.html},
and the \linebreak[4] URL
{\tts \verb*Cmsowww.anu.edu.au/~anton/astroweb/astro_radio.htmlC} presents
links to 65 radio telescopes. Further WWW sites of radio observatories
may be found on AstroWeb ({\tts www.cv.nrao.edu/fits/www/astronomy.html}),
searching for ``Telescopes'' or ``Radio Astronomy''.

A recent inquiry by E.~Raimond of NFRA revealed the following
(priv.comm., cf.\ also \linebreak[4] {\tts www.eso.org/libraries/iau97/commission5.html}).
%\vspace*{-2mm}
\begin{itemize}
\item Most major radio observatories saved their data, which does
not necessarily mean that they are still usable. Only the larger
institutions kept data readable by copying them to modern media.
%\vspace*{-2mm}
\item Data usually become available publicly after 18 months. Outside users
can (sometimes) search a catalogue of observations and/or projects.
%\vspace*{-2mm}
\item Retrieval of archived data often requires the help of observatory
staff. Support is offered in general.
\item Radio observatories with a usable archive, and user support for those
who wish to consult it, typically do not advertise this service well\,!
%\vspace*{-2mm}
\end{itemize}

\subsubsection{Archives of Centimetre- and Metre-wave Telescopes/Arrays}

For the 305-m antenna at Arecibo (Puerto Rico; {\tts www.naic.edu/})
all data were kept, some still on reel-to-reel magnetic tapes, but no
catalogue is available remotely. A catalogue of projects is searchable
with the help of the staff.

Data obtained at the Effelsberg 100-m telescope of the
Max-Planck Institut f\"ur Radioastronomie (MPIfR) at Bonn, Germany
({\tts www.mpifr-bonn.mpg.de/effberg.html}) were originally
archived only for the five years prior to the overwriting of
the storage medium. Presently data are archived
on CD-ROMs, but these are not accessible to the outside world.
There is no public observations catalogue on-line, and help
from staff is required to work with data taken with the 100-m antenna.

The ``Australia Telescope National Facility'' (ATNF) has archived
all raw data of its  ``Compact Array'' (ATCA; {\tts www.narrabri.atnf.csiro.au/}).
Observations and project databases are available at
{\tts www.atnf.csiro.au/observers/data.html}.

The ATNF Parkes 64-m telescope, despite its 27 years of operation, has
no data archive so far. The multibeam H\,I surveys (\S\ref{hi}), just started,
will be archived, and results will be made public.

At the ``Dominion Radio Astronomy Observatory'' (DRAO; {\tts www.drao.nrc.ca})
raw and calibrated data are archived. Observatory staff will assist
in searching the observations catalogue.  Results of the
Galactic Plane Survey (in progress, \S\ref{plannedsurveys}) will be
made available publicly via the CADC.

The Molonglo Observatory ({\tts www.physics.usyd.edu.au/astrop/most/})
has archived its raw data. In general, data can be retrieved via
collaborations with staff. Results of recently started
surveys (\S\ref{sumss}, \S\ref{plannedsurveys}) will be made
publicly available.

The archive of NRAO's {\it Very Large Array}
contains data from 1979 to the present (excluding 1987) and
can be interrogated at
{\tts info.aoc.nrao.edu/doc/vladb/VLADB.html}.
Data are reserved for the observing team for 18 months following the
end of the observations. Archive data after this period must be requested
from either the Assistant Director or the Scheduler at the VLA.

The {\it Netherlands Foundation for Research in Astronomy} (NFRA)
keeps an archive of all the raw data ever taken with the
Westerbork Synthesis Radio Telescope (WSRT), an aperture synthesis telescope
of 14 antennas, of 25-m diameter, in the Netherlands. At URL
{\tts www.nfra.nl/scissor/} one may browse this archive by various
criteria (RA, DEC, frequency, observation date, etc.) and even
formulate a request to obtain the data. Note that you need to specify
a username and password (both ``guest'') before you may query the database.
However, special auxiliary files will be needed to reduce these data
e.g.\ with AIPS. For the processed results of the WENSS survey, see \S\ref{wenss}.

The ``Multi-Element Radio-Linked Interferometer'' (MERLIN), in the UK,
has all raw data archived since MERLIN became a National Facility in
1990. The catalogue of \linebreak[4] observations older than 18 months is
searchable on position and other parameters \linebreak[4] ({\tts www.jb.man.ac.uk/merlin/archive}).
For the actual use of archived data, a visit to Jodrell Bank is
recommended in order to get the processing done properly.

\subsubsection{Archives of (Very) Long Baseline Interferometers}

At the ``Very Long Baseline Array''
(VLBA; {\tts www.nrao.edu/doc/vlba/html/VLBA.html}) in the USA,
all correlated data are archived, and the observations catalogue
may be retrieved from {\tts www.nrao.edu/ftp/cumvlbaobs.txt}.
(When I inquired about the latter URL, the reply came with a
comment\,: ``I'm not sure any URL is sufficiently permanent to be
mentioned in a book.'')

The ``European VLBI Network'' (EVN) has its correlated data archived
at MPIfR Bonn, Germany, so far. A catalogue of observations is available
(\S\ref{vlbi}), but the data are {\it not} in the public domain.
Once the correlator at the ``Joint Institute for VLBI in Europe''
(JIVE) becomes operational, archiving will be done at Dwingeloo
({\tts www.nfra.nl/jive}). A general problem is that in order to
re-analyse archived data, calibration data are also needed, and
these are not routinely archived.

The US Naval Observatory (USNO) has made VLBI observations for
geodetic and reference frame purposes for more than a decade.
Among other results, these lead to the latest estimates of the
precession and nutation constants ({\tts maia.usno.navy.mil/eo/}).
This archive (\S\ref{vlbi}) contains all of the images available from the
Geodetic/Astrometric experiments, including the VLBA.  Almost all
of these VLBA observations have been imaged. However, the USNO
Geodetic/Astrometric database is huge, and the large, continuing project
of imaging has been only partially completed. Examples of archival use
of the Washington VLBI Correlator database can be found in
\cite{1998ApJ...500..810T} and references therein.

% contact Kerry Kingham  {\tt kak@cygx3.usno.navy.mil}
% CDDIS (Crustal Dynamics Data Information System)  S. Britzen

\subsubsection{Millimetre Telescopes and Arrays}

The ``Berkeley-Illinois-Maryland Association'' (BIMA,
{\tts bima.astro.umd.edu/bima/}, \linebreak[4] \cite{1996PASP..108...93W})
maintains a millimetre array which has no formal archiving policy.
However, there is a searchable observatory archive of raw data.
To request authorisation for access, visit the page~
{\tts bima-server.ncsa.uiuc.edu/bima/secure/bima.html}.

At the ``Caltech Millimeter Array'' ({\tts www.ovro.caltech.edu/mm/main.html})
all observed data are in a database. Searching requires the help of
the staff, because it is not easy to protect proprietary data
in another way.

The ``Institut de RadioAstronomie Millim\'etrique'' (IRAM)
maintains the 30-m dish at Pico Veleta (Spain, {\tts ram.fr/PV/veleta.html}),
and the ``Plateau de Bure Interferometer''
(PDBI; {\tts iram.fr/PDBI/bure.html}).
The raw data from the PDBI are archived, but so far the observation catalogues
are made available only via Newsletter and e-mail. Web-access is bein considered
for the future. The P.~de~Bure archive is only accessible within {\tt iram.fr},
the IRAM local network. The only external access to the P.\,de Bure archive
is to pull up the observations list month by month (from 1990 to the present)
from the page {\tts iram.fr/PDBI/project.html}.

The ``James-Clerk-Maxwell Telescope'' (JCMT; {\tts www.jach.hawaii.edu/JCMT/})
on Mauna Kea (Hawaii) has its raw and processed data archived
at the Canadian Astronomy Data Center (CADC; {\tts cadcwww.dao.nrc.ca/jcmt/}).

The ``Caltech Sub-mm Observatory'' (CSO; {\tts \verb*Cwww.cco.caltech.edu/~cso/C})
is a 10.4-m sub-mm dish on Mauna Kea (Hawaii) in operation since 1988.
Its archive can be reached via the URL~ 
{\tts puuoo.submm.caltech.edu/doc\_on\_vax/html/doc/archive.html}.

The ``Nobeyama Millimetre Array'' (NMA; {\tts www.nro.nao.ac.jp/NMA/nma-e.html})
and the Smithsonian Sub-mm Array (SMA; {\tts sma2.harvard.edu/})
have their archives in the software development stage. \\

In conclusion, archiving in radio astronomy is far from optimal, but
not too bad either. Most major radio observatories have an archive of some
sort, and accessibility varies from excellent to usable.
Some observatories could advertise their archives better, e.g.\ on their
own Web-pages, or by registering it in AstroWeb's links.
Most importantly, these archives are {\bf very little} used by
astronomers, and would be well worth many (thesis) projects, e.g.\ to study
source variability over more than a decade. The following section
gives some hints on where to start searching when
software is needed to reduce some of the raw data retrievable from
the Internet.

\subsection{Software for Radio Astronomy}    \label{software}

As are the observing techniques for radio astronomy, its data reduction
methods are much more diverse than those in optical astronomy.
Although AIPS ({\tts www.cv.nrao.edu/aips}) has been the dominating
package for radio interferometer data, many other packages \linebreak[4]
have been developed for special purposes, e.g.\
%NOD2 at MPIfR Bonn ({\tts ftp://fs1.mpifr-bonn.mpg.de/pub/nod}) \\
GILDAS, GREG, \& CLASS at IRAM  \linebreak[4] ({\tts iram.fr/doc/doc/doc/gildas.html}),
Analyz at NAIC ({\tts ftp://naic.edu/pub/Analyz}), \linebreak[4]
% {\tts               ........} : at WSRT \\
GIPSY at Groningen ({\tts ftp://kapteyn.astro.rug.nl/gipsy/}),
Miriad by the BIMA and ATNF staff ({\tts www.atnf.csiro.au/computing/software/miriad/}),
and Karma at ATNF   \linebreak[4]
({\tts ftp://ftp.atnf.csiro.au/pub/software/karma/}).
A comprehensive compilation of links can be found from the
AstroWeb at {\tts www.cv.nrao.edu/fits/www/yp\_software.html}.
The ``Astronomical Software and Documentation Service''
(ASDS; {\tts asds.stsci.edu/asds})  contains links to the major
astronomical software packages and documentation. It allows one to
search for keywords in all the documentation files available.

A complete rewrite of the AIPS package from Fortran to {\tt C++} code,
known as the {\tt AIPS++} project ({\tts aips2.nrao.edu/aips++/docs/html/aips++.html}),
has been under way since mid-1991.

\smallskip

{\it A note on preparing radio-optical overlays with AIPS}\,:~~
With the public availability of 2-dimensional maps from radio (NVSS, FIRST,
WENSS) and optical (DSS) surveys, it is relatively easy to prepare
radio-optical overlays for identification or publication purposes.
Radio and optical maps of similar size should be culled in FITS
format from the WWW. To identify the coordinate system of a map,
AIPS looks for the FITS-header keyword ``EPOCH'' rather
than ``EQUINOX'' (which is one of the very few bugs in the FITS
definition\,!). Maps from SkyView (e.g.\ DSS and NVSS) seem to lack the
EPOCH keyword in their FITS header, thus AIPS {\it assumes}\,(!)
them to be of equinox 1950.0. (Both FIRST and NVSS maps, when
taken from their home institutions, STScI and NRAO, respectively, do have the
EPOCH keyword properly set.)
Thus, for AIPS to work correctly on SkyView maps, it is necessary
to introduce the proper ``EPOCH'' value in the map headers.
This may be done with {\tts gethead} and
{\tts puthead} in AIPS. Then, one of the maps (usually the one
with the coarser pixel size) has to be prepared for re-gridding to
the grid of the map with the finer pixel size.  This preparation
may be done with {\tts EPOSWTCH}, before the actual re-gridding is done
with {\tts HGEOM}. Finally the task {\tts KNTR} permits one to plot
one of the maps (usually the optical) in greyscale, and the other
as contours (usually the radio map). However, for more sophisticated
combined plots of greys and contours (including white contours), other
software packages (\S\ref{software}) allow finer artwork to be produced.

\subsection{Radio Images on the Internet}       \label{images}

Here we have to distinguish between images extracted from
large-scale surveys, and images of individual sources.
Both types will be discussed in the following two subsections.

\subsubsection{Images from Large-scale Surveys}  \label{surveyimages}

I had already mentioned (\S\ref{modsurv}) that the very
large-scale radio surveys like NVSS, FIRST, and WENSS offer
(or are in the process of developing)
so-called ``postage-stamp servers'', i.e.\ WWW interfaces where
desired pieces of the 2-dimensional maps may be extracted, either
in {\tt gif} format, or, if one needs to work with the data, in
the (usually about 10 times larger) FITS format.
For the retrieval of large lists of small images, typically for
identification projects, it is worth noting that several sites
offer scripts (mostly based on {\tt perl}) which allow the
retrieval of these maps ``from the command line'', i.e.\ without
even opening a WWW browser\,!  The source list and map sizes may
be pre-edited locally within a sequence of commands which are run in background
(e.g. during the night, if necessary), and which will save the requested
maps as files with names of the user's choice. For the NVSS,
these may be obtained from W.~Cotton ({\tt bcotton@nrao.edu}) or
from {\tts skyview.gsfc.nasa.gov/batchpage.html}. For FIRST images
look at~ {\tts \verb*Cwww.ast.cam.ac.uk/~rgm/first/collab/first_batch.htmlC},
or use the {\tt lynx} browser from the command line
(consult R.~White at {\tt rlw@stsci.edu} in case of doubt).

FIRST and NVSS have mirror sites for their data products at
the ``Mullard Radio Astronomical Observatory''
(MRAO, Cambridge, UK; {\tts www.mrao.cam.ac.uk/surveys}),
to allow faster access from Europe.
Presently only part of the FIRST maps (and not the FIRST source
catalogue) are available from there. Make sure that the piece
of information you need is included at this site before concluding
that it has not been observed.

A number of large-scale radio surveys are accessible from NASA's {\tt SkyView}
facility \linebreak[4] ({\tts skyview.gsfc.nasa.gov/}).
These are the 34.5\,MHz survey with the GEETEE telescope in India
(\cite{1990JApA...11..323D}),
the 408\,MHz all-sky survey (Haslam et al.\ 1982), % \cite{1982A&AS...47....1H}),
the 1.4\,GHz Stockert 25-m dish surveys
(\cite{1982A&AS...48..219R}, \cite{1986A&AS...63..205R}),
FIRST and NVSS at 1.4\,GHz (see \S\ref{modsurv}), and
the 4.85\,GHz surveys of 1986$+$87 with the Green Bank 300-ft telescope
(\cite{1994AJ....107.1829C}), as well as their southern counterparts
made with the Parkes 64-m dish (PMN; \cite{1993AJ....106.1095C}).
A 4.85\,GHz survey made with the NRAO 140-ft antenna (covering
0$^h<$RA$<$20$^h$, $-$40\deg$<\delta<+$5\deg) is also available
(\cite{1991AJ....102.2041C}).
Descriptions of the surveys accessible from {\tt Skyview} can
be found at URL~ {\tts skyview.gsfc.nasa.gov/cgi-bin/survey.pl}.

An attempt to list some of the survey work at radio wavelengths
in both hemispheres was made with the page ``Radio Surveys of
the Southern and Northern Sky''
({\tts wwwpks.atnf.csiro.au/databases/surveys/surveys.html}).
Links to the data from these surveys are included, where available.

Extractions from the large-scale surveys
made at MPIfR Bonn can be retrieved interactively from
the URL~ {\tts www.mpifr-bonn.mpg.de/survey.html}, including
polarisation maps (Stokes Q and U) of the Galactic plane at 2.7\,GHz.

The WSRT has been used to survey a section of the Galactic plane
at 327\,MHz (\cite{1996ApJS..107..239T}). The region
43\deg$<\ell<$91\deg, $|b|<$1.6\deg\ was covered with 23 overlapping fields.
Each field was observed at two epochs, several years apart, to
identify variable sources. Combined intensity maps from
both epochs, having a sensitivity of typically a few mJy and
angular resolution of 1$'\times$1$'$\,csc($\delta$),
may be viewed or retrieved as FITS images from the URL~
{\tts www.ras.ucalgary.ca/wsrt\_survey.html}.

The Hartebeesthoek Radio Astronomy Observatory (HartRAO) has used
its 26-m dish at 2.3\,GHz to map 67\% of the southern sky
(0$^h<\alpha<$12$^h$,$-$80\deg$<\delta<+$13\deg;
12$^h<\alpha<$24$^h$,$-$83\deg$<\delta<+$32\deg) with an angular
resolution of 20$'$ (\cite{1998MNRAS.297..977J}). Until now this is the
highest frequency at which such large areas have been mapped, while
preserving large-scale emission features.
To see a combination with northern sky surveys, 
go to {\tts www.ru.ac.za/departments/physics/physics.html}, and click
on ``Radio Astronomy Group''.
Survey maps are available at
{\tts ftp://phlinux.ru.ac.za/pub/survey} (or contact J.\,Jonas at
{\tts phjj@hippo.ru.ac.za}).

The southern Galactic plane has been surveyed with the Parkes 64-m dish
at 2.42\,GHz (\cite{1995MNRAS.277...36D}). The region 238\deg$<\ell<$5\deg,
$|b|<$5\deg was mapped with 10.4$'$ resolution. The polarisation
data of that survey have been published in \cite{1997MNRAS.291..279D},
and are accessible from
{\tts \verb*Cwww.rp.csiro.au/~duncan/project.htmlC}.
With a noise level of $\sim$17\,mJy/beam in total intensity, and
5--8\,mJy/beam in Stokes Q and U, it is currently the most sensitive
southern Galactic plane survey. Its sensitivity to
extended ($\gtsim$30$'$) low surface brightness structures is 3--5
times better than a 12-h synthesis with MOST or ATCA. It is
able to detect SNRs of up to 20\deg\ in size. Since the ratio
of the size of the maximal structure detectable to angular resolution
is the same as for the MGPS (\S\ref{plannedsurveys}), they
are complementary surveys.

The IAU (Comm.~9) Working Group on Sky Surveys (formerly ``Wide Field Imaging'')
offers a ``butterfly'' collection of links to sky surveys in radio
and other wavebands at \linebreak[4] {\tts www-gsss.stsci.edu/iauwg/survey\_url.html}.
Some images of outstanding radio sources in the sky are clickable
from an all-sky radio map at {\tts www.ira.bo.cnr.it/radiosky/m.html}
(in Italian).

\subsubsection{Image Galleries of Individual Sources}  \label{galleries}

The NED database has radio images linked to some of their objects,
typically bright radio galaxies from the 3C catalogue.
The ``Astronomy Digital Image Library'' (ADIL; {\tts imagelib.ncsa.uiuc.edu/imagelib})
at the National Center for Supercomputing Applications (NCSA)
offers a search interface by coordinates, object name, waveband, and other
criteria. However, it is not clear from the start how many and what kind
of radio images one may expect.

\begin{figure*}[!h]
%\hspace*{-3mm}
% \epsfig{file=CygAbw.ps,width=13.5cm}
\epsfig{file=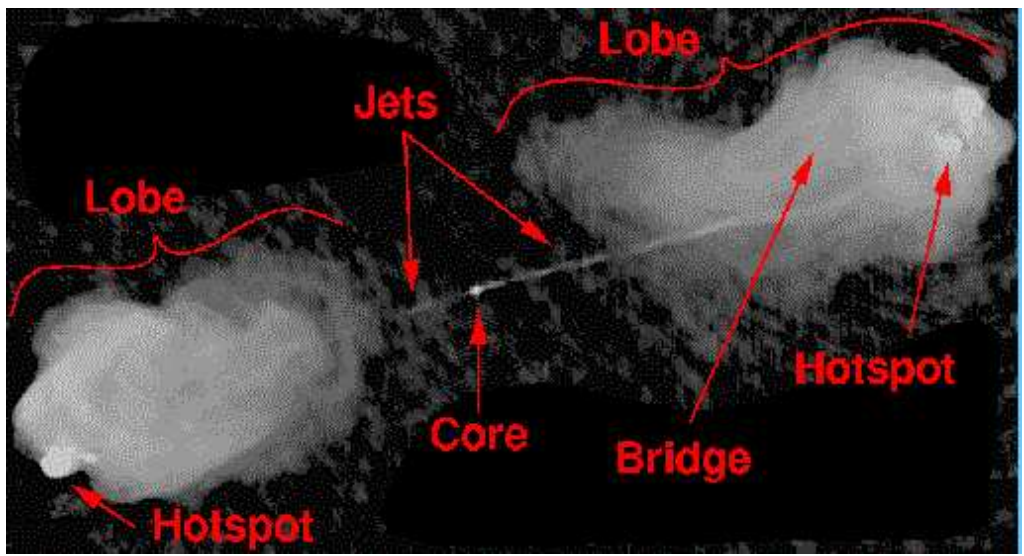,width=13.5cm}   % COLOUR PS FILE, just in case
\caption{A 4.9\,GHz image of Cygnus\,A (3C\,405) at 0.4$''$ resolution, taken
with all four configurations of the VLA, and showing the typical ``ingredients''
of a DRAGN (from Leahy et al.\ (1998), 
% {\tts www.jb.man.ac.uk/atlas/dragns.html} 
courtesy R.\,Perley, C.\,Carilli \& J.P.~Leahy).
The overall size of the source is 2\,arcmin, while the optical galaxy
is $\sim$20$''$ in size on DSS, and coincides with the radio core.
} \label{CygAfig}
\end{figure*}

In an attempt to more adequately describe the phenomenon of classical
double radio sources, \cite{leahy93} coined the term
``Double Radiosource Associated with Galactic Nucleus'' (DRAGN)
for these objects (Fig.~\ref{CygAfig}).
A gallery of images of the 85 nearest DRAGNs is available
in the interactive ``Atlas of DRAGNs'' (\cite{leahy98})
at~ {\tts www.jb.man.ac.uk/atlas/}. Apart from high-resolution
images, the Atlas gives extensive explanations and references on the
physical processes involved. The gallery of icons which has the objects
sorted by their radio luminosity is especially instructive as a demonstration
of the well-established transition from ``Fanaroff-Riley'' class I (FR\,I)
for low-luminosity objects to FR\,II for high-luminosity ones.
The editors of the Atlas are planning to publish their work after
reducing new data of objects for which the published maps are as yet
inadequate. The maps may be downloaded in FITS format, which allows a
number of analyses to be performed on them, at will.

More than a decade after its commissioning, the VLA has been equipped
with low-frequency receivers at 74 and 330\,MHz. At these frequencies,
the field of view is so wide that the sky can no longer be approximated
by a 2-dimensional plane, and 3-dimensional Fourier transforms are
necessary to produce meaningful images. Some examples are given
at {\tts rsd-www.nrl.navy.mil/7214/weiler/4bandfarm.html}.

\subsection{Spectral Lines}        \label{spectrallines}

There is a general scarcity of WWW resources on spectral line data.
Integrated source parameters may be obtained from selected catalogues
in the CDS archive, most conveniently consulted via
{\tts vizier.u-strasbg.fr/cgi-bin/VizieR}.

A list of transition frequencies from 701\,MHz to 3.43\,THz, with
references, has been published by \cite{lovas92}.
A compilation of links to ``Worldwide Molecular Astrophysics Resources'' is
offered at {\tts \verb*Cwww.strw.leidenuniv.nl/~iau34/links.htmlC}.
It provides pointers to several useful mm-wave line databases.
The first of these links (dated June 1997) tells us that 114 molecules
have now been detected in space, containing between two and {\it thirteen} atoms
(HC$_{\rm 11}$N).
One may search for known spectral lines by several parameters
(frequency range, type of molecule, line strength, etc.)
at the URLs {\tts spec.jpl.nasa.gov/ftp/pub/catalog/catform.html}~ and \\
{\tts \verb*Cwww.patnet.caltech.edu/~dmehring/find_lines.1.0.htmlC}.

\subsubsection{Neutral Hydrogen (H\,I)}   \label{hi}

By the 1950s, radio astronomers had already mapped the profiles of the
hyperfine transition of the ground state of the hydrogen atom at
1420\,MHz (21.1\,cm). From this they could infer, using the formulae
by Lindblad \& Oort for differential Galactic rotation, the
spiral structure of the H\,I distribution in the northern
(\cite{1957BAN....13..201W}), and later in both Galactic hemispheres
(e.g.\ \cite{1958MNRAS.118..379O}).

Later these surveys were extended beyond the Galactic plane to the
whole sky, and apart from their importance for the distribution and
kinematics of Galactic H\,I, they are also a necessary tool for X-ray
astronomers to estimate the Galactic absorption (from the H\,I column
density) in X-ray spectra of extragalactic objects, near $\sim$0.1\,keV.
Many of these H\,I surveys are now available in the CDS archive
(e.g.\ catalogues VIII/7, 8, 9, 10, 11, and 47).
Curiously, some of these ``treasures'', e.g.\ the ``Bell Laboratories
H\,I Survey'' by Stark et al.\ (1990; CDS/ADC \#\,VIII/10) were never
really published in an ordinary journal. Others, like VIII/47
(\cite{1982A&AS...49..137W}) are only now being prepared for integration
into the CDS archive, 16 years after publication.
The H\,I survey by \cite{1982MNRAS.201..495S} was tracked down by the
author, motivated by a poster displayed on the current
winter school, and later found to exist in electronic form\,! These authors
used the Parkes 64-m dish to cover a Galactic plane region
($\ell$=245\deg through 12\deg, $|b|<$10\deg), sampled every
0.5\deg\ in $\ell$ and 1\deg\ in $b$ with an
angular resolution of 15$'$.

One of the largest very recent H\,I surveys is indeed available
electronically, but not in a public archive.
Between 1988 through 1993 the sky north of $\delta$=$-$30\deg\ was mapped with the
Dwingeloo 25-m dish (HPBW=36$'$) on a grid of 30$'$ and with a velocity coverage
of 850 km\,s$^{-1}$. The data resulting from this {\it Leiden-Dwingeloo H\,I Survey}
are not presently available on the Internet, but were published as an Atlas
and a CD-ROM (\cite{ldhi97}).
On the CD, there are colour GIF files for all images in the Atlas, as well
as animations running through the data cube in velocity space.

Integrated H\,I fluxes and line widths of galaxies may be obtained
from LEDA, accessible via telnet to {\tt lmc.univ-lyon1.fr}
(login as {\tt leda}), or enter via the WWW at {\tts www-obs.univ-lyon1.fr/leda}.
Currently LEDA offers H\,I fluxes for $\sim$12,400 of a total
of $\sim$170,000 galaxies in LEDA.

H\,I observations of external galaxies (mainly the gas-rich late-type ones)
are important for deriving their rotation curves and detailed kinematics
(e.g.\ \cite{1976ARA&A..14..275B}). These (flat) rotation curves
have led to further evidence for the so-called ``dark matter haloes''
in galaxies. Numerous detailed studies have been published over the
past three decades, very little of which has been preserved in electronic
form, except for compilations of rotation curve parameters (e.g.\
\cite{1983A&AS...53..373B},
\cite{1995ApJS...99..501P} and \cite{1996ApJS..107...97M}).
Also, \cite{1998A&AS..131...73M} gives a compilation of
bibliographic references to H\,I maps of 1439 galaxies published between
1953 and 1995, as well as parameters drawn from these maps.

A gallery of ATCA observations of H\,I in galaxies has been compiled
by B.~Koribalski at {\tts \verb*Cwww.atnf.csiro.au/~bkoribal/atca_obs.htmlC}.

\subsubsection{Molecular and Recombination Lines, and Pulsars}   \label{molecrecomb}

Carbon monoxide (CO) emits one of the most abundant molecular lines in space,
e.g.\ the $^{12}$CO(J=1-0) line at 115\,GHz ($\lambda$=2.6\,mm). These data
allow one to infer the distribution of molecular hydrogen, e.g.\ in the
plane of our Galaxy or throughout other galaxies.  A composite survey of
this line over the entire Milky Way (\cite{1987ApJ...322..706D}) at 8.7$'$ resolution
is available as ADC/CDS catalogue \#\,8039 and at ADIL (\S\ref{galleries}).
It was made with two 1.2-m dishes, one in New York City, the other in Chile,
and provides 720 latitude-velocity maps as FITS files, one for
each 30$'$ of Galactic longitude, and a velocity-integrated map.
The Five College Radio Astronomy Observatory (FCRAO) has mapped 336\,deg$^2$
of the Galactic plane (102.5\deg$<\ell<$141.5\deg, $-$3\deg$<b<+$5.4\deg)
in the $^{12}$CO(1-0) line at 115\,GHz with 45$''$ resolution
(\cite{1998ApJS..115..241H}), available at ADIL (\S\ref{galleries};
{\tts imagelib.ncsa.uiuc.edu/document/97.MH.01}), and at ADC.
A survey of the $^{12}$CO(2-1) line at 230\,GHz was made of the Galactic
plane (20\deg$<\ell<$60\deg\ and $|b|<$1\deg) with 9$'$ resolution, using
the 60-cm sub-mm telescope at Nobeyama (\cite{1995ApJS..100..125S}). It
is available at ADC and CDS (\#\,J/ApJS/100/125/).

The ``Catalog of CO Observations of Galaxies'' (\cite{1985ApJS...57..261V})
(ADC/CDS \#\,7064) is now largely outdated.
The ``FCRAO Extragalactic CO survey'' has measured 1412 positions in 300 galaxies
with the 14-m dish at FCRAO (HPBW=45$''$, \cite{1995ApJS...98..219Y}) in
the $^{12}$CO(1-0) line. It is apparently not available on the WWW, although its
natural home appears to be the URL {\tts \verb*Cdonald.phast.umass.edu/~fcrao/library/C}.
The ``Swedish ESO Submillimetre Telescope'' (SEST) and the 20-m Onsala dish
were used to search for CO in 168 galaxies (\cite{1996A&AS..115..439E}).
The Nobeyama 45-m single dish has surveyed 27 nearby spirals in the
$^{12}$CO(1-0) line (\cite{1997IAUS..184E.132N}).

Other comprehensive molecular line surveys and lists of transition
frequencies have been published (not in electronic form)
in \cite{1985ApJS...58..341S}, \cite{1989ApJS...70..539T},
and \linebreak[4] \cite{1997ApJS..108..301S}.

Radio recombination lines occur through transitions of electrons
between two energy states with very high quantum number $n$.
These lines are named after the atom, the destination quantum
number and the difference in $n$ of the transition
($\alpha$ for $\Delta\,n$=1, $\beta$ for $\Delta\,n$=2,
$\gamma$ for $\Delta\,n$=3, etc.).
Examples are H\,157\,$\alpha$ (transition from $n$=158 to $n$=157 in hydrogen)
or He\,109\,$\beta$ (transition from $n$=111 to $n$=109 in helium).
They are mainly used to map departures from thermal equilibrium,
and the velocity structure in H\,II regions or in planetary nebulae.
To my knowledge there is no WWW site offering data on
recombination lines, but see
{\tts www.hartrao.ac.za/spectra/SP\_Hii.html} for an introduction
to the research done with this type of observation.
See \cite{1968ApJS...16..143L} and \cite{1996ApJS..107..747T} for tables
of radio recombination lines.

A successful search for pulsars over the entire southern sky
with the Parkes 64-m dish, the so-called 70-cm Pulsar Survey or
``Parkes Southern Pulsar Survey'' detected 298 pulsars, 101 of
them previously unknown (\cite{1998MNRAS.295..743L}).

\section{Finding Literature, Addresses, and Proposal Forms on WWW} \label{litpropos}

\subsection{Relevant Literature on the Internet}  \label{literature}

As in other branches of astronomy, much of the most recent publications
can be found on the WWW. The LANL/SISSA
electronic preprint server ({\tts xxx.lanl.gov}) has been described
in some detail in my tutorial.
Radio astronomy topics are becoming increasingly popular on this server,
although some of the most productive radio astronomy institutions
have not yet discovered its efficiency and cost savings for the
distribution of their preprints. Some institutions offer at least
the titles of their preprints (if not full versions) on the WWW.
Those preprints still circulated only on paper may be found in
the STEP- and RAPsheets of the STScI and NRAO
({\tts sesame.stsci.edu/lib/stsci-preprint-db.html} and
{\tts libwww.aoc.nrao.edu/aoclib/rapsheet.html}).
Among the comprehensive collection of astronomy newsletters at
{\tts sesame.stsci.edu/lib/NEWSLETTER.htm}, there are several
of interest to radio astronomers, depending on their area of research.

A few relevant proceedings volumes are also accessible on the WWW, e.g.\
the one on ``Energy Transport in Radio Galaxies and Quasars''
(\cite{hardee96}) discusses a wide variety of phenomena encountered in
extragalactic radio sources, and papers from this volume are available
as PostScript files from {\tts www.cv.nrao.edu/jetworks}.
Three other volumes (\cite{cohenkell95,1995ASPC...82,1998ASPC..144})
bring together recent advances in high-resolution radio imaging of
compact radio sources (see {\tts www.pnas.org/},
{\tts www.cv.nrao.edu/vlbabook}, and
{\tts www.cv.nrao.edu/iau164/164book.html}).

\subsection{Finding Radio Astronomers around the World}  \label{colleagues}

Directories of astronomers in general have been described in section 7
of my tutorial for this winter school. Commission 40 (``Radio Astronomy'')
of the IAU offers a list of its 860 members
({\tts sma-www.harvard.edu/IAU\_Com40/IAU\_scroll.html}),
and 660 of them appear with their email address.

\subsection{Proposing Observations with Radio Telescopes} \label{proposals}

Like in most other parts of the electromagnetic spectrum,
proposals are accepted via email at most radio observatories.
Many of them offer their proposal forms on the WWW, like
e.g.\ for the various NRAO telescopes
at {\tts www.nrao.edu/proposals.html}, for the ATNF telescopes
at {\tts www.atnf.csiro.au/observers/apply/form.html}
(including Parkes and VLBI),
for MERLIN at {\tts www.jb.man.ac.uk/merlin/propsub/},
for the Arecibo dish at {\tts www.naic.edu/vscience/proposal/proposal.htm},
for the BIMA mm array \linebreak[4] at
{\tts \verb*Cwww.astro.uiuc.edu/~bima/call_for_proposals.htmlC},
or for the JCMT at \linebreak[4] {\tts www.jach.hawaii.edu/JCMT/pages/apply.html}.
VLBI proposal forms for the EVN are available at
{\tts www.nfra.nl/jive/evn/proposals/prop.html}, and for the VLBA  see
the NRAO address above.
No web forms were found e.g.\ for the MRAO or IRAM telescopes,
nor for the WSRT.
Some institutions still require the proposals to be sent via regular
mail, e.g.\ the MPIfR Bonn
({\tts www.mpifr-bonn.mpg.de/effelsberg/runkel/info\_pke.html})
for proposing time at the Effelsberg 100-m dish.
It would be difficult here to give a comprehensive list of URLs 
for the many radio
telescopes distributed over the globe.

\section{The near and far Future of Radio Surveys and Telescopes}     \label{future}

In this section I shall present some survey projects currently being carried out
or planned, as well as some telescopes under construction or being
designed.
A good overview of current and planned radio astronomy facilities
and recent research progress up to mid-1996 has been given in the latest
Triennial Report of IAU Commision 40 (``Radio Astronomy''), at the URL
{\tts sma2.harvard.edu/IAU\_Com40/c40rpt/c40report.html}.
The next 3-year report is to become available in late 1999 at
{\tts www.iau.org/div10.html}.
Many such projects have also been described in the proceedings volume
by \cite{jacksondavis97}.

\subsection{Continuing or Planned Large-scale Surveys}  \label{plannedsurveys}

On the island of Mauritius a 151\,MHz survey is being performed with
the ``Mauritius Radiotelescope'' (MRT; \cite{1995Ap&SS.228..373G}), and may
be regarded as the southern continuation of the MRAO 6C survey (cf.\ Table~1).
This T-shaped array of helical antennas provides an angular resolution
of 4$'\times$4.6$'$\,csc(z), where z is the zenith distance.
The aim is to map the sky between declinations $-$10\deg\ and $-$70\deg\ to
a flux limit of $\ltsim$\,200\,mJy, including a map of the Galactic Plane and
studies of pulsars.
A catalogue of $\sim$10$^5$ sources can be expected after
completion of the survey in summer 1998 ({\tts icarus.uom.ac.mu/mrt2.html}).

After the completion of WENSS, the WSRT started in late 1997 the
``WISH'' survey at 350\,MHz ({\tts www.nfra.nl/nfra/projects/index.htm}).
The aim is to survey the region of effective overlap with
ESO's ``Very Large Telescope'' (VLT; {\tts www.eso.org/projects/vlt/}),
which is limited by the WSRT horizon and by the elongation of the
synthesised beam.  In order to have a minimum hour angle coverage of 4\,h,
the area $-$30\deg$\le\delta\le-$10\deg, $|b|>$10\deg, or 5900\,deg$^2$,
will be covered. With an expected noise limit of about 3\,mJy (1-$\sigma$),
and a source density of 20 per deg$^2$, WISH should detect about 120,000 sources.

The DRAO Penticton aperture synthesis array is being used to survey
the northern Galactic plane at 408 and 1420\,MHz in the continuum, and
at 1420\,MHz in the H\,I line ({\tts www.drao.nrc.ca/web/survey.shtml}).
The area covered is 72\deg$<\ell<$\,140\deg, $-$3\deg$<b<+$5\deg, and the
angular resolutions are 1$'$ and 4$'$.
First results of this survey can be viewed
at {\tts www.ras.ucalgary.ca/pilot\_project.html}.

At MPIfR Bonn a 1.4\,GHz Galactic plane survey (4\deg$<|b|<$20\deg),
using the Effelsberg 100-m dish in both total intensity and polarisation,
is under way (\cite{uyaniker98}).
Examples of how this survey will be combined with the NVSS, and with
polarisation data from \cite{1976A&AS...26..129B}, have been
shown by \cite{1998IAUS..179...97F}.

The first-epoch ``Molonglo Galactic Plane Survey''
(MGPS-1; \cite{1998ApJ...5xx..yyyG}) at 843\,MHz,
was obtained with the old, 70$'$ field-of-view MOST and covers the region
245\deg$<\ell<$355\deg, $|b|<$1.5\deg.  The second-epoch
Galactic plane survey (MGPS-2) is being made with the new,
wide-field (2.7\deg) system at 843\,MHz, and will cover the region
240\deg$\le\ell\le$365\deg, $|b|\le$10\deg. With an
angular resolution of 43$''\times$43$''$\,csc\,$\delta$ and a noise
level of 1--2\,mJy/beam it is expected to yield over 80,000 sources
above $\sim$5\,mJy (\cite{1997PASA...14...73G}).
As a part of SUMSS (\S\ref{modsurv}), it is well under way,
and its survey images can be viewed at
{\tts www.physics.usyd.edu.au/astrop/MGPS/}.
Catalogues of sources will be prepared at a later stage.

The Hartebeesthoek Radio Astronomy Group (HartRAO) in South
Africa, after having finished the 2.3\,GHz southern sky survey
(\S\ref{surveyimages}), is planning to use its 26-m dish for an
8.4\,GHz survey of the southern Galactic plane in total intensity
and linear polarisation (\cite{1998IAUS..179...95J})
at $\sim$6$'$ resolution, and for deeper 2.3\,GHz maps of interesting
regions in the afore-mentioned 2.3\,GHz survey.

\subsection{Very Recent Medium-Deep Multi-Waveband Source Surveys}

Between 1995 and 1997, the ``AT-ESP'' continuum survey was carried
out at 1.4\,GHz with ATCA (\cite{prandoni98}).
This survey covers $\sim$\,27 deg$^2$ near the South Galactic Pole
with a uniform sensitivity of $\sim$\,70\,$\mu$Jy (1\,$\sigma$).
About 3\,000 radio sources have been detected, one third of them
being sub-mJy sources. Redshifts from the ``ESO Slice Project''
(ESP) redshift survey for 3342 galaxies down to b$_{\rm J}\sim$\,19.4
(\cite{1998A&AS..130..323V}, \linebreak[4] 
{\tts \verb*Cboas5.bo.astro.it/~cappi/esokp.htmlC})
will allow studies of the population of low-power radio galaxies
and of their 3-dimensional distribution.

The VLA has been used in C-configuration to carry out a sensitive 1.4\,GHz
survey of 4.22\,deg$^2$ of the northern sky that have been surveyed also
in the Far Infra-Red with the ISO satellite, as part of
the ``European Large Area ISO Survey'' (ELAIS; \cite{1998MNRAS.VVV..PPPC},
{\tts \verb*Cwww.ast.cam.ac.uk/~ciliegi/elais/C}).
The 5$\sigma$ flux limit of the survey ranges from
0.14\,mJy (for 0.12\,deg$^2$) to 1.15\,mJy (for the entire 4.22\,deg$^2$).
A careful comparison of the catalogue of 867 detected radio sources
with the FIRST and NVSS catalogues provided insights into the
reliability and resolution-dependent surface brightness effects
that affect interferometric radio surveys. Cross-identification with
IR and optical objects is in progress.

The ``Phoenix Deep Survey'' (\cite{1998MNRAS.296..839H}) has used the ATCA
to map a 2\deg\ diameter region centred on ($\alpha$,$\delta$)=
J\,01$^h$\,14$^m$\,12.2$^s$,$-$45\deg\,44$'$\,08$''$.
A total of 1079 sources were detected above $\sim$\,0.2\,mJy
({\tts \verb*Cwww.physics.usyd.edu.au/~ahopkins/catsC}).
Optical identifications were proposed for half of the sources,
and redshifts were measured for 135 of these.
A comparison with lower resolution 843\,MHz MOST maps is in progress.

\subsection{Extending the Frequency Range of the Radio Window}   \label{extendfreq}

One of the very pioneers of radio astronomy, G.~Reber, has been exploiting
methods to observe cosmic radio emission at $\sim$2\,MHz from the ground, even
very recently. He quite successfully did so from two places in the world
where the ionosphere appears to be exceptionally transparent (see
\cite{1994JRASC..88..297R} and \cite{1995Ap&SS.227...93R}).

The lowest frequency observations regularly being made from the ground
are done with the ``Bruny Island Radio Spectrometer'' % \cite{erick97}
(\cite{1997PASA...14..278E}) on Bruny Island, south of Hobart (Tasmania).
It is used for the study of solar bursts in the rarely observed frequency
range from 3 to 20\,MHz. Successful observations are made down to the
minimum frequency that can propagate through the ionosphere.
This frequency depends upon the zenith distance of the Sun and
is usually between 4 and 8\,MHz.

However, for many years radio astronomers have dreamt of extending the
observing window to frequencies significantly below a few tens of MHz
(where observations can be made more easily from the ground) to
a few tens of kHz (just above the local plasma frequency of the
interplanetary medium). Ionospheric absorption and refraction
requires this to be done from space.
The first radio astronomy at kHz frequencies, and the first radio
astronomy from Space, was the ``Radio Astronomy Explorer'' (RAE; \cite{kaiser87}),
in the late 1960s and early 1970s. It consisted of a V-shaped antenna
450\,m in extent, making it the largest man-made structure in space.
It was equipped with radiometers for 25\,kHz to 13.1\,MHz. Although
no discrete Galactic or extragalactic sources were detected, very crude
all-sky maps were made, and solar system phenomena studied.
Since then none of the various space projects proposed have been realised.
Recent plans for developing low-frequency radio astronomy, both from the
ground and from space, can be viewed at {\tts rsd-www.nrl.navy.mil/7214/weiler}
by following the links to Low Frequency Radio Astronomy (LFRA) and
associated pages, but see also the ALFA project (\S\ref{spaceprojects}).
The proceedings volume by \cite{kassimweiler90} is full of ideas
on technical schemes for very low-frequency radio observatories,
and on possible astrophysical insights from them.

Efforts to extend the radio window to very high frequencies have
been much more serious and successful in the past two decades, and
have led to a whole new branch of ``mm-wave astronomy''.
The multi-feed technique (\S\ref{dishversinterf}) has seen a trend
moving away from just having a single receiver in the focal plane,
towards having multiple receivers there, to help speed up the data collection
(as e.g.\ in the Parkes H\,I multibeam survey, \S\ref{hifuture}).
By building big correlators, and taking
the cross-products between the different beams, the complex field
distribution in the focal plane of a dish may be mapped, and by
transforming that one can correct for pointing, dish deformation, etc.
Arrays of, say, 32 by 32 feeds are able to ``image'' the sky in real time
(see e.g.\ the SEQUOIA system at the FCRAO 14-m dish,
{\tts \verb*Cdonald.phast.umass.edu/~fcrao/instrumentation/C}).
This is only possible at mm wavelengths, where the equipment is small enough
to fit into the focal plane. In perhaps three years such receivers should
exist at $\sim$100\,GHz (3\,mm).

As mentioned in \S\ref{surveycontent} there is a lack of large-area
surveys at frequencies above $\sim$\,5\,GHz.
As \cite{1998IAUS..179...19C} has pointed out, such surveys are made difficult
since the beam solid angle of a telescope scales as~ $\nu^{-2}$ and
system noise generally increases with frequency, so the time needed
to survey a given area
of sky rises very rapidly above 5\,GHz. However, a 7-beam 15\,GHz continuum
receiver being built for the GBT (\S\ref{plannedtelescopes}) could cover
a 1-degree wide strip
along the Galactic plane in one day, with an rms noise of $\sim$2\,mJy.
Repeating it several times would provide the first sensitive and
systematic survey of variable and transient Galactic sources, such as
radio stars, radio-emitting $\gamma$-ray sources, X-ray sources, etc.

More promising for the investigation of possible new source populations
at these high frequencies (\S\ref{surveycontent}) may be the results
from the new CMB satellites.  One is the ``Microwave Astronomy Probe''
(MAP; {\tts map.gsfc.nasa.gov/html/web\_site.html}), expected to be
launched by NASA in 2000. It will operate between
22 and 90\,GHz with a 1.4$\times$1.6\,m diameter primary reflector,
offering angular resolutions between 18$'$ and 54$'$.
The other one is PLANCK ({\tts astro.estec.esa.es/SA-general/Projects/Planck/}),
to be launched by ESA in 2006 (possibly on the same bus as the
``Far InfraRed and Submillimetre Telescope'', FIRST, not to be
confused with the VLA FIRST radio survey).
PLANCK will have a telescope of 1.5\,m aperture, and it will be used with
radiometers for low frequencies \linebreak[4] (30--100\,GHz; 
$\lambda$=3--10\,mm), and
with bolometers for high frequencies (100--857\,GHz; $\lambda$ 0.3--3.0\,mm),
with angular resolutions of $\sim$10$'$ at 100\,GHz.
Both the MAP and PLANCK missions should detect a fair number of
extragalactic sources at 100\,GHz ($\lambda$=3\,mm). In fact,
\cite{1998ApJ...500L..83T} expect PLANCK to detect 40,000 discrete sources
at 857\,GHz.
A highly important by-product will be the compilation of a much denser
grid of calibration sources at mm wavelengths.
The vast majority of the currently known mm-wave calibrators
are variable anyway.

\subsection{Spectral Line and Pulsar Surveys}  \label{hifuture}

The Australia Telescope National Facility (ATNF) has constructed and
commissioned a 21-cm multi-feed system with 13 receivers at the prime
focus of the Parkes 64-m telescope
(\cite{staveley97}; {\tts wwwpks.atnf.csiro.au/people/multi/multi.html}).
The feeds are disposed to form beams with an angular resolution
of 14$'$ and a distance of $\sim$28$'$ between neighbouring feeds.
The on-line correlator measures flux density in all 13 channels
and 2 polarisations simultaneously, with a spectral
resolution of 16\,km\,s$^{-1}$ and a velocity range from
$-$1200\,km\,s$^{-1}$ to $+$12,700\,km\,s$^{-1}$.
The Parkes multi-beam facility commenced regular observing in
1997, and a report on its status is regularly updated at
{\tts www.atnf.csiro.au/research/multibeam/multibeam.html}. Several
major H\,I surveys are planned, including an ``all-sky'' survey
($\delta\,\ltsim\,+$20\deg) with a limiting sensitivity (5$\sigma$, 600\,s) of
$\sim$20\,mJy per channel. The Zone of Avoidance (ZOA, $|b|<$5\deg)
will be covered with the same velocity range and twice the sensitivity.
It will be sensitive to objects with H\,I mass between
10$^6$ and 10$^{10}$~M$_{\odot}$, depending on distance. This will be
the first extensive ``blind" survey of the 21-cm extragalactic sky.
When scheduled, it is possible to watch the signal of all 13 beams
almost {\it in real time} at
{\tts wwwpks.atnf.csiro.au/people/multi/public\_html/live/multibeam\_live.html}.
An extension of this survey to the northern hemisphere
($\delta\,\gtsim\,+$20\deg) will be performed with the
Jodrell-Bank 76-m Mark I antenna, but only 4 receivers will be used.

This Parkes multibeam system is also being used for a sensitive wide-band continuum search
for pulsars at 1.4\,GHz, initially limited to the zone 220\deg$<\ell<$20\deg\ within
5\deg\ from the Galactic plane. A first observing run in August 1997 suggested
that 400 new pulsars may be found in this survey, which is expected to 
take $\sim$100 days of telescope time at Parkes, spread over two years 
(ATNF Newsletter 34, p.~8, 1998).

The ``Westerbork observations of neutral Hydrogen in Irregular and SPiral
galaxies'' (WHISP; {\tts \verb*Cwww.astro.rug.nl/~whispC})
is a survey to obtain WSRT maps of the distribution and velocity
structure of H\,I in 500 to 1000 galaxies, increasing the number of
galaxies with well-studied H\,I observations by an order of magnitude.
By May 1998 about 280 galaxies had been observed, and the data had been
reduced for 160 of them.
H\,I profiles, velocity maps, and optical finding charts
are now available for 150 galaxies. Eventually the data cubes and
(global) parameters of all galaxies will also be made available.

The Dwingeloo 25-m dish is currently pursuing the ``Dwingeloo Obscured Galaxy
Survey'' (DOGS; \cite{1998AJ....115..584H}; {\tts www.nfra.nl/nfra/projects/dogs.htm})
of the area 30\deg$\le\ell\le$220\deg; $|b|\le$\,5.25\deg. This had led to the
discovery of the nearby galaxy Dwingeloo~1 in August 1994 (\cite{1994Natur.372...77K}).
After a shallow survey in the velocity range 0--4000\,km\,s$^{-1}$ with
a noise level of 175\,mJy per channel, a second, deeper survey is
being performed to a noise level of 40\,mJy. The latter has so far discovered
40 galaxies in an area of 790\,deg$^2$ surveyed to date.

The first results of a dual-beam H\,I survey with the Arecibo 305-m dish
have been reported in \cite{1998AAS...192.6601R}. In a 400 deg$^2$ area
of sky 450 galaxies were detected, several of them barely visible on the
Palomar Sky Survey.

Since 1990, the Nagoya University has been executing a $^{13}$CO(1-0) survey at
110\,GHz of the Galactic plane, with a 4-m mm-wave telescope. Since 1996,
this telescope
is operating at La Silla (Chile) to complete the southern Galactic plane
(\cite{1998IAUS..179..165F}).
The BIMA mm-array is currently being used to survey 44 nearby spiral
galaxies in the $^{12}$CO(1-0) line at 6$''$--9$''$ resolution
(\cite{1998AAS...192.7301H}).
% (\cite{1998BAAS...30..928H}, \#\,73.01).

\subsection{CMB and Sunyaev-Zeldovich Effect}    \label{cmb}

The cosmic microwave background (CMB) is a blackbody radiation
of 2.73\,K and has its maximum near $\sim$150\,GHz (2\,mm).
Measurements of its angular distribution on the sky are highly important 
to constrain
cosmological models and structure formation in the early Universe, thus
the mapping of anisotropies of the CMB has become one of the most
important tools in cosmology.
For a summary of current CMB anisotropy experiments, see
\cite{wilkinson98} and \cite{1997PhT....50...32B}; the latter even lists the
relevant URLs (cf.\ also {\tts brown.nord.nw.ru}).
As an example, the Cambridge
``Cosmic Anisotropy Telescope'' (CAT; \cite{1993A&A...277..314R},
{\tts www.mrao.cam.ac.uk/telescopes/cat/}) has started to map such
anisotropies, and it is the prototype for the future, more sensitive
``Very Small Array'' (VSA; \S\ref{mmarrays}).

The ``Sunyaev-Zeldovich'' (SZ) effect is the change of brightness temperature $T_B$
of the CMB towards regions of ``hot'' (T\,$\sim$10$^7$\,K) thermal plasma,
typically in the cores of rich, X-ray emitting clusters of galaxies. The effect is
due to the scattering of microwave photons by fast electrons, and results in
a diminution of $T_B$ below $\sim$\,200\,GHz, and in an excess of $T_B$ above
that frequency. See \cite{birkinshaw98} for a comprehensive review of past
observations and the potential of these for cosmology.
See \cite{liang98} for the status and future plans for observing the
Sunyaev-Zeldovich effect.

\subsection{Radio Telescopes: Planned, under Construction or being Upgraded} \label{plannedtelescopes}

\subsubsection{Low and Intermediate Frequencies}

The Arecibo observatory has emerged in early 1998 from a 2-year upgrading
phase ({\tts www.naic.edu/techinfo/teltech/upgrade/upgrade.htm}).
Thanks to a new Gregorian reflector, the telescope has a
significantly increased sensitivity.

The National Centre for Radio Astrophysics (NCRA) of the Tata Institute for
Fundamental Research (TIFR, India) is nearing the completion of the
``Giant Metrewave Radio Telescope'' (GMRT) at a site about 80\,km north
of Pune, India ({\tts www.ncra.tifr.res.in}).
With 30 fully steerable dishes of 45\,m diameter, spread over distances
of up to 25\,km, it is the world's most powerful radio telescope operating
in the frequency range 50--1500\,MHz with angular resolutions
between 50$''$ and 1.6$''$.
In June 1998, all 30 dishes were controllable from the central electronics
building. Installation of the remaining feeds and front ends is
expected in summer 1998. The digital 30-antenna correlator, combining
signals from all the antennas to produce the complex visibilities over
435 baselines and 256 frequency channels, is being assembled, and
it will be installed at the GMRT site also in summer 1998.
The entire GMRT array should be producing astronomical images
before the end of 1998.

The NRAO ``Green Bank Telescope'' (GBT; {\tts www.gb.nrao.edu/GBT/GBT.html})
is to replace the former 300-ft telescope which collapsed in 1988 from
metal fatigue. The GBT is a 100-m diameter single dish with an unblocked
aperture, to work at frequencies from 300\,MHz to $\sim$\,100\,GHz, with
almost continuous frequency coverage.
It is finishing its construction phase, and is expected to be operational
in 2000 (\cite{vandenbout98}).

The VLA has been operating for 20 years now, and a plan for an upgrade
has been discussed for several years. Some, not very recent, information
may be found at the URL \linebreak[4]
{\tts www.nrao.edu/vla/html/Upgrade/Upgrade\_home.shtml}. Among other things,
larger subreflectors, more antennas, an extension of the A-array,
a super-compact E-array for mosaics of large fields, and continuous
frequency coverage between 1 and 50\,GHz are considered.

An overview of current VLBI technology and outlooks for the future of VLBI
have been given in the proceedings volume by \cite{sasao94}.

For several years the need for and the design of a radio telescope with
a collecting area of one square kilometre have been discussed. The project
is known under different names\,: the ``Square Kilometre Array
Interferometer'' (SKAI; {\tts www.nfra.nl/skai/}; \cite{brown96});
the ``Square Kilometre Array'' (SKA; {\tts www.drao.nrc.ca/web/ska/ska.html}),
and the ``1-km teleskope'' ({\tt 1kT}; {\tts www.atnf.csiro.au/1kT}).
A Chinese version under the name ``Kilometer-square Area Radio Synthesis
Telescope'' (KARST; {\tts www.bao.ac.cn/bao/LT}) was presented by
\cite{1998IAUS..179...93P}, and contemplates the usage of spherical
(Arecibo-type) natural depressions, frequently found in southwest China,
by the placing of $\sim$\,30 passive spherical reflectors, of $\sim$\,300\,m 
diameter, in each of them.
A frequency coverage of 0.2--2\,GHz is aimed at for such an array of
reflectors.

A new design for a large radio telescope, based on several almost flat
primary reflectors, has been recently proposed (\cite{1998A&AS..130..369L}).
The reflectors are adjustable in shape, and are of very long focal length.
The receiver is carried by a powered, helium-filled balloon. Positional errors
of the balloon are corrected either by moving the receiver feed point
electronically, or by adjusting the primary reflector so as to move its
focal point to follow the balloon. The telescope has the wide sky coverage
needed for synthesis observations and an estimated optimum diameter of
100--300\,m. It would operate from decimetre to cm-wavelengths, or, with
smaller panels, mm-wavelengths.

\subsubsection{Where the Action is: Millimetre Telescopes and Arrays} \label{mmarrays}

The ``Smithsonian Submillimeter Wavelength Array'' (SMA; {\tts sma2.harvard.edu/})
on Mauna Kea (Hawaii) consists of eight telescopes of 6\,m aperture, six of
these provided by the Smithsonian Astrophysical Observatory (SAO) and two
by the Astronomica Sinica Institute of Astronomy and Astrophysics (ASIAA, Taiwan).
Eight receivers will cover all bands from 180 to 900\,GHz ($\lambda$=1.7--0.33\,mm).
To achieve an optimised coverage of the uv plane, the antennas will be placed
along the sides of Reuleaux triangles, nested in such a way that they share 
one side, and allow both compact and wide configurations.
Baselines will range from 9 to 460\,m, with angular resolutions as fine as 0.1$''$.
The correlator-spectrometer with
92,160 channels will provide 0.8\,MHz resolution for a bandwidth of 2\,GHz in each of
two bands. One of the SMA telescopes has had ``first light'' in spring 1998,
and the full SMA is expected to be ready for observations in late 1999.

Since April 1998, the ATNF is being upgraded to become the first southern
hemisphere mm-wave synthesis telescope (cf.\ ATNF Newsletter 35, Apr~1998).
The project envisages the ATNF to be equipped with receivers for 12 and 3\,mm 
(\cite{norris98}).

The ``Millimeter Array'' (MMA; {\tts www.mma.nrao.edu/}) is a project by
NRAO to build an array of 40 dishes of 8--10\,m diameter to operate as
an aperture synthesis array at frequencies between 30 and 850\,GHz
($\lambda=$0.35--10\,mm). Array configurations will range from about
80\,m to 10\,km. It will most probably be placed in the Atacama desert
in northern Chile at an altitude near 5000\,m, a site rivalling
the South Pole in its atmospheric transparency (\cite{vandenbout98}).

The ``Large Southern Array'' project (LSA) is coordinated by ESO,
IRAM, NFRA and Onsala Space Observatory (OSO), and it
anticipates the building of a large millimetre array with a collecting
area of up to 10,000\,m$^2$, or roughly 10 times the collecting area of
today's largest millimetre array in the world, the IRAM interferometer at
the Plateau de Bure with five 15-m diameter telescopes. With baselines foreseen
to extend to 10\,km, the angular resolution provided by the new instrument
will be that of a diffraction-limited 4-m optical telescope.
Current plans are to provide the collecting area equivalent to 50--100 dishes
of between 11 and 16\,m diameter, located on a plain
above 3000\,m altitude. Currently only site testing data are available
on the WWW ({\tts puppis.ls.eso.org/lsa/lsahome.html}).

A similar project in Japan, the ``Large Millimeter and Submillimeter Array''
(LMSA; \linebreak[4] {\tts www.nro.nao.ac.jp/LMSA/lmsa.html})
anticipates the building of a mm array of 50 antennas of 10\,m diameter each,
with a collecting area of 3,900\,m$^2$, to operate at frequencies
between 80 and 800\,GHz.

The MMA, LSA and LMSA projects will be so ambitious that negotiations to join
the LMSA and MMA projects, and perhaps all three of them, are under way.
The name ``Atacama Array'' has been coined for such a virtual
instrument (see NRAO Newsletter \#\,73, p.\,1, Oct.~1997).
The MMA will also pose challenging problems for data archiving, and in fact
will rely on a new data storage medium to enable archiving to be feasible
(cf.\ {\tts www.mma.nrao.edu/memos/html-memos/abstracts/abs164.html}).

A comparison of current and future mm arrays is given in Table~\ref{mmtable}.

\begin{table*}[!ht]
\begin{center}
\caption{Comparison of Current and Future mm Arrays~$^a$} \label{mmtable}
% \begin{flushleft}
% \renewcommand{\arraystretch}{1.2}
\begin{tabular}{lllll} \hline
%\noalign{\smallskip}
Array & Completion  & Wavelength & Sensitivity~$^b$ & max baseline\\
& Date & Range (mm) & at 3mm (Jy)  & (km) \\  \hline
% \noalign{\smallskip}
% \hline
% \noalign{\smallskip}
% {\tts bima.astro.umd.edu/bima/}
% {\tts www.ovro.caltech.edu/mm/main.html}
Nobeyama (6\,$\times$\,10\,m)          & $\sim$1986   & 3.0, 2.0      & 1.7   & 0.36 \\
IRAM (5\,$\times$\,15\,m)            & $\sim$1988    & 3.0, 1.5      & 0.3--0.8       & 0.4 \\
OVRO (6\,$\times$\,10.4\,m)             & $\sim$1990 ?         & 3.0, 1.3      & 0.5   & 0.3 \\
BIMA (9\,$\times$\,6\,m)             & 1996  & 3.0 (1.3)     & 0.7   & 1.4 \\
% www.nro.nao.ac.jp/NMA/nma-e.html
SMA CfA (8\,$\times$\,6\,m)           & 1999\,? & 1.7--0.33      & -     & 0.46 \\
% {\tts sma2.harvard.edu/}
ATCA (5\,$\times$\,22\,m)            & 2002\,? & 12.0, 3.0   & 0.5\,?  & 3.0 (6.0\,?) \\
MMA USA (40\,$\times$\,10\,m?)        & 2010\,? & 10.0--0.35     & 0.04\,? & 10.0 \\
LSA Europe (50\,?\,$\times$\,16\,m?)  & 2010\,? & 3.0, 1.3,... & 0.02\,? & 10.0 \\
% (7,2,0.8,0.65,0.45,0.35?)       & 0.02\,? & 10.0 \\
LMSA Japan (50\,$\times$\,10\,m)  & 2010\,? & 3.5--0.35    & 0.03\,? & 10.0 \\  \hline
% \noalign{\smallskip}
% \hline
\end{tabular}
\end{center}
\small
a) adapted from \cite{norris98}, but see also \cite{stark98} for mm-wave single dishes \\
b) rms continuum sensitivity at 100\,GHz to a point source observed for 8\,hours
% \end{flushleft}
\end{table*}

The ``Large Millimeter Telescope'' (LMT)
is a 50-m antenna to be built on the slopes of the highest mountain
in Mexico in the Sierra Negra, $\sim$200\,km east of Mexico City,
at an elevation of 4500\,m.  It will operate at wavelengths between 8.5 and 35\,GHz
($\lambda$=0.85--3.4\,mm) achieving angular resolutions between 5$''$ and 20$''$ (see
{\tts lmtsun.phast.umass.edu/}).

There are plans for a 10-m sub-mm telescope at the South Pole
(\cite{stark98}; {\tts \verb*Ccfa-www.harvard.edu/~aas/tenmeter/tenmeter.htmlC}).
The South Pole has been identified as the best site for sub-mm wave astronomy
from the ground. The 10-m telescope will be suitable for ``large-scale'' (1\,deg$^2$)
mapping of line and continuum from sub-mm sources at mJy flux levels, at
spatial resolutions from 4$''$ to 60$''$, and it will make 
arcminute scale CMB measurements.

The ``Very Small Array'' (VSA; {\tts \verb*Cwww.jb.man.ac.uk/~sjm/cmb_vsa.htmC};
{\tts astro-ph/9804175}), currently in the design phase, consists
of a number of receivers with steerable horn antennas, forming an aperture
synthesis array to work at frequencies around 30\,GHz ($\lambda$=10\,mm).
It will be placed at the Teide Observatory on Tenerife (Spain)
around the year 2000. The VSA will provide images of structures
in the CMB, on angular scales from 10$'$ to 2\deg. Such
structures may be primordial, or due to the SZ effect (\S\ref{cmb})
of clusters of galaxies beyond the limit of current optical sky surveys.

The ``Degree Angular Scale Interferometer'' (DASI; {\tts astro.uchicago.edu/dasi})
is designed to measure anisotropies in the CMB, and consists of 13 closely
packed 20-cm diameter corrugated horns, using cooled High Electron Mobility 
Transistor (HEMT) amplifiers running
between 26 and 36\,GHz. It will operate at the South Pole by late 1999.
A sister instrument, the ``Cosmic microwave Background Interferometer''
(CBI; {\tts \verb*Dastro.caltech.edu/~tjp/CBI/D}) will be located  at high
altitude in northern Chile, and it will probe the CMB on smaller angular scales.

\subsection{Space Projects}     \label{spaceprojects}

Radioastron (\cite{kardashev97}; {\tts www.asc.rssi.ru/radioastron/})
is an international space VLBI project led by the ``Astro Space Center'' of
the Lebedev Physical Institute in Moscow, Russia.  Its key element is an
orbital radio telescope that consists of a deployable 10-m reflector
made of carbon fiber petals. It will have an overall rms surface accuracy of
0.5\,mm, and operate at frequencies of 0.33, 1.66, 4.83 and 22.2\,GHz.
It is planned to be launched in 2000--2002 on a Proton rocket, into
a highly elliptical Earth orbit with an apogee of over 80\,000\,km.

The ``Swedish-French-Canadian-Finnish Sub-mm Satellite'' (ODIN)
will carry a 1.1-m antenna to work in some of the unexplored bands of the
electromagnetic spectrum, e.g.\ around 118, 490 and 560\,GHz.
The main objective is to perform detailed studies of the physics and the
chemistry of the interstellar medium by observing emission from key
species. Among the objects to be studied are comets, planets,
giant molecular clouds and nearby dark clouds, protostars,
circumstellar envelopes, and star forming regions in nearby galaxies
(see {\tts kurp-www.hut.fi/spectroscopy/space-projects.shtml}).

A new space VLBI project, the ``Advanced Radio Interferometry between Space
and Earth'' (ARISE) has recently been proposed by \cite{1998ASPC..144..397U}.
It consists in a 25-m antenna in an elliptical Earth orbit between altitudes
of 5,000 and $\sim$40,000\,km, operating at frequencies from 5 to 90\,GHz.
The estimated launch date is the year 2008.

A space mission called ``Astronomical Low-Frequency Array'' (ALFA) has been
proposed to map the entire sky between 30\,kHz and 10\,MHz.
The project is in the development phase ({\tts sgra.jpl.nasa.gov/html\_dj/ALFA.html})
and no funding exists as yet.

The far side of the Moon has been envisaged for a long time
as an ideal site for radio astronomy, due to the absence of
man-made interference.
Speculations on various kinds of radio observatories on the Moon
can be found in the proceedings volumes by \cite{burns88},
\cite{burns89}, and \cite{kassimweiler90}. Since then, the subject
has been ``dead'' as there has been no sign of interest by
the major space agencies in returning to the Moon within the foreseeable
future.

\subsection{Nomenclature and Databases}  \label{futurenomencl}

More and more astronomers rely on databases like NED, SIMBAD \& LEDA,
assuming they are complete and up-to-date. However,
researchers should make life easier for the managers of these
databases, not only by providing their results and data tables
directly to them, but also by making correct references to
astronomical objects in their publications. According to
IAU recommendations for the designation of celestial objects outside
the solar system ({\tts cdsweb.u-strasbg.fr/iau-spec.html}),
existing names should
neither be changed nor truncated in their number of digits.
Acronyms for newly detected sources, or for large surveys, should
be selected carefully so as to avoid clashes with existing acronyms.
The best way to guarantee this is to register a new acronym
with the IAU at  {\tts vizier.u-strasbg.fr/cgi-bin/DicForm}.
For example, in D.\,Levine's lectures for this winter school
the meaning of FIRST is very different from that in the present
paper (cf.\ \S\ref{extendfreq}).
Together with the Task Group on Designations of Commission~5
of the IAU, the author is currently involved in a project to allow
authors to check their preprints for consistency with current
recommendations. This should {\it not} be seen just as a further obstacle
for authors, but as an offer to detect possible non-conforming designations
which are likely to lead to confusion when it comes to the ingestion of
these data into public databases.

\section{Summary of Practicals}      \label{practicals}

Two afternoons of three to four hours were set aside for exercises
using the WWW facilities listed in these lectures. In the first
practical, the students were offered the names of five
very extended ($\sim$\,20$'$) radio galaxies from the 3C catalogue,
and asked to find out the positions of one or more of them from NED
or SIMBAD, to obtain an optical finding chart from one of the
various DSS servers, and to plot these with sky coordinates along the margins.
The next task was to extract a 1.4\,GHz radio image from the NVSS
survey, and a list of sources from the NVSS catalogue of the same region.
A comparison of the two gave an idea of how well (or less well)
the catalogued sources (or components) represent the real complex structures of
these sources. The students were also asked to look at
higher-resolution images of these sources in the ``Atlas of DRAGNs''.
A further exercise was to find out where the many names under
which these sources were known in NED or SIMBAD come from, by
looking up the acronyms in the On-line Dictionary of Nomenclature.
The optical object catalogues like APM, APS and COSMOS were then
queried for the same regions of sky, which allowed the object
classification as star or galaxy to be checked by comparison
with the optical charts from the DSS server. Eventually, radio images from
the FIRST survey were extracted in order to see to what extent
the large-scale structure of the radio galaxies could still be
recognised.

In the second practical, the students were given a chance to discover
a new, optically bright radio-loud quasar, a radio galaxy,
a starburst galaxy, or even a radio star! Each of the roughly two dozen
participating students was assigned a region of sky of the size
of a Palomar plate (6.5\deg$\times$6.5\deg) in the zone
08$^{h}<$RA$<$16$^{h}$, $+$22\deg$<\,\delta<+$42\deg\
(the region covered by the FIRST 1.4\,GHz survey at that time).
Each student was asked to extract all bright objects
(10\,mag$<$B$<$17\,mag) from the USNO A1.0 catalogue (cf.\ \S\,3 of my tutorial)
in the region assigned to them. This was done by remote
interrogation of the USNO site, using the command {\tt findpmm} from the
CDS client software, which had been installed for the school
on the computers at the IAC. The resulting object list (typically 3,000--10,000
objects per student, depending on the Galactic latitude of the assigned field)
was reformatted so as to serve as input for
interrogation of the NVSS source catalogue, which I had installed at the
IAC for the winter school. It had 1.67 million sources in Nov.~1997.
The FORTRAN program {\tts NVSSlist}, publicly available from NRAO, was used
to search a circle of radius 10$''$ around each optical object in the NVSS
catalogue, limited to 1.4\,GHz fluxes greater than 10\,mJy so as to
assure the positional accuracy of the radio sources.
The students were asked to estimate the chance coincidence rate, and they
found that between 0.5 and 2 matches were to be expected by chance.
Actually each student had between three and 18 ``hits'' and was asked
to concentrate on the optically or radio-brightest objects to
find out whether the identification was correct, whether it was
new, and what was known previously about the object. For promising
candidates, a search in the FIRST 1.4\,GHz image database was suggested,
as well as an extraction of a DSS image. Some students even managed a
radio-optical overlay, and one of them found that the overlay facility
in SkyView had a bug when the pixel sizes of the overlaid images
(like e.g.\ NVSS and DSS) was not identical. This was later reported
to, acknowledged and fixed by the SkyView team. Unfortunately the
FIRST image server went ``out of service'' right during the practical.

With the 23 participating students, a sky area of 800\,deg$^2$ had been
covered, and a cross-identification of altogether $\sim$\,80,000 optical
objects with $\sim$\,12,500 radio sources was accomplished.
Fifteen students sent me their results, of which only the most spectacular
will be mentioned here. The 13.9\,mag IRAS galaxy NGC\,3987 almost filled
the 3$'\times$3$'$ DSS image (which the students were asked to extract) and
gave a splendid appearance with its edge-on orientation and strong
dust lane. It was found to coincide with a 58\,mJy NVSS source extended
along the disk of the galaxy, while FIRST clearly shows a strong compact
source (AGN?) and weak radio emission along the disk. A subsequent search
in NED turned up a few other detailed radio studies (\cite{1987AJ.....94..587B},
\cite{1986AJ.....92...94C}, and \cite{1986AJ.....91..199J}). Comparison of the
609\,MHz flux from the latter reference shows that the compact central source
has an inverted spectrum (rising with frequency), apparently not noticed before
in literature. Another student came across
UGC~5146 (Arp\,129), an interacting pair. While the FIRST image server
was unavailable, the FIRST catalogue showed it to be a very complex source,
aligned along the connecting line between the pair, and very extended.
A third student ``rediscovered'' the well-known Seyfert galaxy NGC~4151 with
its 600\,mJy nuclear point source. At first sight no bright radio star
had been discovered, not surprising given the more thorough searches for
radio stars now available (\S\ref{gplansurv}).

\section{Conclusions}

This was a most unusual and rewarding winter school, and exhausting for
lecturers, students and organisers. The organisers are to be
congratulated for the excellent planning and running of the school,
and the IAC for its vision and courage to choose such a topic,
which, at least when it comes to requests for funding, is often claimed to lack
merit, and to be regarded as {\it not scientific}. The school has clearly
proven that scientific expertise is a prerequisite for constructing
and maintaining data archives, databases, and WWW interfaces, so as to make
them user-friendly and reliable at the same time. Too much effort
is often spent on fancy user interfaces, rather than on the
content or its adequate documentation in an archive or database.

I have tried to show that more concerted effort is necessary
to avoid duplication of similar WWW facilities, and the deterioration
of WWW pages with outdated information. Much effort is being spent
by individuals, without an institutional support or obligation, in
providing useful WWW pages. The advantage is that these are
often highly motivated and qualified researchers, but also with the
disadvantage that the service will likely be discontinued
with personal changes.

I also hope that my lectures will stimulate the use of archives
both for advanced research as well as for thesis projects.
Nevertheless, the warnings and pointers to possible pitfalls
I have tried to strew about my lectures cannot replace a sound
observational experience during the first years of research.

Clearly, the educational
possibilities of the Internet have not been fully exploited. In fact,
this winter school, gathering 50 students in one place and putting
them in front of 25 computer terminals, to go and try what they had been
taught during the lectures, may have been an interim between
a classical school without hands-on exercises and a fully
distributed and interactive one, where students would follow
lectures over the WWW and perform exercises at home.

\begin{acknowledgments}
I am grateful to the organisers for inviting me to give these lectures,
and for their financial support. The persistence and excitement
of the students during the practicals was truly impressive.
These practicals would have been impossible without the excellent
computing facilities prepared for the school by R.\,Kroll and
his team of the ``Centro de Calculo'' of the IAC.
I would like to thank the many people who provided useful information,
enriching and completing these lecture notes\,:
D.~Banhatti, E.\,Brinks, S.\,Britzen, J.~Burns, J.J.\,Condon,
W.R.~Cotton, G.\,Dulk, W.\,Erickson, L.\,Feretti, G.~Giovannini,
Gopal~Krishna, L.\,Gurvits, S.E.G.~Hales, R.W.~Hunstead, S.\,J.~Katajainen,
K.\,Kingham, N.~Loiseau, V.~Migenes, R.~Norris, E.~Raimond, W.~Sherwood,
S.\,A.~Trushkin, K.~Weiler, and Rick L.~White.
Thanks also go to all authors of useful WWW pages with compilations of links
which I came across while surfing the WWW for this contribution.
This paper has made use of NASA's Astrophysics Data System
Abstract Service.
A.~Koekemoer provided help to print Figure~\ref{CenAfig} successfully,
E.~Tago kindly helped to produce Figure~\ref{radhistfig}, and special
thanks also go to A.C.\,Davenhall, A.~Fletcher, S.~Kurtz, and O.B.~Slee for their
careful reading of the manuscript at the very last moment.
All of them strengthened my belief that most of the information
given here must have been correct at least at some point in time.
Last, but not least, the Editors of this volume are thanked
for their eternal patience with the delivery of this report.
\end{acknowledgments}


\begin{thebibliography}{}   %  346 references (28-Jul-98)

\bibitem[Abell et al.\ (1989)]{1989ApJS...70....1A}
    {\sc Abell, G. O., Corwin, H. G. Jr, \& Olowin, R. P.} 1989, {\it ApJS} {\bf 70}, 1--138

\bibitem[Acker et al.\ (1991)]{1991A&AS...87..499A}
  {\sc Acker, A., Stenholm, B., \& V\'eron, P.} 1991,
  {\it A\&AS} {\bf 87}, 499

\bibitem[Acker \& Stenholm (1990)]{1990A&AS...86..219A}
  {\sc Acker, A., \& Stenholm, B.} 1990, {\it A\&AS} {\bf 86}, 219--225

\bibitem[Altenhoff et al.\ (1979)]{1979A&AS...35...23A}
    {\sc Altenhoff, W. J., Downes, D., Pauls, T., \& Schraml, J.} 1979, {\it A\&AS} {\bf 35}, 23--54

\bibitem[Altschuler (1986)]{1986A&AS...65..267A}
    {\sc Altschuler, D. R.} 1986, {\it A\&AS} {\bf 65}, 267--283

\bibitem[Amirkhanyan et al.\ (1989)]{1989MIRpubl.......A}
    {\sc Amirkhanyan, V. R., Gorshkov, A. G., Larionov, M. G.,
   Kapustkin, A. A., Konnikova, V. K., Lazutkin, A. N.,
   Nikanorov, A. S., Sidorenkov, V. N., \& Ugol'kova, L. S.} 1989,
   The Zelenchuk Survey of Radio Sources between
  Declinations 0\deg and $+$14\deg, MIR Publ., Moscow, ISBN 5-211-01151-1

\bibitem[Andernach (1989)]{1989BICDS..37..139A}
    {\sc Andernach, H.} 1989, {\it Bull.\ Inf.\ CDS} {\bf 37}, 139--141

\bibitem[Andernach (1990)]{1990BICDS..38...69A}
    {\sc Andernach, H.} 1990, {\it Bull.\ Inf.\ CDS} {\bf 38}, 69--94

\bibitem[Andernach (1992)]{andern92}
    {\sc Andernach H.} 1992,
   in {\it Astronomy from Large Data Bases -- II},
   eds.\ A.\,Heck \& F.\,Murtagh, ESO Conf.\ \& Workshop Proceedings,
  {\bf 43}, 185--190, ESO, Garching

\bibitem[Andernach et al.\ (1992)]{1992A&AS...93..331A}
  {\sc Andernach, H., Feretti, L., Giovannini, G., Klein, U.,
  Rossetti, E., \& Schnaubelt, J.} 1992, {\it A\&AS} {\bf 93}, 331--357

\bibitem[Baade \& Minkowski (1954)]{1954ApJ...119..206B}
    {\sc Baade, W., \& Minkowski, R.} 1954, {\it ApJ} {\bf 119}, 206--214

\bibitem[Baars et al.\ (1977)]{1977A&A....61...99B}
    {\sc Baars, J. W. M., Genzel, R., Pauliny-Toth, I. I. K., \& Witzel, A.}
   1977, {\it A\&A} {\bf 61}, 99--106

\bibitem[Baiesi-Pillastrini et al.\ (1983)]{1983A&AS...53..373B}
    {\sc Baiesi-Pillastrini, G. C., Palumbo, G. G. C., \& Vettolani, G.}
   1983, {\it A\&AS} {\bf 53}, 373--381

\bibitem[Baldwin et al.\ (1985)]{1985MNRAS.217..717B}
    {\sc Baldwin, J. E., Boysen, R. C., Hales, S. E. G., Jennings, J. E.,
   Waggett, P. C., Warner, P. J., \& Wilson, D. M. A.}
  1985, {\it MNRAS} {\bf 217}, 717--730, microfiche

\bibitem[Baleisis et al.\ (1998)]{1998MNRAS.297..545B}
   {\sc Baleisis, A., Lahav, O., Loan, A. J., \& Wall, J. V.} 1998,
   {\it MNRAS} {\bf 297}, 545--558

\bibitem[Becker et al.\ (1991)]{1991ApJS...75....1B}
    {\sc Becker, R. H., White, R. L., \& Edwards, A. L.} 1991, {\it ApJS} {\bf 75}, 1--229

\bibitem[Becker et al.\ (1994)]{1994ApJS...91..347B}
    {\sc Becker, R. H., White, R. L., Helfand, D. J., \& Zoonematkermani, S.}
   1994, {\it ApJS} {\bf 91}, 347--387

\bibitem[Bennett et al.\ (1986)]{1986ApJS...61....1B}
    {\sc Bennett, C. L., Lawrence, C. R., Burke, B. F., Hewitt, J. N.,
   \& Mahoney, J.} 1986, {\it ApJS} {\bf 61}, 1--104

\bibitem[Bennett et al.\ (1997)]{1997PhT....50...32B}
    {\sc Bennett, C. L., Turner, M. S., \& White M.} 1997, {\it Physics Today}, {\bf 50}, 32--38

\bibitem[Birkinshaw (1998)]{birkinshaw98}
    {\sc Birkinshaw, M.} 1998, {\it Phys.\,Rep.}, in press;
    {\tts www.star.bris.ac.uk/mb1/Export.html}

\bibitem[Bischof \& Becker (1997)]{1997AJ....113.2000B}
    {\sc Bischof, O. B., \& Becker, R. H.} 1997, {\it AJ} {\bf 113}, 2000--2005

\bibitem[Boller et al.\ (1998)]{1998A&AS..129...87B}
    {\sc Boller, T., Bertoldi, F., Dennefeld, M., \& Voges, W.}
   1998, {\it A\&AS} {\bf 129}, 87--145

\bibitem[Bolton et al.\ (1949)]{1949Natur.164..101B}
    {\sc Bolton, J. G., Stanley, G., \& Slee, O. B.} 1949,
   {\it Nature}, {\bf 164}, 101

\bibitem[Bolton et al.\ (1964)]{1964AuJPh..17..340B}
    {\sc Bolton, J. G., Gardner, F. F., \& Mackey, M. B.} 1964,
   {\it Aust.\ J.\ Phys.}  {\bf 17}, 340

\bibitem[Bozyan (1992)]{1992ApJS...82....1B}
    {\sc Bozyan, E. P.} 1992, {\it ApJS} {\bf 82}, 1--92

\bibitem[Braude et al.\ (1979)]{1979Ap&SS..64...73B}
    {\sc Braude, S. Ia., Megn, A. V., Sokolov, K. P.,
    Tkachenko, A. P., \& Sharykin, N. K.} 1979, {\it Ap\&SS} {\bf 64}, 73--126

\bibitem[Braude et al.\ (1995)]{1995Ap&SS.226..245B}
    {\sc Braude, S. Y., Sokolov, K. P., Sharykin, N. K., \& Zakharenko, S. M.}
  1995, {\it Ap\&SS} {\bf 226}, 245--271

\bibitem[Bremer et al.\ (1998)]{obscos98}
    {\sc Bremer, M., Jackson, N., \& P\'erez-Fournon, I. (eds.)} 1998,
  {\it Observational Cosmology with the New Radio Surveys},
  Kluwer Acad.\,Publ., ASSL vol.\ 226

\bibitem[Brinkmann \& Siebert (1994)]{1994A&A...285..812B}
   {\sc Brinkmann, W., \& Siebert, J.} 1994, {\it A\&A} {\bf 285}, 812--818

\bibitem[Brinks \& Shane (1984)]{1984A&AS...55..179B}
    {\sc Brinks, E., \& Shane, W. W.} 1984, {\it A\&AS} {\bf 55}, 179--251

\bibitem[Broten et al.\ (1988)]{1988Ap&SS.141..303B}
   {\sc Broten, N. W., MacLeod, J. M., \& Vall\'ee J.P.}
   1988, {\it Ap\&SS} {\bf 141}, 303--331

\bibitem[Brouw \& Spoelstra (1976)]{1976A&AS...26..129B}
   {\sc Brouw, W. N., \& Spoelstra, T. A. Th.} 1976,
  {\it A\&AS} {\bf 26}, 129--144

\bibitem[Brown (1996)]{brown96}
    {\sc Brown, R. L.} 1996, in
   {\it The Westerbork Observatory, Continuing Adventure in Radio Astronomy},
  Kluwer Acad. Publ., Dordrecht, p. 167--183

\bibitem[Burn (1966)]{1966MNRAS.133...67B}
    {\sc Burn, B. J.} 1966, {\it MNRAS} {\bf 133}, 67--83

\bibitem[Burns et al.\ (1983)]{1983ApJ...273..128B}
    {\sc Burns, J. O., Feigelson, E. D., \& Schreier, E. J.} 1983,
  {\it ApJ} {\bf 273}, 128--153

\bibitem[Burns et al.\ (1987)]{1987AJ.....94..587B}
   {\sc Burns, J. O., Hanisch, R. J., White, R. A., Nelson, E. R.,
    Morrisette, K. A., \& Ward, M. J.} 1987, {\it AJ} {\bf 94}, 587--617

\bibitem[Burns \& Mendell (1988)]{burns88}
    {\sc Burns, J. O., \& Mendell, W. W. (eds.)} 1988,
   Future Astronomical Observatories on the Moon, NASA Conf.\ Publ.\ 2489,
   Proc.\ of Workshop held in Houston, TX, USA, Jan.\ 1986

\bibitem[Burns et al.\ (1989)]{burns89}
    {\sc Burns, J. O., Duric, N., Johnson, S., \& Taylor, G. J. (eds.)} 1989,
  A Lunar Far-Side Very Low Frequency Array, NASA Conf.\ Publ.\ 3039,
   Proc.\ of Workshop held in Albuquerque, NM, USA, Feb.\ 1988

\bibitem[Bursov et al.\ (1997)]{1997BSAO...42....5B}
    {\sc Bursov, N. N., Lipovka, N. M., Soboleva, N. S., Temirova, A. V.,
   Gol'neva, N. E., Parijskaya, E. Yu., \& Savastenya, A. V.}
   1997, {\it Bull.\ SAO} {\bf 42}, 5

\bibitem[Burton (1976)]{1976ARA&A..14..275B}
    {\sc Burton, W. B.} 1976, {\it ARA\&A} {\bf 14}, 275--306

\bibitem[Bystedt et al.\ (1984)]{1984A&AS...56..245B}
    {\sc Bystedt, J. E. V., Brinks, E., de\,Bruyn, A.G.,
   Israel, F. P., Schwering, P. B. W., Shane, W. W., \& Walterbos, R. A. M.}
    1984, {\it A\&AS} {\bf 56}, 245--280

\bibitem[Christiansen \& H\"ogbom (1985)]{christhoeg85}
    {\sc Christiansen, W. N. \& H\"ogbom, J. A.} 1985,
   {\it Radio Telescopes}, 2nd ed., Cambridge University Press, Cambridge, UK

\bibitem[Ciliegi et al.\ (1998)]{1998MNRAS.VVV..PPPC}
   {\sc Ciliegi, P., McMahon, R. G., Miley, G., Gruppioni, C.,
    Rowan-Robinson, M., Cesarsky, C., Danese, L., Franceschini, A.,
    Genzel, R., Lawrence, A., Lemke, D., Oliver, S., Puget, J.-L.,
    Rocca-Volmerange, B.} 1998, subm.\ to {\it MNRAS}, {\tts astro-ph/9805353}

\bibitem[Cioffi \& Jones (1980)]{1980AJ.....85..368C}
    {\sc Cioffi, D. F., \& Jones, T. W.} 1980, {\it AJ} {\bf 85}, 368--375

\bibitem[Clarke et al.\ (1976)]{1976AuJPA..40....1C}
    {\sc Clarke, M. E., Little, A. G., \& Mills, B. Y.} 1976,
   {\it Aust.\ J.\ Phys.\ Ap.\ Suppl.} {\bf 40}, 1--71

\bibitem[Clarke et al.\ (1980)]{1980MNRAS.190..205C}
    {\sc Clarke, J. N., Kronberg, P. P., \& Simard-Normandin, M.}
   1980, {\it MNRAS} {\bf 190}, 205--215

\bibitem[Cohen \& Kellermann (1995)]{cohenkell95}
    {\sc  Cohen, M. H., \& Kellermann, K. I.} 1995,
  {\it Quasars and Active Galactic Nuclei\,: High Resolution Radio Imaging},
  Proc.\ Natl.\ Acad. Sci., USA, {\bf 92}, 11339--11450

\bibitem[Colla et al.\ (1970)]{1970A&AS....1..281C}
    {\sc Colla, G., Fanti, C., Fanti, R., Ficarra, A.,
   Formiggini, L., Gandolfi, E., Grueff, G., Lari, C.,
   Padrielli, L., Roffi, G., Tomasi, P., \& Vigotti, M.}
  1970, {\it A\&AS} {\bf 1}, 281--317

\bibitem[Colla et al.\ (1972)]{1972A&AS....7....1C}
    {\sc Colla, G., Fanti, C., Fanti, R., Ficarra, A.,
   Formiggini, L., Gandolfi, E., Lari, C., Marano, B.,
   Padrielli, L., \& Tomasi, P.} 1972, {\it A\&AS} {\bf 7}, 1--34

\bibitem[Colla et al.\ (1973)]{1973A&AS...11..291C}
    {\sc Colla, G., Fanti, C., Fanti, R., Ficarra, A.,
  Formiggini, L., Gandolfi, E., Gioia, I., Lari, C.,
   Marano, B., Padrielli, L., \& Tomasi, P.}
  1973, {\it A\&AS} {\bf 11}, 291--325

\bibitem[Combi et al.\ (1998)]{1998A&A...333..298C}
   {\sc Combi, J. A., Romero, G. E., \& Arnal, E. M.} 1998,
   {\it A\&A} {\bf 333}, 298--304

\bibitem[Condon (1974)]{1974ApJ...188..279C}
    {\sc Condon, J. J.} 1974, {\it ApJ} {\bf 188}, 279--286

\bibitem[Condon \& Broderick (1986a)]{1986AJ....91.1051C}
    {\sc Condon, J. J., \& Broderick, J. J.} 1986a, {\it AJ} {\bf 91}, 1051--1057

\bibitem[Condon \& Broderick (1986b)]{1986AJ.....92...94C}
  {\sc Condon, J. J., \& Broderick, J. J.} 1986b, {\it AJ} {\bf 92}, 94--102

\bibitem[Condon et al.\ (1989)]{1989AJ.....97.1064C}
    {\sc Condon, J. J., Broderick, J. J., \& Seielstad, G. A.}
  1989, {\it AJ} {\bf 97}, 1064--1073

\bibitem[Condon \& Lockman (1990)]{condonlockman90}
    {\sc Condon, J. J., \& Lockman, F. J.} 1990,
   Large-Scale Surveys of the Sky, Proc.\ NRAO Green-Bank Workshop No.\ 20,
  held Sept.\ 1987, publ.\ by NRAO, Charlottesville, WV, USA

\bibitem[Condon et al.\ (1991)]{1991AJ....102.2041C}
    {\sc Condon, J. J., Broderick, J. J., \& Seielstad, G. A.} 1991,
  {\it AJ} {\bf 102}, 2041--2046

\bibitem[Condon et al.\ (1993)]{1993AJ....106.1095C}
    {\sc Condon, J. J., Griffith, M. R., \& Wright, A. E.} 1993,
  {\it AJ} {\bf 106}, 1095--1100

\bibitem[Condon et al.\ (1994)]{1994AJ....107.1829C}
    {\sc Condon, J. J., Broderick, J. J., Seielstad, G. A.,
  Douglas, K., \& Gregory, P. C.} 1994,
  {\it AJ} {\bf 107}, 1829--1833

\bibitem[Condon et al.\ (1995)]{1995AJ....109.2318C}
    {\sc Condon, J. J., Anderson, E., \& Broderick, J. J.} 1995,
  {\it AJ} {\bf 109}, 2318--2354

\bibitem[Condon (1997)]{1997PASP..109..166C}
    {\sc Condon, J. J.} 1997, {\it PASP} {\bf 109}, 166--172

\bibitem[Condon et al.\ (1997)]{1997AAS...191.1402C}
    {\sc Condon, J. J., Kaplan, D. L., \& Yin, Q. F.}
   1997, {\it BAAS} {\bf 29}, 1231

\bibitem[Condon (1998)]{1998IAUS..179...19C}
    {\sc Condon, J. J.} 1998, {\it IAU Symp.} {\bf 179}, 19--25,
  eds.\ B.J.~McLean et al., Kluwer Acad.\ Publ., Dordrecht

\bibitem[Condon et al.\ (1998)]{1998AJ....115.1693C}
    {\sc Condon, J. J., Cotton, W. D., Greisen, E. W., Yin, Q. F.,
  Perley, R. A., Taylor, G. B., \& Broderick, J. J.} 1998,
  {\it AJ} {\bf 115}, 1693--1716;  {\tts \verb*Cwww.cv.nrao.edu/~jcondon/nvss.htmlC}

\bibitem[Condon \& Kaplan (1998)]{1998ApJS..11x..yyyC}
    {\sc Condon, J. J., \& Kaplan, D. L.} 1998, {\it ApJS}, in press

\bibitem[Cooray et al.\ (1998)]{1998AJ....115.1388C}
    {\sc Cooray, A. R., Grego, L., Holzapfel, W. L., Joy, M.,
   \& Carlstrom, J. E.} 1998, {\it AJ} {\bf 115}, 1388--1399

\bibitem[Cornwell (1988)]{1988A&A...202..316C}
   {\sc Cornwell, T. J.} 1988, {\it A\&A} {\bf 202}, 316--321

\bibitem[Cornwell (1989)]{cornwell89}
    {\sc Cornwell, T.} 1989, in {\it Synthesis Imaging in Radio Astronomy},
   eds.\ Perley, R. A., Schwab, F. R. \& Bridle, A. H.,
   ASP Conf.\ Ser.\ {\bf 6}, ASP, San Francisco, p.~277--286

\bibitem[Cornwell \& Perley (1991)]{1991ASPC...19}
    {\sc Cornwell, T. J. \& Perley, R. A. (eds.)} 1991,
   {\it Radio Interferometry: Theory, Techniques, and Applications},
  ASP Conf.\ Ser.\ {\bf 19}, ASP, San Francisco

\bibitem[CRAF Handbook for Radio Astronomy (1997)]{craf97}
    {\sc CRAF Handbook for Radio Astronomy, 2nd ed.} 1997,
  Committee on Radio Astronomy Frequencies (CRAF),
  Netherlands Foundation for Research in Astronomy (NFRA), Dwingeloo,
   The Netherlands

\bibitem[Crawford et al.\ (1996)]{1996ApJ...460..225C}
    {\sc Crawford, T., Marr, J., Partridge, B., \& Strauss, M. A.}
   1996, {\it ApJ} {\bf 460}, 225--243

\bibitem[Dame et al.\ (1987)]{1987ApJ...322..706D}
    {\sc Dame, T. M., Ungerechts, H., Cohen, R. S., De Geus, E. J.,
   Grenier, I. A., May, J., Murphy, D. C., Nyman, L.-A., \& Thaddeus, P.}
  1987, {\it ApJ} {\bf 322}, 706--720

\bibitem[Davies et al.\ (1973)]{1973AuJPA..28....1D}
    {\sc Davies, I. M., Little, A. G., \& Mills, B. Y.}
  1973, {\it Aust.\ J.\ Phys.\ Ap.\ Suppl.} {\bf 28}, 1--59

\bibitem[de\,Bruyn \& Sijbring (1993)]{debrusij93}
    {\sc de\,Bruyn, A. G., \& Sijbring, D.} 1993,
  chap.~2 of Sijbring, D., PhD Thesis, Groningen, 1993

\bibitem[de\,Ruiter et al.\ (1977)]{1977A&AS...28..211d}
  {\sc de\,Ruiter, H. R., Arp, H. C., \& Willis, A. G.} 1977,
   {\it A\&AS} {\bf 27}, 211--293

\bibitem[Dixon (1970)]{1970ApJS...20....1D}
    {\sc Dixon, R. S.} 1970, {\it ApJS} {\bf 20}, 1

\bibitem[Douglas et al.\ (1996)]{1996AJ....111.1945D}
    {\sc Douglas, J. N., Bash, F. N., Bozyan, F. A., Torrence, G. W.,
  \& Wolfe, C.} 1996, {\it AJ} {\bf 111}, 1945--1963

\bibitem[Dreher et al.\ (1987)]{1987ApJ...316..611D}
    {\sc Dreher, J. W., Carilli, C. L., \& Perley, R. A.} 1987, {\it ApJ} {\bf 316}, 611--625

\bibitem[Dressel \& Condon (1978)]{1978ApJS...36...53D}
    {\sc Dressel, L. L., Condon, J. J.} 1978, {\it ApJS} {\bf 36}, 53--75

\bibitem[Duncan et al.\ (1995)]{1995MNRAS.277...36D}
    {\sc Duncan, A. R., Stewart, R. T., Haynes, R. F., \& Jones, K. L.}
   1995, {\it MNRAS} {\bf 277}, 36--52

\bibitem[Duncan et al.\ (1997)]{1997MNRAS.291..279D}
    {\sc Duncan, A. R., Haynes, R. F., Jones, K. L., \& Stewart, R. T.}
  1997, {\it MNRAS} {\bf 291}, 279--295

\bibitem[Durdin et al.\ (1975)]{1975NAICR..45.....D}
    {\sc Durdin, J. M., Pleticha, D., Condon, J. J.,
   Yerbury, M. J., Jauncey, D. L., \& Hazard, C.} 1975,
  The NAIC 611-MHz Multi-Beam Sky Survey Source List,
  NAIC Report {\bf 45}, March 1975; available from {\tts cats.sao.ru/doc/NAIC.html}

\bibitem[Dwarakanath \& Udaya Shankar (1990)]{1990JApA...11..323D}
    {\sc Dwarakanath, K. S., \& Udaya Shankar, N.} 1990,
   {\it J.\ Ap.\ Astron.} {\bf 11}, 323--410

\bibitem[Edge \& Mulkay (1976)]{edgemulkay76}
   {\sc Edge, D. O., \& Mulkay, M. J.} 1976,
  {\it  Astronomy Transformed: The Emergence of Radio Astronomy in Britain},
   John Wiley \& Sons Inc., New York, NY, USA

\bibitem[Egger \& Aschenbach (1995)]{1995A&A...294L..25E}
  {\sc Egger, R.J., \& Aschenbach, B.} 1995, {\it A\&A} {\bf 294}, L25--L28

\bibitem[Elfhag et al.\ (1996)]{1996A&AS..115..439E}
    {\sc Elfhag, T., Booth, R. S., H\"oglund, B.,
    Johansson, L. E. B., \& Sandqvist, Aa.} 1996, {\it A\&AS} {\bf 115}, 439--468

\bibitem[Emerson et al.\ (1979)]{1979A&A....76...92E}
    {\sc Emerson, D. T., Klein, U., \& Haslam, C. G. T.}
  1979, {\it A\&A}, {\bf 76}, 92--105

\bibitem[Emerson \& Gr\"ave (1988)]{1988A&A...190..353E}
   {\sc Emerson, D. T., \& Gr\"ave, R.} 1988, {\it A\&A}, {\bf 190}, 353--358

\bibitem[Emerson \& Payne (1995)]{1995ASPC...75}
    {\sc Emerson, D. T. \& Payne, J. M. (eds.)} 1995,
   {\it Multi-Feed Systems for Radio Telescopes},
   ASP Conf.\ Ser.\  {\bf 75}, ASP, San Francisco

\bibitem[Erickson (1997)]{1997PASA...14..278E}
    {\sc Erickson, W. C.} 1997, {\it Proc.~ASA} {\bf 14}, 278--282

\bibitem[Fanti et al.\ (1974)]{1974A&AS...18..147F}
    {\sc Fanti, C., Fanti, R., Ficarra, A., \& Padrielli, L.} 1974,
   {\it A\&AS} {\bf 18}, 147--156

\bibitem[Feretti et al.\ (1997)]{1997NewA....2..501F}
    {\sc Feretti, L., Giovannini, G., \& B\"ohringer, H.} 1997,
   {\it New Astron.} {\bf 2}, 501--515

\bibitem[Ficarra et al.\ (1985)]{1985A&AS...59..255F}
    {\sc Ficarra, A., Grueff, G., \& Tomassetti, G.} 1985,
   {\it A\&AS} {\bf 59}, 255--347

\bibitem[Fomalont et al.\ (1997)]{1997ApJ...475L..5F}
  {\sc Fomalont, E. B., Kellermann, K. I., Richards, E. A.,
   Windhorst, R. A., \& Patridge R. B.} 1997,
   {\it ApJ} {\bf 475}, L5--L8

\bibitem[F\"urst et al.\ (1990)]{1990A&AS...85..805F}
    {\sc F\"urst E., Reich, W., Reich, P., \& Reif, K.} 1990,
   {\it A\&AS} {\bf 85}, 805--811

\bibitem[F\"urst et al.\ (1998)]{1998IAUS..179...97F}
    {\sc  F\"urst, E., Reich, W., Reich, P., Uyan{\i}ker, B., \& Wielebinski, R.} 1998,
  {\it IAU Symp.} {\bf 179}, 165--171, eds.\ B.J.~McLean et al., Kluwer Acad.\ Publ., Dordrecht

\bibitem[Fukui \& Yonekura (1998)]{1998IAUS..179..165F}
    {\sc  Fukui, Y., \& Yonekura, Y.} 1998,
   {\it IAU Symp.} {\bf 179}, 165--171,
   eds.\ B.J.~McLean et al., Kluwer Acad.\ Publ., Dordrecht

\bibitem[Gardner et al.\ (1966)]{1966ARA&A...4..245G}
    {\sc Gardner, F. F., Whiteoak, J. B., \& Morris, D.}
   1966, {\it ARA\&A} {\bf 4}, 245

\bibitem[Giovannini et al.\ (1991)]{1991A&A...252..528G}
    {\sc Giovannini, G., Feretti, L., \& Stanghellini, C.}
  1991, {\it A\&A} {\bf 252}, 528--537

\bibitem[Golap et al.\ (1995)]{1995Ap&SS.228..373G}
    {\sc Golap, K., Issur, N. H., \& Somanah, R.} 1995,
   {\it Ap\&SS} {\bf 228}, 373--377

\bibitem[Gower et al.\ (1967)]{1967MmRAS..71...49G}
    {\sc Gower, J. F. R., Scott, P. F., \& Wills, D.}
   1967, {\it MmRAS} {\bf 71}, 49--144

\bibitem[Graham (1970)]{1970MNRAS.149..319G}
   {\sc Graham, I.} 1970, {\it MNRAS} {\bf 149}, 319--339

\bibitem[Graham-Smith (1974)]{grahamsmith74}
  {\sc Graham-Smith, F.} 1974, {\it Radio Astronomy}, 4$^{th}$ edition,
   Penguin Books Ltd., UK

\bibitem[Gray (1994a)]{1994MNRAS.270..822G}
    {\sc Gray, A. D.} 1994a, {\it MNRAS} {\bf 270}, 822--834

\bibitem[Gray (1994b)]{1994MNRAS.270..861G}
    {\sc Gray, A. D.} 1994b, {\it MNRAS} {\bf 270}, 861--870

\bibitem[Gray et al.\ (1998)]{1998Natur.393..660G}
    {\sc Gray, A. D., Landecker, T. L., Dewdney, P. E., \& Taylor, A. R.}
   1998, {\it Nature}, {\bf 393}, 660 % {\tts ftp.drao.nrc.ca:pub/agray/nature.ps}

\bibitem[Green (1997)]{1997PASA...14...73G}
    {\sc Green, A. J.} 1997, {\it Proc.~ASA} {\bf 14}, 73--76

\bibitem[Green et al.\ (1998)]{1998ApJ...5xx..yyyG}
    {\sc Green, A. J., Cram, L. E., \& Large, M. I.} 1998, {\it ApJ}, in press

\bibitem[Gregory \& Taylor (1986)]{1986AJ.....92..371G}
    {\sc Gregory, P. C., \& Taylor, A. R.} 1986, {\it AJ} {\bf 92}, 371--411

\bibitem[Gregory \& Condon (1991)]{1991ApJS...75.1011G}
    {\sc Gregory, P. C., \& Condon, J. J.} 1991,
   {\it ApJS} {\bf 75}, 1011--1291

\bibitem[Gregory et al.\ (1994)]{1994ApJS...90..173G}
    {\sc Gregory, P. C., Vavasour, J. D., Scott, W. K., \& Condon, J. J.} 1994,
   {\it ApJS} {\bf 90}, 173--177

\bibitem[Gregory et al.\ (1996)]{1996ApJS..103..427G}
    {\sc Gregory, P. C., Scott, W. K., Douglas, K., \& Condon, J. J.} 1996,
   {\it ApJS} {\bf 103}, 427--432

\bibitem[Gregory et al.\ (1998)]{1998ASPC..144..283G}
  {\sc Gregory, P. C., Scott, W. K., \& Poller, B.} 1998, in IAU Coll.\ 164,
  {\it Radio Emission from Galactic and Extragalactic Compact Sources},
  ASP Conf.\ Ser.\ {\bf 144}, 283--284, eds.\ J.\,A.~Zensus, G.\,B.~Taylor, \&
  J.\,M.~Wrobel, ASP, San Francisco.

\bibitem[Griffith et al.\ (1990)]{1990ApJS...74..129G}
    {\sc Griffith, M., Langston, G., Heflin, M., Conner, S., Leh\'ar J., \& Burke, B.} 1990,
   {\it ApJS} {\bf 74}, 129--180

\bibitem[Griffith et al.\ (1991)]{1991ApJS...75..801G}
    {\sc Griffith, M., Langston, G., Heflin, M., Conner, S., \& Burke, B.} 1991,
   {\it ApJS} {\bf 75}, 801--833

\bibitem[Griffith et al.\ (1994)]{1994ApJS...90..179G}
    {\sc Griffith, M. R., Wright, A. E., Burke, B. F., \& Ekers, R. D.} 1994,
   {\it ApJS} {\bf 90}, 179--295

\bibitem[Griffith et al.\ (1995)]{1995ApJS...97..347G}
    {\sc Griffith, M. R., Wright, A. E., Burke, B. F., \& Ekers, R. D.} 1995,
   {\it ApJS} {\bf 97}, 347--453

\bibitem[Gubanov \& Andernach (1997)]{1997BaltA...6..263G}
    {\sc Gubanov, A. G., \& Andernach, H.} 1997,
   {\it Balt.\ Astron.} {\bf 6}, 263--266

\bibitem[Hacking et al.\ (1989)]{1989ApJ...339...12H}
    {\sc Hacking, P., Condon, J. J., Houck, J. R., \& Beichman, C. A.} 1989,
   {\it ApJ} {\bf 339}, 12--26

\bibitem[Haigh et al.\ (1997)]{1997PASA...14..221H}
    {\sc Haigh, A. J., Robertson, J. G., \& Hunstead, R. W.} 1997,
   {\it Proc.~ASA} {\bf 14}, 221--229

\bibitem[Hales et al.\ (1988)]{1988MNRAS.234..919H}
    {\sc Hales, S. E. G., Baldwin, J. E., \& Warner, P. J.} 1988,
   {\it MNRAS} {\bf 234}, 919--936

\bibitem[Hales et al.\ (1990)]{1990MNRAS.246..256H}
    {\sc Hales, S. E. G., Masson, C. R., Warner, P. J., \& Baldwin, J. E.} 1990,
   {\it MNRAS} {\bf 246}, 256--262

\bibitem[Hales et al.\ (1991)]{1991MNRAS.251...46H}
    {\sc Hales, S. E. G., Mayer, C. J., Warner, P. J., \& Baldwin, J. E.} 1991,
   {\it MNRAS} {\bf 251}, 46--53

\bibitem[Hales et al.\ (1993a)]{1993MNRAS.262.1057H}
    {\sc Hales, S. E. G., Masson, C. R., Warner, P. J., Baldwin, J. E., \& Green, D. A.} 1993a,
   {\it MNRAS} {\bf 262}, 1057--1061

\bibitem[Hales et al.\ (1993b)]{1993MNRAS.263...25H}
    {\sc Hales, S. E. G., Baldwin, J. E., \& Warner, P. J.} 1993b,
   {\it MNRAS} {\bf 263}, 25--30

\bibitem[Hales et al.\ (1995)]{1995MNRAS.274..447H}
    {\sc Hales, S. E. G., Waldram, E. M., Rees, N., \& Warner, P. J.} 1995,
    {\it MNRAS} {\bf 274}, 447--451

\bibitem[Hardee et al.\ (1996)]{hardee96}
    {\sc Hardee, P. E., Bridle, A. H., \& Zensus, A. (eds.)}
    1996, {\it Energy Transport in Radio Galaxies and Quasars},
  ASP Conf.\ Ser.\  {\bf 100}, ASP, San Francisco, ({\tts www.cv.nrao.edu/jetworks})

\bibitem[Harris \& Miley (1978)]{1978A&AS...34..117H}
  {\sc Harris, D. E., \& Miley, G. K.} 1978,
    {\it A\&AS} {\bf 34}, 117--128

\bibitem[Harris et al.\ (1995)]{harris95}
    {\sc Harris, D. E., Grant C. P. S., \& Andernach, H.} 1995,
   {\it Astronomical Data Analysis Software and Systems -- IV},
   ASP~Conf.\ Ser.\ {\bf 77}, 48--51, eds.\ R.\,A.~Shaw, H.\,E.~Payne, 
   J.J.E.\,Hayes; ASP, San Francisco, {\tts astro-ph/94011021}

\bibitem[Hartmann \& Burton (1997)]{ldhi97}
    {\sc Hartmann, D. \& Burton, W. B.} 1997, {\it Atlas of Galactic Neutral Hydrogen},
   Cambridge University Press, ISBN 0-521-47111-7 (no URL available)

\bibitem[Haslam (1974)]{1974A&AS...15..333H}
   {\sc Haslam, C. G. T.} 1974, {\it A\&AS} {\bf 15}, 333--

\bibitem[Haslam et al.\ (1981)]{1981A&A...100..209H}
    {\sc Haslam, C. G. T., Klein, U., Salter, C. J., Stoffel, H., Wilson, W. E.,
   Cleary, M. N., Cooke, D. J., \& Thomasson, P.} 1981,
   {\it A\&A} {\bf 100}, 209--219

\bibitem[Haslam et al.\ (1982)]{1982A&AS...47....1H}
    {\sc Haslam, C. G. T., Stoffel, H., Salter, C. J., \& Wilson, W. E.}
  1982, {\it A\&AS} {\bf 47}, 1--142

\bibitem[Hauschildt (1987)]{1987A&A...184...43H}
   {\sc Hauschildt, M.} 1987, {\it A\&A} {\bf 184}, 43--56

\bibitem[Haynes et al.\ (1979)]{1979AuJPA..48....1H}
    {\sc Haynes, R. F., Caswell, J. L., \& Simons, L. W. J.} 1979,
  {\it Aust.\ J.\ Phys.\ Ap.\ Suppl.} {\bf 48}, 1--30

\bibitem[Haynes et al.\ (1996)]{hhmm96}
    {\sc Haynes, Raymond; Haynes, Roslynn; Malin, D., \& McGee, R.}
   1996, {\it Explorers of the Southern Sky: A History of Australian Astronomy},
   Cambridge University Press, Cambridge, UK

\bibitem[Helfand et al.\ (1992)]{1992ApJS...80..211H}
    {\sc Helfand, D. J., Zoonematkermani, S., Becker, R. H.,
  \& White, R. L.} 1992, {\it ApJS} {\bf 80}, 211--255

\bibitem[Helfand et al.\ (1997)]{1997AAS...191.1403H}
    {\sc Helfand, D. J., Schnee, S., Becker, R. H.,
    White, R. L., \& McMahon, R. G.} 1997, {\it BAAS} {\bf 29}, 1231

\bibitem[Helfer et al.\ (1998)]{1998AAS...192.7301H}
    {\sc Helfer, T. T., Thornley, M. D., Regan, M. W., Sheth, K.,
   Vogel, S. N., Wong, T., Blitz, L., \& Bock, D.} 1998, {\it BAAS} {\bf 30}, 928

\bibitem[Henning et al.\ (1998)]{1998AJ....115..584H}
    {\sc Henning, P. A., Kraan-Korteweg, R. C., Rivers, A. J., Loan, A. J.,
     Lahav, O., \& \linebreak[4] Burton, W. B.} 1998, {\it AJ} {\bf 115}, 584--591

\bibitem[Henstock et al.\ (1995)]{1995ApJS..100....1H}
 {\sc Henstock, D. R., Browne, I. W. A., Wilkinson, P. N., Taylor, G. B.,
     Vermeulen, R. C., Pearson, T. J., \& Readhead, A. C. S.} 1995,
     {\it ApJS} {\bf 100}, 1--36

\bibitem[Herbig \& Readhead (1992)]{1992ApJS...81...83H}
    {\sc Herbig, T., \& Readhead, A. C. S.} 1992, {\it ApJS} {\bf 81}, 83--66

\bibitem[Hewitt \& Burbidge (1989)]{1989ApJS...69....1H}
    {\sc Hewitt, A., \& Burbidge, G.} 1989, {\it ApJS} {\bf 69}, 1--63

\bibitem[Hey (1971)]{hey71}
  {\sc Hey, J. S.} 1971, {\it The Radio Universe}, 1$^{st}$ edition,
   Pergamon Press Ltd,, UK

\bibitem[Hey (1973)]{hey73}
  {\sc Hey, J. S.} 1973, {\it The Evolution of Radio Astronomy},
   1$^{st}$ edition, Neale Watson Academic Publications, New York, NY 10010, USA

\bibitem[Heyer et al.\ (1998)]{1998ApJS..115..241H}
    {\sc Heyer, M. H., Brunt, C., Snell, R. L., Howe, J. E.,
   Schloerb, F. P., \& Carpenter, J. M.} 1998, {\it ApJS} {\bf 115}, 241--258

\bibitem[Hooper et al.\ (1996)]{1996ApJ...473..746H}
    {\sc Hooper, E. J., Impey, C. D., Foltz, C. B.,
   \& Hewett, P. C.} 1996, {\it ApJ} {\bf 473}, 746--759

\bibitem[Hopkins (1998)]{hopkins98}
   {\sc Hopkins, A. M.} 1998, Ph.\,D.\ thesis, University of Sydney

\bibitem[Hopkins et al.\ (1998)]{1998MNRAS.296..839H}
    {\sc Hopkins, A. M., Mobasher, B., Cram, L., \& Rowan-Robinson, M.}
   1998, {\it MNRAS} {\bf 296}, 839--846

\bibitem[Hughes \& MacLeod (1989)]{1989AJ.....97..786H}
    {\sc Hughes, V. A., \& MacLeod, G. C.} 1989, {\it AJ} {\bf 97}, 786--800

\bibitem[Hughes et al.\ (1992)]{1992ApJ...396..469H}
    {\sc Hughes, P. A., Aller, H. D., \& Aller, M. F.} 1992,
    {\it ApJ} {\bf 396}, 469--486

\bibitem[Hunstead et al.\ (1998)]{hunst98}
    {\sc Hunstead, R. W., Cram, L. E., \& Sadler, E. M.} 1998,
   Proc.\ Observational Cosmology with the New Radio Surveys,
   eds.\ M.\,Bremer, N.\,Jackson \& I.\,P\'erez-Fournon,
   Kluwer Acad.\,Publ., p.\,55--62

\bibitem[Hutchings et al.\ (1991)]{1991PASP..103...21H}
  {\sc Hutchings, J. B., Durand, D., \& Pazder, J.} 1991,
   {\it PASP} {\bf 103}, 21--25

\bibitem[Jackson \& Davis (1997)]{jacksondavis97}
    {\sc Jackson, N. \& Davis, R. J. (eds.)} 1997,
   {\it High Sensitivity Radio Astronomy},
  Cambridge University Press, Cambridge, UK

\bibitem[Jaffe et al.\ (1986)]{1986AJ.....91..199J}
   {\sc Jaffe, W., Gavazzi, G., \& Valentijn, E.} 1986, {\it AJ} {\bf 91}, 199--203

\bibitem[Jauncey (1977)]{IAUS74}
    {\sc Jauncey, D. L. (ed.)} 1977, {\it Radio Astronomy and Cosmology},
   {\it IAU Symp.} {\bf 74}, 398 pp., D.~Reidel, Dordrecht

\bibitem[Johnston et al.\ (1995)]{1995AJ....110..880J}
    {\sc Johnston, K. J., Fey, A. L., Zacharias, N., Russell, J. L.,
   Ma, C., De~Vegt, C., Reynolds, J. E., Jauncey, D. L.,
   Archinal, B. A., Carter, M. S., Corbin, T. E., Eubanks, T. M.,
   Florkowski, D. R., Hall, D. M., McCarthy, D. D., McCulloch, P. M.,
   King, E. A., Nicolson, G., \& Shaffer, D. B.} 1995, {\it AJ} {\bf 110}, 880--915

\bibitem[Jonas et al.\ (1985)]{1985A&AS...62..105J}
    {\sc Jonas, J. L., de\,Jager, G., \& Baart, E. E.} 1985,
   {\it A\&AS} {\bf 62}, 105--128

\bibitem[Jonas (1998)]{1998IAUS..179...95J}
    {\sc Jonas, J. L.} 1998, {\it IAU Symp} {\bf 179}, 95--96,
   eds.\ B.J.~McLean et al., Kluwer Acad.\ Publ., Dordrecht

\bibitem[Jonas et al.\ (1998)]{1998MNRAS.297..977J}
    {\sc Jonas, J. L., Baart, E. \& Nicolson, G.} 1998, 
    {\it MNRAS}, {\bf 297}, 977--989

\bibitem[Joncas et al.\ (1992)]{1992ApJ...387..591J}
    {\sc Joncas, G., Durand, D., \& Roger, R. S.} 1992, {\it ApJ} {\bf 387}, 591--611

\bibitem[Junkes et al.\ (1987)]{1987A&AS...69..451J}
   {\sc Junkes, N., F\"urst, E., \& Reich, W.} 1987,
   {\it A\&AS} {\bf 69}, 451--464

\bibitem[Junkes et al.\ (1993)]{1993A&A...269...29J}
    {\sc Junkes, N., Haynes, R. F., Harnett, J. I., \& Jauncey, D. L.} 1993,
     {\it A\&A} {\bf 269}, 29--38

\bibitem[Kaiser (1987)]{kaiser87}
    {\sc Kaiser, M. L.} 1987, in Radio Astronomy from Space,
   Proc.\ Workshop \#\,18 held at NRAO Green Bank, WV, USA,
   Oct.\ 1986, ed.\ K.\,W.~Weiler, p.\ 227--238, publ.\ by NRAO

\bibitem[Kallas \& Reich (1980)]{1980A&AS...42..227K}
    {\sc Kallas, E., \& Reich, W.} 1980, {\it A\&AS} {\bf 42}, 227

\bibitem[Kaplan et al.\ (1998)]{1998ApJS..11x..yyyK}
    {\sc Kaplan, D. L., Condon, J. J., Arzoumanian, Z., \&
    Cordes, J. M.} 1998, {\it ApJS}, in press

\bibitem[Kardashev (1997)]{kardashev97}
    {\sc Kardashev, N. S.} 1997, {\it Exp. Astron.} {\bf 7}, 329--343

\bibitem[Kassim (1988)]{1988ApJS...68..715K}
    {\sc Kassim, N. E.} 1988, {\it ApJS} {\bf 68}, 715--733

\bibitem[Kassim \& Weiler (1990)]{kassimweiler90}
    {\sc Kassim, N. E., \& Weiler, K. W. (eds.)} 1990,
   Low Frequency Astrophysics from Space,
   Proc.\ of Workshop held in Crystal City, VA, USA, Jan.\ 1990,
   Lecture Notes in Physics, Springer Verlag, Berlin

\bibitem[Katz-Stone \& Rudnick (1994)]{1994ApJ...426..116K}
    {\sc Katz-Stone, D. M., \& Rudnick, L.} 1994,
  {\it ApJ} {\bf 426}, 116--122

\bibitem[Kellermann \& Sheets (1984)]{kellerm84}
    {\sc Kellermann, K. I., \& Sheets, B., eds.} 1984,
   {\it Serendipitous discoveries in radio astronomy}, Proc.\ NRAO Workshop 7,
   Green Bank, WV, USA, May 1983; publ.\ by National Radio Astronomy Observatory

\bibitem[Kellermann et al.\ (1998)]{1998AJ....115.1295K}
    {\sc Kellermann, K. I., Vermeulen, R. C., Zensus, J. A., \& Cohen, M. H.} 1998,
  {\it AJ} {\bf 115}, 1295--1318

\bibitem[Klein \& Emerson (1981)]{1981A&A....94...29K}
    {\sc Klein, U., \& Emerson, D. T.} 1981,
    {\it A\&A} {\bf 94}, 29--44

\bibitem[Klein et al.\ (1989)]{1989A&A...211..280K}
    {\sc Klein, U., Wielebinski, R., Haynes, R. F., \& Malin, D. F.} 1989,
   {\it A\&A} {\bf 211}, 280--292

\bibitem[Klein \& Mack (1995)]{1995ASPC...75..318K}
    {\sc Klein, U. \& Mack, K.-H.} 1995, in
  {\it Multi-Feed Systems for Radio Telescopes}, eds.\ Emerson, D. T. \& Payne, J. M.,
   ASP Conf.\ Ser.\ {\bf 75}, ASP, San Francisco, p.~318

\bibitem[Klein et al.\ (1996)]{1996A&A...313..417K}
   {\sc Klein U., Vigotti M., Gregorini L., Reuter H.-P., Mack K.-H., \& Fanti R.}
   1996, {\it A\&A} {\bf 313}, 417--422

\bibitem[Kohoutek (1997)]{1997AN....318...35K}
   {\sc Kohoutek, L.} 1997, {\it AN} {\bf 318}, 35--44

\bibitem[Kollgaard et al.\ (1994)]{1994ApJS...93..145K}
    {\sc Kollgaard, R. I., Brinkmann, W., McMath Chester, M.,
   Feigelson, E. D., Hertz, P., Reich, P., \& Wielebinski, R.} 1994,
   {\it ApJS} {\bf 93}, 145--159

\bibitem[Kouwenhoven et al.\ (1996)]{1996ASPC..105...15K}
    {\sc Kouwenhoven, M., Berger, M., Deich, W., \& de\,Bruyn, G.} 1996,
   Pulsars: Problems \& Progress, eds.\ S.~Johnston, M.\,A.~Walker, \& M.~Bailes,
   ASP Conf.\ Ser.\ {\bf 105}, 15--16, ASP, San Francisco

\bibitem[Kraan-Korteweg et al.\ (1994)]{1994Natur.372...77K}
    {\sc Kraan-Korteweg, R. C., Loan, A. J., Burton, W. B., Lahav, O.,
    Ferguson, H. C., Henning, P. A., \& Lynden-Bell, D.} 1994,
    {\it Nature} {\bf 372}, 77--79

\bibitem[Kraan-Korteweg et al.\ (1997)]{1997PASA...14...15K}
    {\sc Kraan-Korteweg, R. C., Woudt, P. A., \& Henning, P. A.} 1997,
   {\it Proc.~ASA} {\bf 14}, 15--20

\bibitem[Kuchar \& Clark (1997)]{1997ApJ...488..224K}
    {\sc Kuchar, T. A., \& Clark, F. O.} 1997,
   {\it ApJ} {\bf 488}, 224--233

\bibitem[K\"uhr et al.\ (1979)]{kuhr79}
    {\sc K\"uhr, H., Nauber, U., Pauliny-Toth, I. I. K., \& Witzel, A.}
   1979, A Catalogue of Radio Sources, MPIfR preprint no.~55

\bibitem[K\"uhr et al.\ (1981)]{1981A&AS...45..367K}
    {\sc K\"uhr, H., Witzel, A., Pauliny-Toth, I. I. K., \& Nauber, U.}
   1981, A\&AS, {\bf 45}, 367

\bibitem[Kurtz et al.\ (1994)]{1994ApJS...91..659K}
    {\sc Kurtz, S., Churchwell, E., \& Wood, D. O. S.}
   1994, {\it ApJS} {\bf 91}, 659--712

\bibitem[Landecker et al.\ (1990)]{1990A&A...232..207L}
    {\sc Landecker, T. L., Clutton-Brock, M., \& Purton, C. R.}
  1990, {\it A\&A} {\bf 232}, 207--214

\bibitem[Landecker et al.\ (1992)]{1992A&A...258..495L}
    {\sc Landecker, T. L., Anderson, M. D., Routledge, D.,
   \& Vaneldik, J. F.} 1992, {\it A\&A} {\bf 258}, 495--506

\bibitem[Langer et al.\ (1995)]{1995ApJ...453..293L}
    {\sc Langer, W. D., Velusamy, T., Kuiper, T. B. H.,
   Levin, S., Olsen, E., \& Migenes, V.} 1995,
   {\it ApJ} {\bf 453}, 293--307

\bibitem[Langston et al.\ (1990)]{1990ApJS...72..621L}
    {\sc Langston, G. I., Heflin, M. B., Conner, S. R.,
  Leh\'ar J., Carrilli, C. L., \& Burke, B. F.} 1990,
  {\it ApJS} {\bf 72}, 621--691

\bibitem[Large et al.\ (1981)]{1981MNRAS.194..693L}
    {\sc Large, M. I., Mills, B. Y., Little, A. G., Crawford, D. F.,
  \& Sutton, J. M.} 1981, {\it MNRAS} {\bf 194}, 693--704

\bibitem[Large et al.\ (1991)]{1991Obs...111...72L}
    {\sc Large, M. I., Cram, L. E., \& Burgess, A. M.} 1991,
  {\it The Observatory} {\bf 111}, 72--75

\bibitem[Larionov (1991)]{1991SoSAO..68...14L}
    {\sc Larionov, M. G.} 1991,
   {\it Soobshch.\ Spets.\ Astrof.\ Obs.} {\bf 68}, 14--46

\bibitem[Laurent-Muehleisen et al.\ (1997)]{1997A&AS..122..235L}
    {\sc Laurent-Muehleisen, S. A., Kollgaard, R. I., Ryan, P. J.,
    Feigelson, E. D., Brinkmann, W., \& Siebert J.} 1997, {\it A\&AS} {\bf 122}, 235--247

\bibitem[Lawrence et al.\ (1983)]{1983ApJS...51...67L}
    {\sc Lawrence, C. R., Bennett, C. L., Garcia-Barreto, J. A.,
   Greenfield, P. E., \& Burke, B. F.} 1983, {\it ApJS} {\bf 51}, 67--114

\bibitem[Leahy (1993)]{leahy93}
    {\sc Leahy, J. P.} 1993, in Jets in Extragalactic Radio Sources,
   Lecture Notes in Physics {\bf 421}, p.\ 1--13,
   eds.\ H.-J.~R\"oser \& K.~Meisenheimer, Springer Verlag, Berlin

\bibitem[Leahy et al.\ (1998)]{leahy98}
    {\sc Leahy, J. P., Bridle, A. H., \& Strom, R. G. (eds.)} 1998,
   An Atlas of DRAGNs, see URL~ {\tts www.jb.man.ac.uk/atlas/}

\bibitem[Ledlow et al.\ (1998)]{1998ApJ...495..227L}
    {\sc Ledlow, M. J., Owen, F. N., \& Keel, W. C.} 1998,
   {\it ApJ} {\bf 495}, 227--238

\bibitem[Legg (1998)]{1998A&AS..130..369L}
   {\sc Legg, T. H.} 1998, {\it A\&AS} {\bf 130}, 369--379

\bibitem[Lewis (1995)]{lewis95}
    {\sc Lewis, J. W.} 1995, Astronomical Data Analysis Software and Systems -- IV,
    eds.\ R.\,A.~Shaw, H.\,E. Payne, \& J.\,J.\,E.~Hayes,
   ASP Conf.\ Ser.\ {\bf 77}, p.\ 327, ASP, San Francisco

\bibitem[Liang \& Birkinshaw (1998)]{liang98}
    {\sc Liang, H. \& Birkinshaw, M.} 1998, Proc.\ ESO/ATNF Workshop
    ``Looking Deep in the Southern Sky'', Sydney, Dec.~1997,
   eds.\ R.~Morganti \& W.~Couch, Springer-Verlag, Berlin, in press

\bibitem[Lilley \& Palmer (1968)]{1968ApJS...16..143L}
    {\sc Lilley, A. E., \& Palmer, P.} 1968, {\it ApJS} {\bf 16}, 143

\bibitem[Lorimer et al.\ (1998)]{1998A&AS..128..541L}
    {\sc Lorimer, D. R., Jessner, A., Seiradakis, J. H., Lyne, A. G.,
   D'Amico, N., Athanasopoulos, A., Xilouris, K. M.,
   Kramer, M., \& Wielebinski, R.} 1998, {\it A\&AS} {\bf 128}, 541--544

\bibitem[Lortet et al.\ (1994)]{lortet94}
  {\sc Lortet, M.-C., Borde, S., \& Ochsenbein, F.} 1994,
  {\it A\&AS} {\bf 107}, 193--218

\bibitem[Lovas (1992)]{lovas92}
    {\sc Lovas, F. J.} 1992, {\it J.\ Phys.\ Chem.\ Ref.\ Data} {\bf 21}, 181

\bibitem[Lyne et al.\ (1998)]{1998MNRAS.295..743L}
    {\sc Lyne, A. G., Manchester, R. N., Lorimer, D. R.,
   Bailes, M., D'Amico, N., Tauris, T. M., Johnston, S.,
   Bell, J. F., Nicastro, L.} 1998, {\it MNRAS} {\bf 295}, 743--755

\bibitem[Machalski (1978)]{1978AcA....28..367M}
    {\sc Machalski, J.} 1978, {\it Acta Astron.} {\bf 28}, 367--440

\bibitem[Mack et al.\ (1997)]{1997A&AS..123..423M}
   {\sc Mack, K.-H., Klein U., O'Dea, C. P., \& Willis A. G.}
   1997, {\it A\&AS} {\bf 123}, 423--444

\bibitem[Malofeev (1996)]{malofeev96}
    {\sc  Malofeev, V. M.} 1996, Pulsars: Problems \& Progress,
   eds.\ S.~Johnston, M.\,A.~Walker, \& M.~Bailes,
   ASP Conf.\ Ser.\ {\bf 105}, 271--277, ASP, San Francisco

\bibitem[Marsalkova (1974)]{1974Ap&SS..27....3M}
    {\sc Marsalkova, P.} 1974, {\it Ap\&SS} {\bf 27}, 3--110

\bibitem[Mart\'\i n (1998)]{1998A&AS..131...73M}
 {\sc Mart\'\i n, M. C.} 1998, {\it A\&AS} {\bf 131}, 73--76

\bibitem[Maslowski (1972)]{1972AcA....22..227M}
    {\sc Maslowski, J.} 1972, {\it Acta Astron.} {\bf 22}, 227--260

\bibitem[Mathewson \& Ford (1996)]{1996ApJS..107...97M}
    {\sc Mathewson, D. S., \& Ford, V. L.} 1996, {\it ApJS} {\bf 107}, 97--102

\bibitem[McAdam (1991)]{1991PASAu...9..255M}
    {\sc McAdam, W. B.} 1991, {\it Proc.~ASA} {\bf 9}, 255--256

\bibitem[McGilchrist et al.\ (1990)]{1990MNRAS.246..110M}
    {\sc McGilchrist, M. M., Baldwin, J. E., Riley, J. M.,
   Titterington, D. J., Waldram, E. M., \& Warner, P. J.}
  1990, {\it MNRAS} {\bf 246}, 110--122

\bibitem[McKay \& McKay (1998)]{vri98}
    {\sc McKay, N. P. F. \& McKay, D. J.} 1998, in
   {\it ADASS VII}, ASP Conf.\ Ser.\ {\bf 145}, p.~240
  ({\tts www.stsci.edu/stsci/meetings/adassVII})

\bibitem[Miley et al.\ (1972)]{1972Natur.237..269M}
    {\sc Miley, G. K., Perola, G. C., van der Kruit, P. C., \& van der Laan, H.}
   1972, {\it Nature} {\bf 237}, 269

\bibitem[Miley (1980)]{1980ARA&A..18..165M}
    {\sc Miley, G. K.} 1980, {\it ARA\&A} {\bf 18}, 165

\bibitem[Moran et al.\ (1996)]{1996ApJ...461..127M}
    {\sc Moran, E. C., Helfand, D. J., Becker, R. H.,
   \& White, R. L.} 1996, {\it ApJ} {\bf 461}, 127--145

\bibitem[Nicholls (1987)]{zipflaw}
    {\sc Nicholls, P. N.} 1987, {\it J.\,Amer.\,Soc.\, Information Science} {\bf 38}, 443

\bibitem[Nishiyama \& Nakai (1998)]{1997IAUS..184E.132N}
    {\sc Nishiyama, K., \& Nakai, N.} 1998, {\it IAU Symp.} {\bf 184},
   in press; ADS {\tts 1997IAUS..184E.132N}

\bibitem[Nodland \& Ralston (1997)]{1997PhRvL..78.3043N}
    {\sc Nodland, B., \& Ralston, J. P.} 1997,
   {\it Phys.\ Rev.\ Lett.}  {\bf 78}, 3043; {\tts astro-ph/9704196}

\bibitem[Noordam \& de\,Bruyn (1982)]{1982Natur.299..597N}
    {\sc Noordam, J. E., \& de\,Bruyn. A. G.} 1982, {\it Nature} {\bf 299}, 597--600

\bibitem[Normandeau et al.\ (1992)]{1992A&AS...92...63N}
    {\sc Normandeau, M., Joncas, G., \& Green, D. A.} 1992,
  {\it A\&AS} {\bf 92}, 63--83

\bibitem[Normandeau et al.\ (1996)]{1996Natur.380..687N}
    {\sc Normandeau, M., Taylor, A. R., Dewdney, P. E.} 1996,
   {\it Nature} {\bf 380}, 687--689

\bibitem[Norris (1998)]{norris98}
    {\sc Norris, R. P.} 1998, Proc.\ ESO/ATNF Workshop
   ``Looking Deep in the Southern Sky'', Sydney, Dec.~1997,
   eds.\ R.~Morganti \& W.~Couch, Springer-Verlag, Berlin, in press

\bibitem[Notni \& Fr\"ohlich (1975)]{1975AN....296..197N}
    {\sc Notni, P., \& Fr\"ohlich, H.-E.} 1975,
  {\it AN} {\bf 296}, 197

\bibitem[O'Dea (1998)]{1998PASP..110..493O}
    {\sc O'Dea, C. P.} 1998, {\it PASP} {\bf 110}, 493--532

\bibitem[Oort et al.\ (1958)]{1958MNRAS.118..379O}
    {\sc Oort, J. H., Kerr, F. J., \& Westerhout, G.} 1958,
  {\it MNRAS} {\bf 118}, 379

\bibitem[Oort \& van\,Langevelde (1987)]{1987A&AS...71...25O}
   {\sc Oort, M. J. A., \& van\,Langevelde, H. J.}
  1987, {\it A\&AS} {\bf 71}, 25--38

\bibitem[Otrupcek \& Wright (1991)]{1991PASAu...9..170O}
    {\sc Otrupcek, R. E., \& Wright, A. E.} 1991,
   {\it Proc.~ASA} {\bf 9}, 170 
   (PKSCAT90; {\tts ftp://ftp.atnf.csiro.au/pub/data/pkscat90/})

\bibitem[Ott et al.\ (1994)]{1994A&A...284..331O}
    {\sc Ott, M., Witzel, A., Quirrenbach, A., Krichbaum, T. P., Standke, K. J.,
    Schalinski, C. J., \& Hummel, C. A.} 1994, {\it A\&A} {\bf 284}, 331--339

\bibitem[Pacholczyk (1970)]{pach70}
    {\sc Pacholczyk, A. G.} 1970, {\it Radio Astrophysics},
   W.H.\,Freeman \& Co., San Francisco

\bibitem[Pacholczyk (1977)]{pach77}
    {\sc Pacholczyk, A. G.} 1977, {\it Radio Galaxies},
   Pergamon Press, Oxford

\bibitem[Padrielli \& Conway (1977)]{1977A&AS...27..171P}
   {\sc Padrielli, L., \& Conway, R. G.}, 1977,
   {\it A\&AS} {\bf 27}, 171--180

\bibitem[Parijskij et al.\ (1991)]{1991A&AS...87....1P}
    {\sc Parijskij, Yu. N., Bursov, N. N., Lipovka, N. M.,
   Soboleva, N. S., \& Temirova, A. V.} 1991, {\it A\&AS} {\bf 87}, 1--32

\bibitem[Parijskij et al.\ (1992)]{1992A&AS...96..583P}
    {\sc Parijskij, Yu. N., Bursov, N. N., Lipovka, N. M.,
   Soboleva, N. S., Temirova, A. V., \& Chepurnov, A. V.} 1992,
  {\it A\&AS} {\bf 96}, 583--592

\bibitem[Parijskij et al.\ (1993)]{1993A&AS...98..391P}
    {\sc Parijskij, Yu. M., Bursov, N. M., Lipovka, N. M.,
    Soboleva, M. S., Temirova, A. V., \& Chepurnov, A. V.} 1993,
   {\it A\&AS} {\bf 98}, 391

\bibitem[Patnaik et al.\ (1992)]{1992MNRAS.254..655P}
    {\sc Patnaik, A. R., Browne, I. W. A., Wilkinson, P. N.,
   \& Wrobel, J. M.} 1992, {\it MNRAS} {\bf 254}, 655--676

\bibitem[Pearson \& Readhead (1988)]{1988ApJ...328..114P}
   {\sc Pearson, T. J., \& Readhead, A. C. S.} 1988, {\it ApJ}
      {\bf 328}, 114--142

\bibitem[Peng \& Nan (1998)]{1998IAUS..179...93P}
    {\sc Peng, B., \& Nan, R.} 1998, {\it IAU Symp.} {\bf 179}, 93--94,
   eds.\ B.J.~McLean et al., Kluwer Acad.\ Publ., Dordrecht

\bibitem[Perley (1989)]{perley89}
    {\sc Perley, R.} 1989, in {\it Synthesis Imaging in Radio Astronomy},
   eds.\ Perley, R. A., Schwab, F. R. \& Bridle, A. H., ASP Conf.\,Ser.
  {\bf 6}, ASP, San Francisco, p.~287--313

\bibitem[Perley et al.\ (1989)]{1989ASPC....6}
    {\sc Perley, R. A., Schwab, F. R. \& Bridle, A. H. (eds.)} 1989,
   {\it Synthesis Imaging in Radio Astronomy} ASP Conf.\ Ser.\ {\bf 6}, ASP, San Francisco

\bibitem[Persic \& Salucci (1995)]{1995ApJS...99..501P}
    {\sc Persic, M., \& Salucci, P.} 1995, {\it ApJS} {\bf 99}, 501--541

\bibitem[Pilkington \& Scott (1965)]{1965MmRAS..69..183P}
    {\sc Pilkington, J. D. H., \& Scott, P. F.} 1965, {\it MmRAS} {\bf 69}, 183

\bibitem[Porcas et al.\ (1980)]{1980MNRAS.191..607P}
    {\sc Porcas, R. W., Urry, C. M., Browne, I. W. A., Cohen, A. M.,
   Daintree, E. J., \& Walsh, D.} 1980, {\it MNRAS} {\bf 191}, 607--614

\bibitem[Pound et al.\ (1997)]{pound97}
    {\sc Pound, M. W., Gruendl, R., Lada, E. A. \& Mundy, L.} 1997,
  Star Formation Near and Far, eds.\ S.~Holt \& L.~Mundy, p.~395--397

\bibitem[Prandoni et al.\ (1998)]{prandoni98}
    {\sc Prandoni, I., Gregorini, L., Parma, P., de\,Ruiter, R. H.,
    Vettolani, G., Wieringa, M. H., \& Ekers, R. D.} 1998,
   Proc.\ ESO/ATNF Workshop ``Looking Deep in the Southern Sky'',
   Sydney, Dec.~1997, eds.\ R.~Morganti \& W.~Couch,
   Springer-Verlag, Berlin, in press

\bibitem[Purton \& Durrell (1991)]{purtondurrell91}
   {\sc Purton, C., \& Durrell, P.} 1991, {\it Program SURSEARCH:
  SEARCHing radio continuum source SURveys for overlap with a given area of sky},
  FORCE software package available from C.\,R.~Purton (DRAO);
  see also {\tts cats.sao.ru/doc/SURSEARCH.html}

\bibitem[Purvis et al.\ (1987)]{1987MNRAS.229..589P}
    {\sc Purvis, A., Tappin, S. J., Rees, W. G., Hewish, A.,
  \& Duffett-Smith, P. J.} 1987, {\it MNRAS} {\bf 229}, 589--619

\bibitem[Quiniento et al.\ (1988)]{1988A&AS...76...21Q}
   {\sc Quiniento, Z. M., Cersosimo, J. C., \& Colomb, F. R.} 1988,
   {\it A\&AS} {\bf 76}, 21--34

\bibitem[Reber (1994)]{1994JRASC..88..297R}
    {\sc Reber, G.} 1994, {\it J.Roy.Astr.Soc.\ Canada}, {\bf 88}, 297--302

\bibitem[Reber (1995)]{1995Ap&SS.227...93R}
    {\sc Reber, G.} 1995, {\it Ap\&SS} {\bf 227}, 93--96

\bibitem[Rees (1990a)]{1990MNRAS.243..637R}
    {\sc Rees, N.} 1990a, {\it MNRAS} {\bf 243}, 637--639

\bibitem[Rees (1990b)]{1990MNRAS.244..233R}
    {\sc Rees, N.} 1990b, {\it MNRAS} {\bf 244}, 233--246

\bibitem[Reference Data for Radio Engineers, 1975]{refradio75}
    {\sc Reference Data for Radio Engineers} 1975,
   publ.\ by Howard W.~Sams \& Co., ITT, ISBN 0-672-21218-8, p.~1-3

\bibitem[Reich (1982)]{1982A&AS...48..219R}
   {\sc Reich, W.} 1982, {\it A\&AS} {\bf 48}, 219--297

\bibitem[Reich et al.\ (1984)]{1984A&AS...58..197R}
   {\sc Reich, W., F\"urst, E., Haslam, C. G. T.,
    Steffen, P., \& Reif, K.} 1984, {\it A\&AS} {\bf 58}, 197--199

\bibitem[Reich \& Reich (1986)]{1986A&AS...63..205R}
   {\sc Reich, P., \& Reich, W.} 1986,
   {\it A\&AS} {\bf 63}, 205--288

\bibitem[Reich \& Reich (1988)]{1988A&AS...74....7R}
    {\sc Reich, P., \& Reich, W.} 1988, {\it A\&AS} {\bf 74}, 7--23

\bibitem[Reich (1991)]{1991IAUS..144..187R}
    {\sc Reich W.} 1991, {\it IAU Symp.} {\bf 144}, 187--196,
   ed.\ H.~Bloemen, Kluwer Acad.\ Publ., Dordrecht

\bibitem[Reich et al.\ (1990)]{1990A&AS...83..539R}
    {\sc Reich, W., Reich, P., \& F\"urst E.} 1990, {\it A\&AS} {\bf 83}, 539

\bibitem[Reich et al.\ (1997)]{1997A&AS..126..413R}
    {\sc Reich, P., Reich, W., \& F\"urst E.} 1997,
   {\it A\&AS} {\bf 126}, 413--435

\bibitem[Rengelink et al.\ (1997)]{1997A&AS..124..259R}
    {\sc Rengelink, R. B., Tang, Y., De, Bruyn, A. G., Miley, G. K.,
   Bremer, M. N., R\"ottgering H. J. A., \& Bremer, M. A. R.}
   1997, {\it A\&AS} {\bf 124}, 259--280

\bibitem[Richter (1975)]{1975AN....296...65R}
    {\sc Richter, G. A.} 1975, {\it AN} {\bf 296}, 65--81

\bibitem[Righetti et al.\ (1988)]{1988A&AS...74..315R}
    {\sc Righetti, G., Giovannini, G., \& Feretti, L.}
    1988, {\it A\&AS} {\bf 74}, 315--324

\bibitem[Robertson (1992)]{robertson92}
    {\sc Robertson, P.} 1992, {\it Beyond Southern Skies\,:
   Radio Astronomy and the Parkes Telescope},
   Cambridge University Press, Cambridge, UK

\bibitem[Robson et al.\ (1993)]{1993A&A...277..314R}
   {\sc Robson, M., Yassin, G., Woan, G., Wilson, D. M. A.,
   Scott, P. F., Lasenby, A. N., Kenderdine, S., \& Duffett-Smith, P. J.}
   1993, {\it A\&A} {\bf 277}, 314--320

\bibitem[R\"ottgering et al.\ (1997)]{1997MNRAS.290..577R}
    {\sc R\"ottgering H. J. A., Wieringa, M. H., Hunstead, R. W.,
   \& Ekers, R. D.} 1997, {\it MNRAS} {\bf 290}, 577--584

\bibitem[Rohlfs \& Wilson (1996)]{rohwil96}
    {\sc Rohlfs, K. \& Wilson, T. L.} 1996 {\it Tools of Radio Astronomy},
  2nd ed., Springer Verlag, Berlin

\bibitem[Rosenberg \& Schneider (1998)]{1998AAS...192.6601R}
    {\sc Rosenberg, J. L., \& Schneider, S. E.} 1998,
    {\it BAAS} {\bf 30}, 914

\bibitem[Sadler (1998)]{sadler98}
    {\sc Sadler, E. M.} 1998, Proc.\ ESO/ATNF Workshop
   ``Looking Deep in the Southern Sky'', Sydney, Dec.~1997,
   eds.\ R.~Morganti \& W.~Couch, Springer-Verlag, Berlin, in press

\bibitem[Sakamoto et al.\ (1995)]{1995ApJS..100..125S}
    {\sc Sakamoto, S., Hasegawa, T., Hayashi, M., Handa, T., \& Oka, T.}
  1995, {\it ApJS} {\bf 100}, 125

\bibitem[Salter (1983)]{1983BASI...11....1S}
    {\sc Salter, C. J.} 1983, {\it Bull.\ Astron.\ Soc.\ India} {\bf 11}, 1--142

\bibitem[Salter \& Brown (1988)]{salter88}
    {\sc Salter, C. J., \& Brown, R. L.} 1988, in
   {\it Galactic and Extragalactic Radio Astronomy},
   eds.\ G.\,L.~Verschuur \& K.\,I.~Kellermann,
   2nd.\ edn., Springer-Verlag, Berlin

\bibitem[Sasao et al.\ (1994)]{sasao94}
    {\sc Sasao, T., Manabe, S., Kameya, O., \& Inoue, M. (eds.)} 1994,
  {\it VLBI Technology}, Terra Science Publ. Company, Tokyo

\bibitem[Schilke et al.\ (1997)]{1997ApJS..108..301S}
   {\sc Schilke, P., Groesbeck, T. D., Blake G. A., \& Phillips, T.G}
   1997, {\it ApJS} {\bf 108}, 301--337

\bibitem[Schoenmakers et al.\ (1998)]{1998A&A...VVV..pppS}
   {\sc  Schoenmakers, A. P., Mack, K.-H., Lara, L.,
   R\"ottgering, H. J. A., de\,Bruyn, A.G., van~der~Laan, H., Giovannini, G.}
   1998, {\it A\&A}, in press; {\tts astro-ph/9805356}

\bibitem[Seiradakis et al.\ (1985)]{1985A&A...143..478S}
    {\sc Seiradakis, J. H., Reich, W., Sieber, W.,
   Schlickeiser, R. \& K\"uhr, H.} 1985, {\it A\&A} {\bf 143}, 478--480

\bibitem[Shakeshaft et al.\ (1955)]{1955MmRAS..67..106S}
    {\sc Shakeshaft, J. R., Ryle, M., Baldwin, J. E.,
   Elsmore, B., \& Thomson, J. H.} 1955, {\it MmRAS} {\bf 67}, 106--154

\bibitem[Sharpless (1959)]{1959ApJS....4..257S}
    {\sc Sharpless, S.} 1959, {\it ApJS} {\bf 4}, 257

\bibitem[Sieber et al.\ (1979)]{1979A&A....74..361S}
    {\sc Sieber, W., Haslam, C. G. T., \& Salter, C. J.} 1979,
  {\it A\&A} {\bf 74}, 361--368

\bibitem[Sijbring \& de\,Bruyn (1998)]{1998A&A...331..901S}
    {\sc Sijbring, D., de\,Bruyn, A. G.} 1998,
    {\it A\&A} {\bf 331}, 901--915

\bibitem[Simard-Normandin et al.\ (1981)]{1981ApJS...45...97S}
    {\sc Simard-Normandin, M., Kronberg, P. P., \& Button, S.}
   1981, {\it ApJS} {\bf 45}, 97--111

\bibitem[Slee \& Higgins (1973)]{1973AuJPA..27....1S}
    {\sc Slee, O. B., \& Higgins, C. S.} 1973,
   {\it Aust.\ J.\ Phys.\ Ap.\ Suppl.} {\bf 27}, 1--43

\bibitem[Slee \& Higgins (1975)]{1975AuJPA..36....1S}
    {\sc Slee, O. B., \& Higgins, C. S.} 1975,
  {\it Aust.\ J.\ Phys.\ Ap.\ Suppl.} {\bf 36}, 1--60

\bibitem[Slee (1977)]{1977AuJPA..43....1S}
    {\sc Slee, O. B.} 1977, {\it Aust.\ J.\ Phys.\ Ap.\ Suppl.} {\bf 43}, 1--123

\bibitem[Slee (1995)]{1995AuJPh..48..143S}
    {\sc Slee, O. B.} 1995, {\it Aust.\ J.\ Phys.} {\bf 48}, 143--186

\bibitem[Snellen et al.\ (1996)]{snellen96}
    {\sc Snellen, I. A. G., Schilizzi, R. T., R\"ottgering, H. J. A., \& 
   Bremer, M. N. (eds.)} 1996, 
  {\it Second Workshop on GPS and CSS Radio Sources}, Leiden Observatory

\bibitem[Sovers et al.\ (1998)]{sovers98}
    {\sc Sovers, O. J., Fanselow, J. L., \& Jacobs, C. S.} 1998,
   {\it Rev.~Mod.~Phys.}, {\bf 70}, in press (Oct. issue),
  {\tts astro-ph/9712238}

\bibitem[Stark et al.\ (1990)]{stark90}
    {\sc Stark, A. A., et al.} 1990,
   The Bell Laboratories H\,I Survey, ``Preliminary Draft'',
   available as ADC/CDS catalogue \#\,8010

\bibitem[Stark et al.\ (1998)]{stark98}
    {\sc Stark, A. A., Carlstrom, J. E., Israel, F. P., Menten, K. M.,
  Peterson, J. B., Phillips, T. G., Sironi, G., \& Walker, C. K.} 1998,
  SPIE, in press ({\tts astro-ph/9802326})

\bibitem[Staveley-Smith (1997)]{staveley97}
    {\sc Staveley-Smith, L.} 1997, {\it Proc.~ASA} {\bf 14}, 111--116

\bibitem[Strong et al.\ (1982)]{1982MNRAS.201..495S}
    {\sc Strong, A. W., Riley, P. A., Osborne, J. L., \& Murray, J. D.}
   1982, {\it MNRAS} {\bf 201}, 495--501

\bibitem[Sullivan III (1982)]{sulliv82}
    {\sc Sullivan III, W. T.} 1982,
   {\it Classics in Radio Astronomy}, Reidel, Dordrecht, The Netherlands

\bibitem[Sullivan III (1984)]{sulliv84}
    {\sc Sullivan III, W. T. (ed.)} 1984,
   {\it The early years of radio astronomy: reflections
  fifty years after Jansky's discovery},
   Cambridge University Press, Cambridge, UK

\bibitem[Sutton et al.\ (1985)]{1985ApJS...58..341S}
  {\sc Sutton, E. C., Blake G. A., Masson, C. R., \& Phillips T.G}
   1985, {\it ApJS} {\bf 58}, 341--378

\bibitem[Tabara \& Inoue (1980)]{1980A&AS...39..379T}
    {\sc Tabara, H., \& Inoue, M.} 1980, {\it A\&AS} {\bf 39}, 379--393

\bibitem[Tashiro et al.\ (1998)]{1998ApJ...499..713T}
    {\sc Tashiro, M., Kaneda, H., Makishima, K., Iyomoto, N., Idesawa, E.,
    Ishisaki, Y., Kotani, T., Takahashi, T., \& Yamashita, A.} 1998,
  {\it ApJ} {\bf 499}, 713--718

\bibitem[Tateyama et al.\ (1998)]{1998ApJ...500..810T}
    {\sc Tateyama, C. E., Kingham, K. A., Kaufmann, P. Piner, B. G.,
   De Lucena, A. M. P., \& Botti, L. C. L.} 1998, {\it ApJ} {\bf 500}, 810--815

\bibitem[Tasker \& Wright (1993)]{1993PASAu..10..320T}
    {\sc Tasker, N., \& Wright, A.} 1993, {\it Proc.~ASA} {\bf 10}, 320

\bibitem[Taylor et al.\ (1993)]{1993ApJS...88..529T}
    {\sc Taylor, J. H., Manchester, R. N., \& Lyne, A. G.} 1993,
  {\it ApJS} {\bf 88}, 529--568

\bibitem[Taylor et al.\ (1996)]{1996ApJS..107..239T}
    {\sc Taylor, A. R., Goss, W. M., Coleman, P. H., Van Leeuwen, J.,
   \& Wallace, B. J.} 1996, {\it ApJS} {\bf 107}, 239--254

\bibitem[Tegmark \& De Oliveira-Costa (1998)]{1998ApJ...500L..83T}
    {\sc Tegmark, M., \& De Oliveira-Costa, A.} 1998,
    {\it ApJ} {\bf 500}, L83--L86

\bibitem[Towle et al.\ (1996)]{1996ApJS..107..747T}
  {\sc Towle, J. P., Feldman, P. A., \& Watson, J. K. G.}
   1996, {\it ApJS} {\bf 107}, 747--760

\bibitem[Trimble \& McFadden (1998)]{1998PASP..110..223T}
    {\sc Trimble, V., \& McFadden, L.-A.} 1998, {\it PASP} {\bf 110}, 223--267

\bibitem[Trushkin (1996)]{1996A&ATr..11..225T}
    {\sc Trushkin, S. A.} 1996, {\it A\&ATr} {\bf 11}, 225--233

\bibitem[Trushkin (1996)]{1996BSAO...41...64T}
    {\sc Trushkin, S. A.} 1997, {\it Bull.\ SAO} {\bf 41}, p. 64--79

\bibitem[Trushkin (1998)]{trushkin98}
    {\sc Trushkin, S. A.} 1998, SAO Preprint, N131, 30 pp.; ({\tts ftp://cats.sao.ru/SNR\_spectra/})

\bibitem[Turner (1989)]{1989ApJS...70..539T}
    {\sc Turner, B. E.} 1989, {\it ApJS} {\bf 70}, 539--622

\bibitem[Ulvestad \& Linfield (1998)]{1998ASPC..144..397U}
  {\sc Ulvestad, J. S., \& Linfield, R. P.} 1998, in IAU Coll.\ 164, {\it
 Radio Emission from Galactic and Extragalactic Compact Sources}, ASP
 Conf.\ Ser.\ {\bf 144}, 397--398, eds.\ J.\,A.~Zensus, G.\,B.~Taylor, \&
 J.\,M.~Wrobel, ASP, San Francisco.

\bibitem[Uyan{\i}ker et al.\ (1998)]{uyaniker98}
    {\sc Uyan{\i}ker, B., F\"urst, E., Reich, W., Reich, P., \& Wielebinski, R.} 1998, {\it A\&A},
 in press, {\tts astro-ph/9807013}

\bibitem[Vanden Bout (1998)]{vandenbout98}
    {\sc Vanden Bout, P. A.} 1998, SPIE Proc.~3357, NRAO preprint 98/038

\bibitem[Verkhodanov et al.\ (1997)]{verkho97}
    {\sc Verkhodanov, O.V., Trushkin, S.A., Andernach, H., \&
   Chernenkov, V.N.} 1997, ASP~Conf.\,Ser., {\bf 125}, 322--325,
  eds.~G.\,Hunt \& H.E.\,Payne, ASP, San Francisco ({\tts astro-ph/9610262})

\bibitem[V\'eron-Cetty \& V\'eron (1983)]{1983A&AS...53..219V}
    {\sc V\'eron-Cetty, M. P., \& V\'eron, P.} 1983, {\it A\&AS}, {\bf 53}, 219;
    ADC/CDS catalogue \#\,7054

\bibitem[V\'eron-Cetty \& V\'eron (1989)]{1989ESOSR...7....1V}
   {\sc V\'eron-Cetty, M. P., \& V\'eron, P.} 1989,
   {\it Catalogue of Quasars and Active Galactic Nuclei}, 4$^{\rm th}$ Edition,
{\it ESO Sci. Rep.} {\bf 7}; ADC/CDS catalogue \#\,7126

\bibitem[V\'eron-Cetty \& V\'eron (1998)]{1998ESOSR..18....1V}
    {\sc V\'eron-Cetty, M. P., \& V\'eron, P.} 1998,
   {\it Catalogue of Quasars and Active Galactic Nuclei}, 8$^{\rm th}$ Edition,
 {\it ESO Sci. Rep.} {\bf 18}; ADC/CDS catalogue \#\,7207

\bibitem[Verschuur \& Kellermann (1988)]{verkel88}
    {\sc Verschuur, G. L. \& Kellermann, K. I.} (eds.) 1988,
   {\it Galactic and Extragalactic Radio Astronomy}, 2nd. edn., Springer-Verlag, Berlin

\bibitem[Verter (1985)]{1985ApJS...57..261V}
   {\sc Verter, F.} 1985, {\it ApJS} {\bf 57}, 261--285, ADC/CDS \#\,7064

\bibitem[Vessey \& Green (1998)]{1998MNRAS.294..607V}
    {\sc Vessey, S. J., \& Green, D. A.} 1998, {\it MNRAS} {\bf 294}, 607--614

\bibitem[Vettolani et al.\ (1998)]{1998A&AS..130..323V}
   {\sc Vettolani, G., Zucca, E., Merighi, R., Mignoli, M., Proust, D.,
   Zamorani, G., Cappi, A., Guzzo, L., Maccagni, D., Ramella, M.,
   Stirpe, G. M., Blanchard, A., Cayatte, V., Collins, C.,
   MacGillivray, H., Maurogordato, S., Scaramella, R., Balkowski, C.,
   Chincarini, G., \& Felenbok, P.} 1998, {\it A\&AS} {\bf 130}, 323--332

\bibitem[Visser et al.\ (1995)]{1995A&AS..110..419V}
    {\sc Visser, A. E., Riley, J. M., R\"ottgering H. J. A., \& Waldram, E. M.} 1995,
   {\it A\&AS} {\bf 110}, 419--439

\bibitem[Waldram et al.\ (1996)]{1996MNRAS.282..779W}
    {\sc Waldram, E. M., Yates, J. A., Riley, J. M., \& Warner, P. J.} 1996,
   {\it MNRAS} {\bf 282}, 779--787, erratum in {\it MNRAS} {\bf 284}, 1007

\bibitem[Walsh et al.\ (1997)]{1997MNRAS.291..261W}
    {\sc Walsh, A. J., Hyland, A. R., Robinson, G., \& Burton, M. G.} 1997,
   {\it MNRAS} {\bf 291}, 261--278

\bibitem[Walterbos et al.\ (1985)]{1985A&AS...61..451W}
   {\sc Walterbos, R. A. M., Brinks, E., \& Shane, W. W.}
   1985, {\it A\&AS} {\bf 61}, 451--471

\bibitem[Wardle et al.\ (1997)]{1997PhRvL..79.1801W}
    {\sc Wardle, J. F. C., Perley, R. A., \& Cohen, M. H.} 1997,
   {\it Phys.\ Rev.\ Lett.} {\bf 79}, 1801

\bibitem[Welch et al.\ (1996)]{1996PASP..108...93W}
    {\sc Welch, W. J., Thornton, D. D., Plambeck, R. L., Wright, M. C. H.,
 Lugten, J., Urry, L., Fleming, M., Hoffman, W., Hudson, J.,
  Lum, W. T., Forster, J. R., Thatte, N., Zhang, X., Zivanovic, S., Snyder, L.,
  Crutcher, R., Lo, K. Y., Wakker, B., Stupar, M., Sault, R., Miao, Y.,
  Rao, R., Wan, K., Dickel, H. R., Blitz, L., Vogel, S. N., Mundy, L.,
  Erickson, W., Teuben, P. J., Morgan, J., Helfer, T., Looney, L., De~Gues, E.,
 Grossman, A., Howe, J. E., Pound, M., \& Regan, R.} 1996, {\it PASP} {\bf 108}, 93--103

\bibitem[Wendker (1995)]{1995A&AS..109..177W}
   {\sc Wendker, H. J.} 1995, {\it A\&AS} {\bf 109}, 177--179

\bibitem[Westerhout (1957)]{1957BAN....13..201W}
    {\sc Westerhout, G.} 1957, {\it Bull. Astron.\ Inst.\ Netherlands} {\bf 13}, 201--246

\bibitem[Westerhout et al.\ (1982)]{1982A&AS...49..137W}
   {\sc Westerhout, G., Mader, G. L. \& Harten, R. H.} 1982,
   {\it A\&AS} {\bf 49}, 137--141

\bibitem[White (1984)]{1984PASAu...5..290W}
    {\sc White, G. L.} 1984, {\it Proc.~ASA} {\bf 5}, 290--340

\bibitem[White \& Becker (1992)]{1992ApJS...79..331W}
    {\sc White, R. L., \& Becker, R. H.} 1992, {\it ApJS} {\bf 79}, 331--467

\bibitem[White et al.\ (1997)]{1997ApJ...475..479W}
    {\sc White, R. L., Becker, R. H., Helfand, D. J., \& Gregg, M. D.}
   1997, {\it ApJ} {\bf 475}, 479--493;  see also {\tts http://sundog.stsci.edu}

\bibitem[Wieringa (1991)]{wieringa91}
   {\sc Wieringa, M. H.} 1991, Ph.\,D.\ thesis, Leiden University

\bibitem[Wieringa (1993a)]{1993BICDS..43...17W}
    {\sc  Wieringa, M. H.} 1993a, {\it Bull.\ Inf.\ CDS} {\bf 43}, 17;
    and PhD Thesis, Leiden Univ. (1991)

\bibitem[Wieringa et al.\ (1993b)]{1993A&A...268..215W}
    {\sc Wieringa, M. H., de\,Bruyn, A. G., Jansen, D.,
    Brouw, W. N., \& Katgert P.} 1993b, {\it A\&A} {\bf 268}, 215--229

\bibitem[Wilkinson (1998)]{wilkinson98}
    {\sc Wilkinson D.} 1998, {\it Proc.\ Natl.\ Acad.\ Sci.\ USA}
    {\bf 95}, 29--34; ({\tts www.pnas.org/all.shtml})

\bibitem[Wills (1975)]{1975AuJPA..38....1W}
    {\sc Wills, B. J.} 1975, {\it Aust.\ J.\ Phys.\ Ap.\ Suppl.} {\bf 38}, 1--65

\bibitem[Wood \& Churchwell (1989)]{1989ApJ...340..265W}
   {\sc Wood, D. O. S., \& Churchwell, E.}
   1989, {\it ApJ} {\bf 340}, 265--272

\bibitem[Wright et al.\ (1994)]{1994ApJS...91..111W}
    {\sc Wright, A. E., Griffith, M. R., Burke, B. F., \& Ekers, R. D.}
   1994, {\it ApJS} {\bf 91}, 111--308

\bibitem[Wright et al.\ (1996)]{1996ApJS..103..145W}
    {\sc Wright, A. E., Griffith, M. R., Hunt, A. J., Troup, E.,
   Burke, B. F., \& Ekers, R. D.} 1996, {\it ApJS} {\bf 103}, 145--172

\bibitem[Xu et al.\ (1995)]{1995ApJS...99..297X}
  {\sc Xu, W., Readhead, A. C. S., Pearson, T. J., Polatidis, A. G., \&
     Wilkinson, P. N.} 1995, {\it ApJS}, {\bf 99}, 297--348

\bibitem[Young et al.\ (1995)]{1995ApJS...98..219Y}
    {\sc Young, J. S., Xie, S., Tacconi, L., Knezek, P., Viscuso, P.,
   Tacconi-Garman, L., Scoville, N., Schneider, S., Schloerb, F. P.,
  Lord, S., Lesser, A., Kenney, J., Huang, Y.-L., Devereux, N., Claussen, M.,
  Case, J., Carpenter, J., Berry, M., \& Allen, L.} 1995,
  {\it ApJS} {\bf 98}, 219--257

\bibitem[Zensus et al.\ (1995)]{1995ASPC...82}
    {\sc Zensus, J. A., Diamond, P. J. \& Napier, P. J. (eds.)} 1995,
   {\it Very Long Baseline Interferometry and the VLBA},
   ASP Conf.\ Ser.\ {\bf 82}, ASP, San Francisco; ({\tts www.cv.nrao.edu/vlbabook})

\bibitem[Zensus et al.\ (1998)]{1998ASPC..144}
   {\sc Zensus, J. A., Taylor, G. B., \& Wrobel, J. M. (eds.)} 1998,
   {\it Radio Emission from Galactic and Extragalactic Compact Sources},
   Proc.\ IAU Coll.\ 164, ASP Conf.\ Ser.\ {\bf 144}, ASP, San Francisco;
  ({\tts http://www.cv.nrao.edu/iau164/164book.html})

\bibitem[Zhang et al.\ (1997)]{1997A&AS..121...59Z}
    {\sc Zhang, X., Zheng, Y., Chen, H., Wang, S., Cao, A.,
    Peng, B., \& Nan, R.} 1997, {\it A\&AS} {\bf 121}, 59--63

\bibitem[Zoonematkermani et al.\ (1990)]{1990ApJS...74..181Z}
    {\sc Zoonematkermani, S., Helfand, D. J., Becker, R. H.,
   White, R. L., \& Perley, R. A.} 1990, {\it ApJS} {\bf 74}, 181--224

\end{thebibliography}
\end{document}